\documentclass[aps,pre,superscriptaddress]{revtex4}
\usepackage[dvips]{graphics}
\usepackage{graphicx}
\usepackage{amsfonts}
\usepackage{amssymb}
\usepackage{amsmath}
\usepackage{subfigure}

\begin{document}

\title{Two-Component Nonlinear Schr{\"{o}}dinger Models with a Double-Well
Potential}
\author{C. Wang, P. G.\ Kevrekidis, N. Whitaker}
\affiliation{Department of Mathematics and Statistics, University of Massachusetts,
Amherst MA 01003-4515, USA}
\author{B. A.\ Malomed}
\affiliation{Department of Physical Electronics, Faculty of Engineering, Tel Aviv
University, Tel Aviv 69978, Israel}

\begin{abstract}
We introduce a model motivated by studies of Bose-Einstein condensates
(BECs) trapped in double-well potentials. We assume that a mixture of two
hyperfine states of the same atomic species is loaded in such a trap.The
analysis is focused on symmetry-breaking bifurcations in the system,
starting at the linear limit and gradually increasing the nonlinearity.
Depending on values of the chemical potentials of the two species, we find
numerous states, as well as symmetry-breaking bifurcations, in addition to
those known in the single-component setting. These branches, which include
all relevant stationary solutions of the problem, are predicted analytically
by means of a two-mode approximation, and confirmed numerically. For
unstable branches, outcomes of the instability development are explored in
direct simulations.
\end{abstract}

\maketitle


\section{Introduction}

The nonlinear Schr{\"{o}}dinger (NLS) equation is a ubiquitous partial
differential equation (PDE) with a broad spectrum of applications, including
nonlinear optics in the temporal and spatial domains, matter waves, plasmas,
water waves, and some biophysical models \cite{NLS,trubatch}. One of the
most fundamental applications of the NLS equation stems from its direct
relevance as an accurate mean-field model (known as the Gross-Pitaevskii
equation, GPE, in that context) of Bose-Einstein Condensates (BECs) \cite%
{book1,book2}. The GPE usually includes an external potential accounting for
the magnetic, optical or combined confinement of dilute alkali vapors that
constitute the BEC \cite{reviewsbec}. Basic types of such trapping
potentials include parabolic and spatially periodic ones (the latter is
created, as an \textit{optical lattice}, by the interference of
counter-propagating laser beams). The NLS equations with similar potentials
are also relevant models for optical beams in graded-index waveguides and
periodic waveguiding arrays \cite{kivshar,reviewsopt}.

A setting that has recently drawn much interest in the context of BECs is
based on a double-well potential (DWP),\ which originates from a combination
of the two above-mentioned types of potentials. It was realized
experimentally in \cite{markus1}, leading to particularly interesting
phenomena including tunneling and Josephson oscillations (for a small number
of atoms), or macroscopic quantum self-trapping, with an asymmetric division
of atoms between the two wells, for a large number of atoms. Numerous
theoretical studies of DWP\ settings have been performed in parallel to the
experimental work \cite%
{smerzi,kiv2,mahmud,bam,Bergeman_2mode,infeld,todd,theo,carr}. They
addressed finite-mode reductions, analytical results for specially designed
shapes of the potential, quantum effects, and other aspects of the theory
(in particular, tunneling between vortex and antivortex states in BEC
trapped in a two-dimensional (2D) anisotropic harmonic potential \cite%
{Pethik} belongs to the latter category).

An interesting generalization of the DWP is provided by multi-well
potentials. In particular, nonequilibrium BEC states and generation of
quantum entanglement in such settings were recently studied in detail
theoretically in \cite{Yuk}.

Also considered were 2D \cite{2DArik,2DMichal} and 3D extensions of DWP
settings, which add one or two extra dimensions to the model, either without
any additional potential, or with an independent optical lattice acting in
these directions. The extended geometry makes it possible to consider
solitons, self-trapped in the extra dimension(s), which therefore emerge as
effectively 1D \cite{2DArik,2DMichal} or 2D \cite{3D} dual-core states. The
solitons may be symmetric and antisymmetric, as well as ones breaking their
symmetry between the wells through bifurcations. These states may be
described by simplified systems of linearly coupled 1D \cite{2DArik} or 2D
\cite{2DArik,3D} PDEs, or, in a more accurate form, effectively 1D solitons
can be found as solutions to the full two-dimensional PDE, that takes into
regard a particular form of the DWP (which depends on the transverse
coordinate and is uniform in the longitudinal direction, allowing the
solitons to self-trap in that free direction) \cite{2DMichal}.

DWPs are also relevant to nonlinear-optics settings, such as the twin-core
self-guided laser beams in Kerr media \cite{HaeltermannPRL02}, optically
induced dual-core waveguiding structures in a photorefractive crystal \cite%
{zhigang}, and DWP configurations for trapped light beams in a structured
annular core of an optical fiber \cite{Longhi}. As concerns DWPs with extra
dimensions, their counterparts in fiber optics are twin--core fibers \cite%
{dual-core-fiber} and fiber Bragg gratings \cite{dual-core-FBG} that were
shown to support symmetric and asymmetric solitons. In addition, also
investigated was the symmetry breaking of solitons in models of twin-core
optical waveguides with non-Kerr nonlinear terms, \textit{viz}, quadratic
\cite{dual-core-quadratic} and cubic-quintic (CQ) \cite{dual-core-CQ}. All
these optical model are based on systems of linearly coupled 1D PDEs. Also
in the context of nonlinear optics, a relevant model is based on a system of
linearly coupled complex Ginzburg-Landau equations with the CQ nonlinearity,
that gives rise to stable dissipative solitons with broken symmetry \cite%
{Ariel}.

In addition to the linearly coupled systems of two nonlinear PDEs motivated
by the DWP transverse structures, several models of triangular
configurations, which admit their own specific modes of the symmetry
breaking, have also been introduced in optics, each model based on a system
of three nonlinear PDEs with symmetric linear couplings between them. These
include tri-core nonlinear fibers \cite{Buryak} and fiber Bragg gratings
\cite{Arik}, as well as a system of three coupled Ginzburg-Landau equations
with the intra-core nonlinearity of the CQ type \cite{Skryabin}.

The significant interest in the DWP settings has also motivated the
appearance of rigorous mathematical results regarding the symmetry-breaking
bifurcations and the stability of nonlinear stationary states \cite{kirr}. A
rigorous treatment was also developed for the dynamical evolution in such
settings \cite{bambusi1,bambusi2}.

Our objective in this work is to extend the realm of DWPs to multi-component
settings. This is of particular relevance to experimental realizations of
BEC -- e.g.,, in a mixture of different hyperfine states of $^{87}$Rb \cite%
{book1,book2} (see also very recent experiments reported in Ref.
\cite{hall} and references therein). This extension is relevant to
optics too, where multi-component dynamics can be, for instance,
realized in photorefractive crystals \cite{zhigang2}. In the
one-component setting, the analysis of the weakly nonlinear regime
has given considerable insight into the emergence of asymmetric
branches from symmetric or anti-symmetric ones, in the models with
self-focusing and self-defocusing nonlinearity, respectively), and
the destabilization of the ``parent" branches through the respective
symmetry-breaking bifurcations, as well as the dynamics initiated by
the destabilization
\cite{smerzi,kiv2,Bergeman_2mode,todd,theo,kirr}.

In the present work, we aim to extend the analysis of the DWP
setting to two-component systems. In addition to the finite-mode
approximation, which can be justified rigorously \cite{kirr} under
conditions that remain valid in the present case, we use numerical
methods to follow the parametric evolution of solution branches that
emerge from bifurcations in the two-component system. We observe
that, in addition to the branches inherited from the
single-component model, new branches appear with the growth of the
nonlinearity. We dub these new solutions ``combined" ones, as they
involve both components. In fact, the new branches connect some of
the single-component ones. Furthermore, these new branches may
undergo their own symmetry-breaking bifurcations. The solutions
depend on chemical potentials of the two species, $\mu _{1}$ and
$\mu _{2}$ (in terms of BEC),
and, accordingly, loci of various bifurcations will be identified in plane $%
\left( \mu _{1},\mu _{2}\right) $.

The paper is structured as follows. In Section II, we present the framework
of the two-component problem, including the two-mode reduction (in each
component), that allows us to considerably simplify the existence and
stability problems. The formulation includes also a physically possible
linear coupling between the components, but the actual analysis is performed
without the linear coupling. In section III, we report numerical results
concerning the existence and stability of the new states, as well as the
evolution of unstable ones. In section IV, we summarize the findings and
discuss directions for further studies.

\section{Analytical consideration}

As a prototypical model that is relevant both to BECs \cite{book1,book2} and
optics \cite{kivshar}, we take the normalized two-component NLS equation of
the following form:

\begin{equation}
\begin{split}
iu_{t}& =Lu+\kappa v+\sigma (|u|^{2}+g|v|^{2})u-\mu _{1}u, \\
iv_{t}& =Lv+\kappa u+\sigma (|v|^{2}+g|u|^{2})v-\mu _{2}v,
\end{split}
\label{eq1}
\end{equation}%
where $u$ and $v$ are the wave functions of the two BEC components, or local
amplitudes of the two optical modes, $\mu _{1,2}$ are chemical potentials in
BEC or propagation constants in the optical setting, $\kappa $ is the
coefficient of the linear coupling between the components, and
\begin{equation}
L=-(1/2)\partial _{x}^{2}+V(x)  \label{L}
\end{equation}%
is the usual single-particle operator with trapping potential $V(x)$. The
repulsive or attractive character of the nonlinearity in BEC\
(self-defocusing or self-focusing, in terms of optics) is defined by $\sigma
=+1$ and $\sigma =-1$, respectively, while $\sigma g$ is the coefficient of
the inter-species interactions in BEC, or cross-phase modulation (XPM) in
optics.

In the case of (for instance) two hyperfine states with $|F,m_{F}>=|1,-1>$
and $|2,1>$ in $^{87}$Rb, both coefficients of the self-interaction and the
cross-interaction coefficient are very close, although their slight
difference is critical in accounting for the immiscibility of the two
species (see, e.g.,, \cite{hall} and references therein). However, for
problems considered below, the difference does not play a significant role,
hence we set $g=1$ in Eqs. (\ref{eq1}), which corresponds to the Manakov's
system, i.e., one which is integrable, in the absence of the potential, $%
V(x)=0$ in Eq. (\ref{L}) \cite{manakov,kivshar}. In fact, $V(x)$ is the DWP,
which we compose of a parabolic trap (with corresponding frequency $\Omega $%
) and a $\mathrm{sech}^{2}$-shaped barrier (of strength $V_{0}$ and width $w$%
) in the middle:
\begin{equation}
V(x)=(1/2)\Omega ^{2}x^{2}+V_{0}\mathrm{sech}^{2}\left( x/W\right) .
\label{omega0}
\end{equation}%
The calculations presented below have been performed for $\Omega =0.1$, $%
V_{0}=1$ and $W=0.5$, in which case the first two eigenvalues of linear
operator (\ref{L}) are $\omega _{0}=0.1328$ and $\omega _{1}=0.1557$.
However, we have checked that the generic picture presented below is
insensitive to specific details of the potential, provided that it is
symmetric.

The linear coupling in Eqs. (\ref{eq1}), via the terms proportional to $%
\kappa $, accounts for the possibility of interconversion between the two
hyperfine states in the BEC mixture, which may be induced, e.g.,, by
appropriate two-photon pulses in the situation considered in Ref. \cite{hall}%
. In optics, the linear coupling is induced by a twist of the optical
waveguide, if the two wave components correspond to orthogonal linear
polarizations of light.


To define an appropriate basis for the finite-mode expansion of solutions to
Eqs. (\ref{eq1}), we first consider the following eigenvalue problem,
\begin{equation}
\begin{pmatrix}
L & \kappa  \\
\kappa  & L%
\end{pmatrix}%
\begin{pmatrix}
u \\
v%
\end{pmatrix}%
=\omega
\begin{pmatrix}
u \\
v%
\end{pmatrix}%
.  \label{E}
\end{equation}%
We denote the ground and first excited eigenstates of the matrix operator in
Eq. (\ref{E}), which appertain to the two lowest eigenvalues, as $\left\{
u_{0},v_{0}\right\} $ and $\left\{ u_{1,}v_{1}\right\} $ (these two states
are, as usual, spatially even and odd ones, respectively). The Hermitian
nature of the linear operator in Eq. (\ref{E}) ensures that the eigenvalues
are real, and the eigenfunctions can also be chosen in a real form. In fact,
in the case that we will examine here, we will have $u_{0}=v_{0}$ and $%
u_{1}=v_{1}$, but the different notations will be kept for the $u$ and $v$
components, to demonstrate how the nonlinear analysis can be carried out in
the general case. Then, the two-mode approach is based on the assumption
that, in the weakly nonlinear case, the solution is decomposed as a linear
combination of these eigenfunctions, i.e.,
\begin{equation}
\begin{split}
u(x,t)& =c_{0}(t)u_{0}(x)+c_{1}(t)u_{1}(x), \\
v(x,t)& =c_{2}(t)v_{0}(x)+c_{3}(t)v_{1}(x).
\end{split}
\label{ansatz}
\end{equation}%
This assumption can be rigorously substantiated, with an estimate for the
accuracy, upon imposing suitable conditions on the smallness of the
nonlinearity, and properties of the rest of the linear spectrum \cite{kirr}.
Therefore, as shown in \cite{kirr}, such an approach can be controllably
accurate for small 
$N=\int_{-\infty }^{+\infty }\left( |u|^{2}+|v|^{2}\right) dx$,
although below we examine its comparison with numerical results
even for values of $N$ of O$(1)$.

It is relevant to mention that an alternative approach to the decomposition
of the two-component wave functions could be based on using the basis of
symmetric and antisymmetric wave functions, $u_{0}\pm v_{0}$ and $u_{1}\pm
v_{1}$. However, the resort to this basis does not make the final equations
really simpler. On the other hand, unlike the approach developed below, the
use of the alternative basis would make it much harder to estimate the error
of the finite-mode approximation, and thus rigorously substantiate the
approximation, cf. Ref. \cite{kirr}.

Substituting ansatz (\ref{ansatz}) in Eqs. (\ref{eq1}), and projecting the
resulting equations onto eigenfunctions $\left\{ u_{0},v_{0}\right\} $ and $%
\left\{ u_{1,}v_{1}\right\} $, one can derive at the following nonlinear
ODEs for the temporal evolution of the complex amplitudes of the
decomposition, $c_{0,1,2,3}$:
\begin{equation}
\begin{split}
i\dot{c_{0}}=(\omega _{0}-\mu _{1})c_{0}+\kappa c_{2}& +\sigma \lbrack
\Gamma _{0000}|c_{0}|^{2}c_{0}+\Gamma _{0011}(c_{1}^{2}c_{0}^{\ast
}+2c_{0}|c_{1}|^{2})] \\
& +\sigma g[\Gamma _{0000}c_{0}|c_{2}|^{2}+\Gamma
_{0011}(c_{0}|c_{3}|^{2}+c_{1}c_{2}c_{3}^{\ast }+c_{1}c_{2}^{\ast }c_{3})],
\end{split}
\label{1}
\end{equation}

\begin{equation}
\begin{split}
i\dot{c_{1}}=(\omega _{1}-\mu _{1})c_{1}+\kappa c_{3}& +\sigma \lbrack
\Gamma _{1111}|c_{1}|^{2}c_{1}+\Gamma _{0011}(c_{0}^{2}c_{1}^{\ast
}+2|c_{0}|^{2}c_{1})] \\
& +\sigma g[\Gamma _{1111}c_{1}|c_{3}|^{2}+\Gamma
_{0011}(c_{1}|c_{2}|^{2}+c_{0}c_{2}c_{3}^{\ast }+c_{0}c_{2}^{\ast }c_{3})],
\end{split}
\label{2}
\end{equation}

\begin{equation}
\begin{split}
i\dot{c_{2}}=(\omega _{0}-\mu _{2})c_{2}+\kappa c_{0}& +\sigma \lbrack
\Gamma _{0000}|c_{2}|^{2}c_{2}+\Gamma _{0011}(c_{3}^{2}c_{2}^{\ast
}+2c_{2}|c_{3}|^{2})] \\
& +\sigma g[\Gamma _{0000}|c_{0}|^{2}c_{2}+\Gamma
_{0011}(|c_{1}|^{2}c_{2}+c_{0}c_{1}^{\ast }c_{3}+c_{0}^{\ast }c_{1}c_{3})],
\end{split}
\label{3}
\end{equation}

\begin{equation}
\begin{split}
i\dot{c_{3}}=(\omega _{1}-\mu _{2})c_{3}+\kappa c_{1}& +\sigma \lbrack
\Gamma _{1111}|c_{3}|^{2}c_{3}+\Gamma _{0011}(c_{2}^{2}c_{3}^{\ast
}+2|c_{2}|^{2}c_{3})] \\
& +\sigma g[\Gamma _{1111}|c_{1}|^{2}c_{3}+\Gamma
_{0011}(|c_{0}|^{2}c_{3}+c_{0}c_{1}^{\ast }c_{2}+c_{0}^{\ast }c_{1}c_{2})].
\end{split}
\label{4}
\end{equation}%
In these equations, $\Gamma _{ijkl}\equiv \int_{-\infty }^{+\infty
}u_{i}(x)u_{j}(x)u_{k}(x)u_{l}(x)dx$ are the so-called matrix elements of
the nonlinear four-wave interactions, with $\Gamma _{ijkl}=0$ when $i$+$j$+$%
k $+$l$ is odd.

Seeking for real fixed points to Eqs. (\ref{1}) - (\ref{4}), $a_{j}(t)\equiv
\rho _{j}$, we reduce the equations to an algebraic system,
\begin{equation}
\begin{split}
\mu _{1}\rho _{0}=(\omega _{0}\rho _{0}+\kappa \rho _{2})& +\sigma (\Gamma
_{0000}\rho _{0}^{3}+3\Gamma _{0011}\rho _{0}\rho _{1}^{2}) \\
& +\sigma g(\Gamma _{0000}\rho _{0}\rho _{2}^{2}+\Gamma _{0011}\rho _{0}\rho
_{3}^{2}+2\Gamma _{0011}\rho _{1}\rho _{2}\rho _{3}),
\end{split}
\label{rho0}
\end{equation}

\begin{equation}
\begin{split}
\mu_1\rho_1 = (\omega_1\rho_1+\kappa\rho_3)
&+\sigma(\Gamma_{1111}\rho_1^3+3\Gamma_{0011}\rho_0^2\rho_1) \\
&+\sigma
g(\Gamma_{1111}\rho_1\rho_3^2+\Gamma_{0011}\rho_1\rho_2^2+2\Gamma_{0011}%
\rho_0\rho_2\rho_3),
\end{split}
\label{rho1}
\end{equation}

\begin{equation}
\begin{split}
\mu_2\rho_2 = (\omega_0\rho_2+\kappa\rho_0)
&+\sigma(\Gamma_{0000}\rho_2^3+3\Gamma_{0011}\rho_2\rho_3^2) \\
&+\sigma
g(\Gamma_{0000}\rho_0^2\rho_2+\Gamma_{0011}\rho_1^2\rho_2+2\Gamma_{0011}%
\rho_0\rho_1\rho_3),
\end{split}
\label{rho2}
\end{equation}

\begin{equation}
\begin{split}
\mu _{2}\rho _{3}=(\omega _{1}\rho _{3}+\kappa \rho _{1})& +\sigma (\Gamma
_{1111}\rho _{3}^{3}+3\Gamma _{0011}\rho _{2}^{2}\rho _{3}) \\
& +\sigma g(\Gamma _{1111}\rho _{1}^{2}\rho _{3}+\Gamma _{0011}\rho
_{0}^{2}\rho _{3}+2\Gamma _{0011}\rho _{0}\rho _{1}\rho _{2}),
\end{split}
\label{rho3}
\end{equation}%
The simplest solution to the above equations can be obtained in the form of $%
\rho _{2}=\rho _{0}$ and $\rho _{3}=\rho _{1}$, for $\mu _{1}=\mu _{2}$.
With this substitution, the remaining equations are
\begin{equation}
\begin{split}
\mu _{1}-\tilde{\omega}_{0}& =\tilde{\sigma}(\Gamma _{0000}\rho
_{0}^{2}+3\Gamma _{0011}\rho _{1}^{2}), \\
\mu _{1}-\tilde{\omega}_{1}& =\tilde{\sigma}(\Gamma _{1111}\rho
_{1}^{2}+3\Gamma _{0011}\rho _{0}^{2}),
\end{split}
\label{remaining}
\end{equation}%
where $\tilde{\omega}_{0}\equiv \omega _{0}+\kappa ,\;\tilde{\omega}%
_{1}\equiv \omega _{1}+\kappa ,\;\tilde{\sigma}\equiv \sigma (1+g).$ System (%
\ref{remaining}) can be solved as a linear one for $\rho _{0,1}^{2}$. For
given values of $\Gamma $ and $\omega $, a physical range of chemical
potential $\mu _{1}$ is that which provides physical solutions for $\rho
_{0,1}^{2}$.

In what follows, we focus on the case of $\kappa =0$, since $\kappa \neq 0$
imposes the condition $\mu _{1}=\mu _{2}$, while we are interested in more
general solutions. With $\kappa =0$, chemical potentials $\mu _{1}$ and $\mu
_{2}$ may be different, allowing us to explore the two-parameter solution
space $(\mu _{1},\mu _{2})$, for both signs of the nonlinearity,
self-attractive ($\sigma =-1$) and the self-repulsive ($\sigma =+1$).

The simplest possible branches of solutions are single-mode ones, with only
one of amplitudes $\rho _{j}$ different from zero. Branches of these
solutions can be easily found from Eqs. (\ref{rho0}) - (\ref{rho3}),
\begin{equation}
\begin{split}
\rho _{0}^{2}& =\frac{\mu _{1}-\omega _{0}}{\sigma \Gamma _{0000}}\neq
0,\quad \rho _{1}=\rho _{2}=\rho _{3}=0 \\
\rho _{1}^{2}& =\frac{\mu _{1}-\omega _{1}}{\sigma \Gamma _{1111}}\neq
0,\quad \rho _{0}=\rho _{2}=\rho _{3}=0 \\
\rho _{2}^{2}& =\frac{\mu _{2}-\omega _{0}}{\sigma \Gamma _{0000}}\neq
0,\quad \rho _{0}=\rho _{1}=\rho _{3}=0 \\
\rho _{3}^{2}& =\frac{\mu _{2}-\omega _{1}}{\sigma \Gamma _{1111}}\neq
0,\quad \rho _{0}=\rho _{1}=\rho _{2}=0
\end{split}
\label{simplest}
\end{equation}%
The first one of these branches contains the projection only onto the even
(symmetric0 eigenfunction of the first component, and will hereafter be
accordingly denoted S1. Similarly, the second branch has a projection onto
the odd (antisymmetric) mode of the first component, and will be named AN1.
The third and fourth solutions are similar modes in the second component, to
be denoted S2 and AN2, respectively.

In addition to these modes, there exist asymmetric states in each of the
components (similar to the one-component model \cite{theo,kirr}),
\begin{eqnarray}
\rho _{0}^{2} &=&\frac{\Gamma _{1111}(\mu _{1}-\omega _{0})-3\Gamma
_{0011}(\mu _{1}-\omega _{1})}{\sigma (\Gamma _{1111}\Gamma _{0000}-9\Gamma
_{0011}^{2})},\quad \rho _{2}=0,  \label{as1} \\
\rho _{1}^{2} &=&\frac{\Gamma _{0000}(\mu _{1}-\omega _{1})-3\Gamma
_{0011}(\mu _{1}-\omega _{0})}{\sigma (\Gamma _{1111}\Gamma _{0000}-9\Gamma
_{0011}^{2})},\quad \rho _{3}=0.  \label{as2}
\end{eqnarray}%
Since these solutions do not exist in the linear limit, they have to
bifurcate at a non-zero value of the amplitude, from either symmetric or
antisymmetric branches (\ref{simplest}). In fact, such solutions bifurcate,
depending on the sign of $\sigma $ and coefficients $\Gamma $, either from
S1 at
\begin{equation}
\mu _{1}^{\mathrm{(cr)}}={\omega _{1}}-\frac{3\Gamma _{0011}({\omega _{0}}-{%
\omega _{1}})}{\Gamma _{0000}-3\Gamma _{0011}},
\end{equation}%
or from AN1 at
\begin{equation}
\mu _{1}^{\mathrm{(cr)}}={\omega _{0}}-\frac{3\Gamma _{0011}({\omega _{1}}-{%
\omega _{0}})}{\Gamma _{1111}-3\Gamma _{0011}}.
\end{equation}%
The asymmetric solutions given by Eqs. (\ref{as1}) and (\ref{as2}) will be
denoted AS1. Similarly, there is an asymmetric branch of solutions in the
second component, hereafter denoted AS2, with
\begin{eqnarray}
\rho _{2}^{2} &=&\frac{\Gamma _{1111}(\mu _{2}-\omega _{0})-3\Gamma
_{0011}(\mu _{2}-\omega _{1})}{\sigma (\Gamma _{1111}\Gamma _{0000}-9\Gamma
_{0011}^{2})},\quad \rho _{0}=0  \label{as3} \\
\rho _{3}^{2} &=&\frac{\Gamma _{0000}(\mu _{2}-\omega _{1})-3\Gamma
_{0011}(\mu _{2}-\omega _{0})}{\sigma (\Gamma _{1111}\Gamma _{0000}-9\Gamma
_{0011}^{2})},\quad \rho _{1}=0,  \label{as4}
\end{eqnarray}%
and conclusions that can be deduced about its bifurcation are similar to
those concerning AS1.

In addition to these obvious solutions, there exist others, that can be
obtained as solutions to Eqs. (\ref{rho0})-(\ref{rho3}) and involve both
components (obviously, they have no counterparts in the single-component
model). They will be denoted as ``combined" ones (C1, C2, C3 and C4) in what
follows. Although they cannot be expressed by simple analytical formulas,
unlike the S, AN and AS branches that were defined above, they can be easily
found as numerical solutions of Eqs. (\ref{rho0})-(\ref{rho3}), and will be
compared to full numerical results in the following section.

\section{Numerical Results}

We start the numerical analysis by examining the self-focusing case. Because
of the complexity of the bifurcation diagrams in the plane of the chemical
potentials of the two components, $\left( \mu _{1},\mu _{2}\right) $, we
will illustrate the bifurcation phenomenology by showing cross-sections of
the full two-parameter bifurcation diagram for varying $\mu _{1}$, at
different fixed values of $\mu _{2}$.

In connection to the single-component states presented above, four distinct
scenarios have been found, when the second chemical potential, $\mu _{2}$,
appears. The first (most complex) scenario takes place at $\mu _{2}<\mu
_{2}^{\mathrm{(cr)}}$ (with some critical value $\mu _{2}^{\mathrm{(cr)}}$),
when branch AS2 exists (in addition to S2 and AN2). The second case is
observed in region $\mu _{2}^{\mathrm{(cr)}}<\mu _{2}<\omega _{0}$, in which
case only S2 and AN2 exist (recall $\omega _{0}$ and $\omega _{1}$ are two
lowest eigenvalues of operator (\ref{L})). The third scenario is found in
the interval of $\omega _{0}<\mu _{2}<\omega _{1}$, where only branch AN2
may exist, and, finally, the fourth one takes place at $\omega _{1}<\mu
_{2}, $ where there are no single-component solutions for the second field
(in the case of the self-attractive nonlinearity).

\begin{figure}[tbph]
\centering
\includegraphics[width=.4\textwidth]{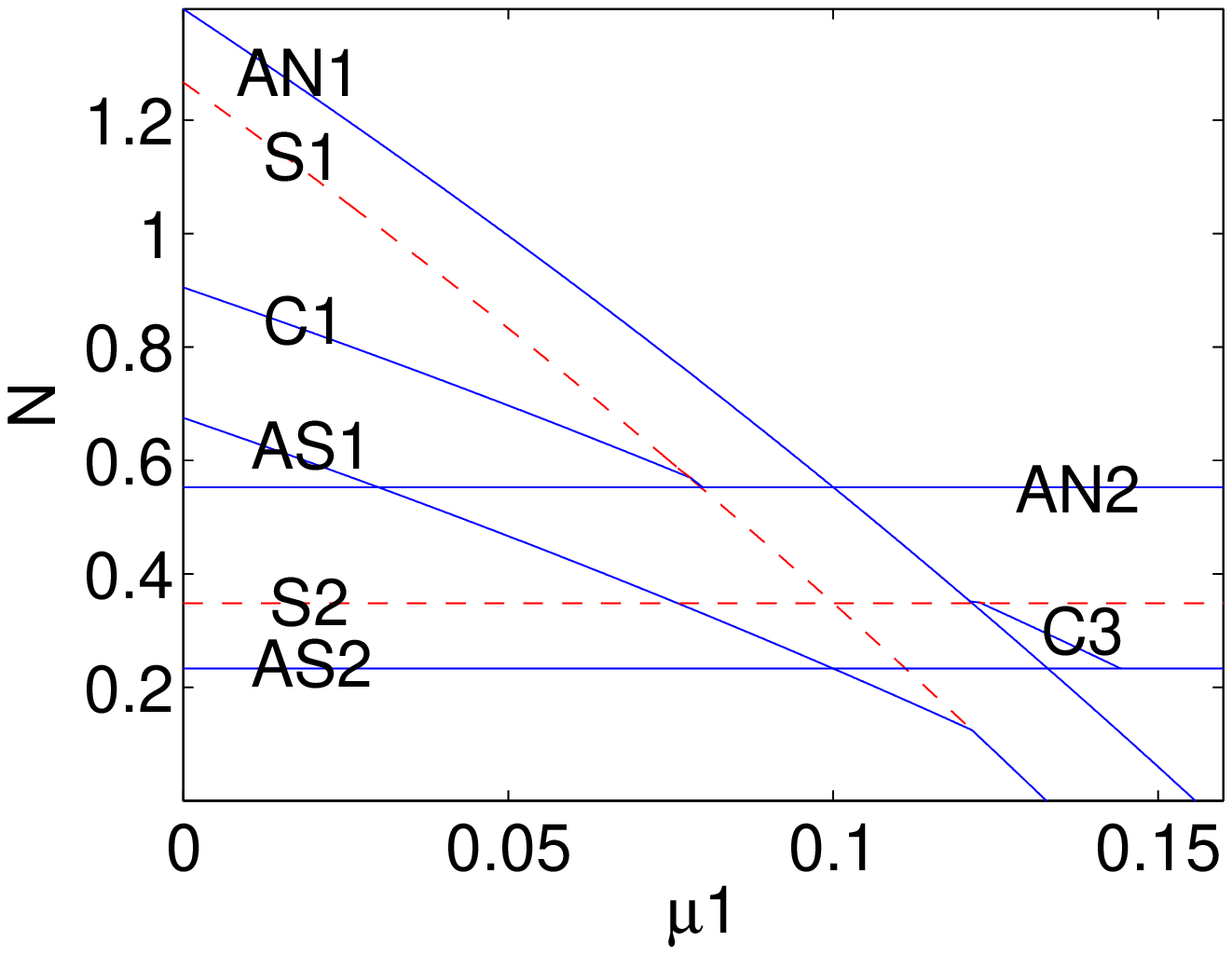} %
\includegraphics[width=.4\textwidth]{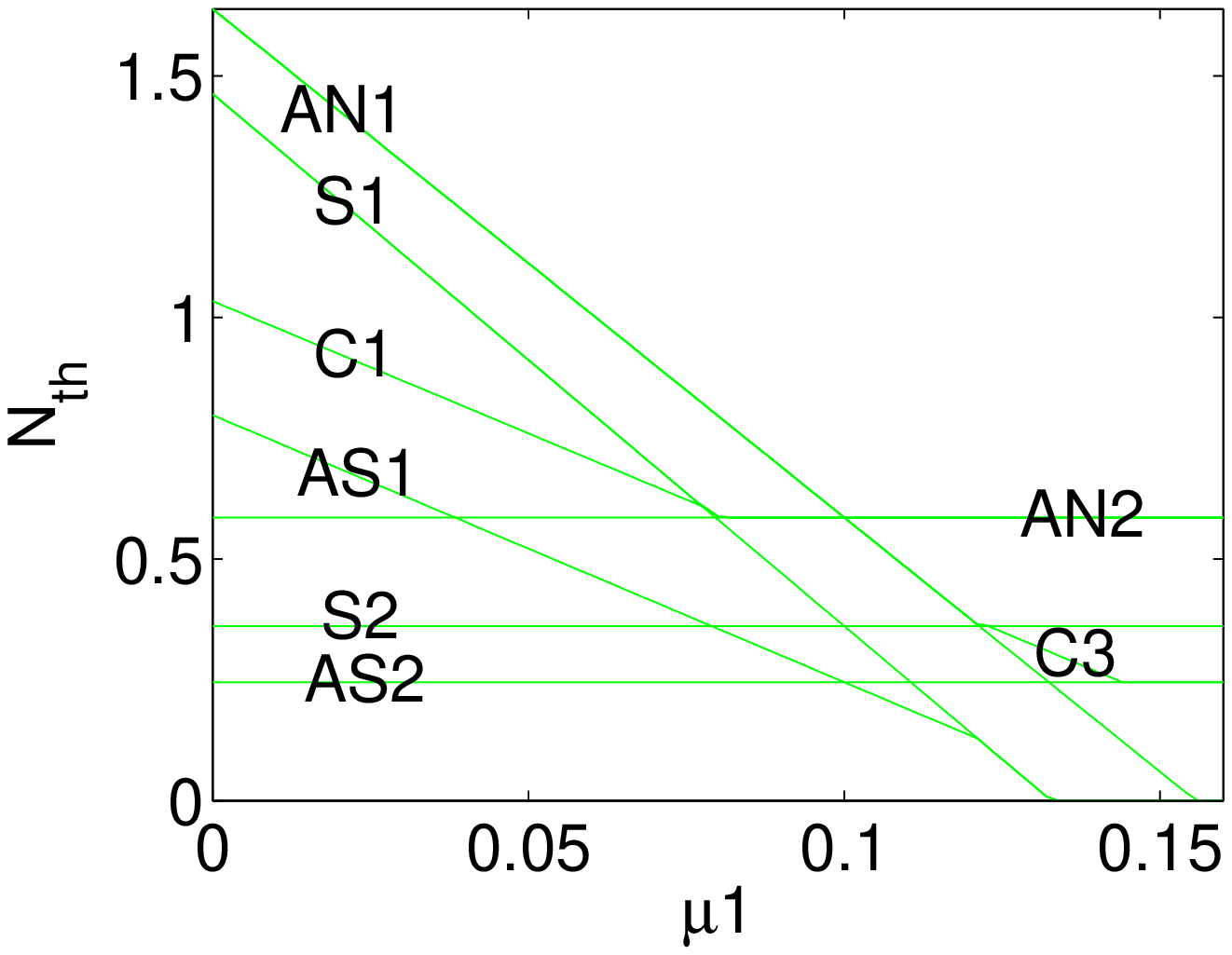}\newline
\includegraphics[width=.4\textwidth]{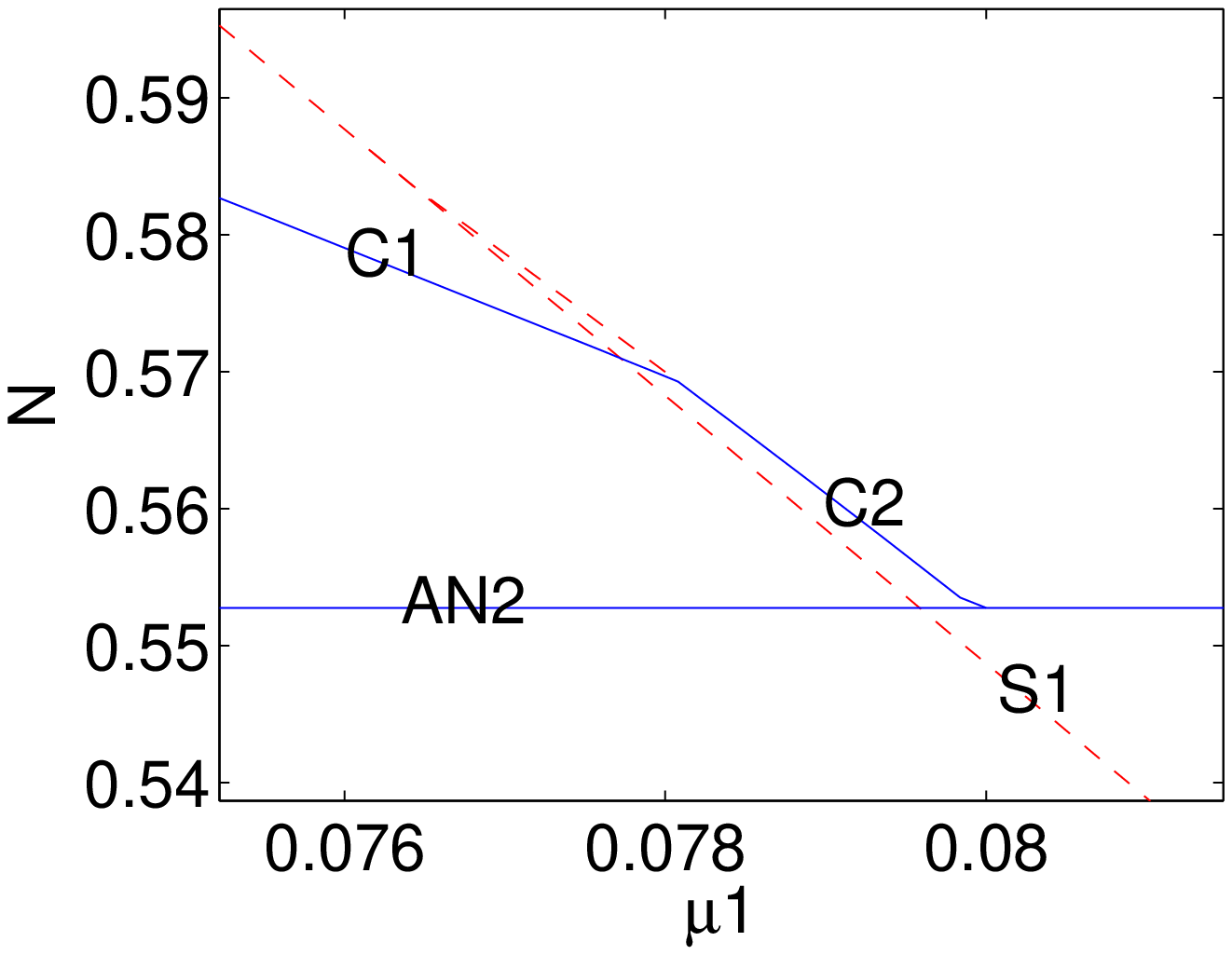} %
\includegraphics[width=.4\textwidth]{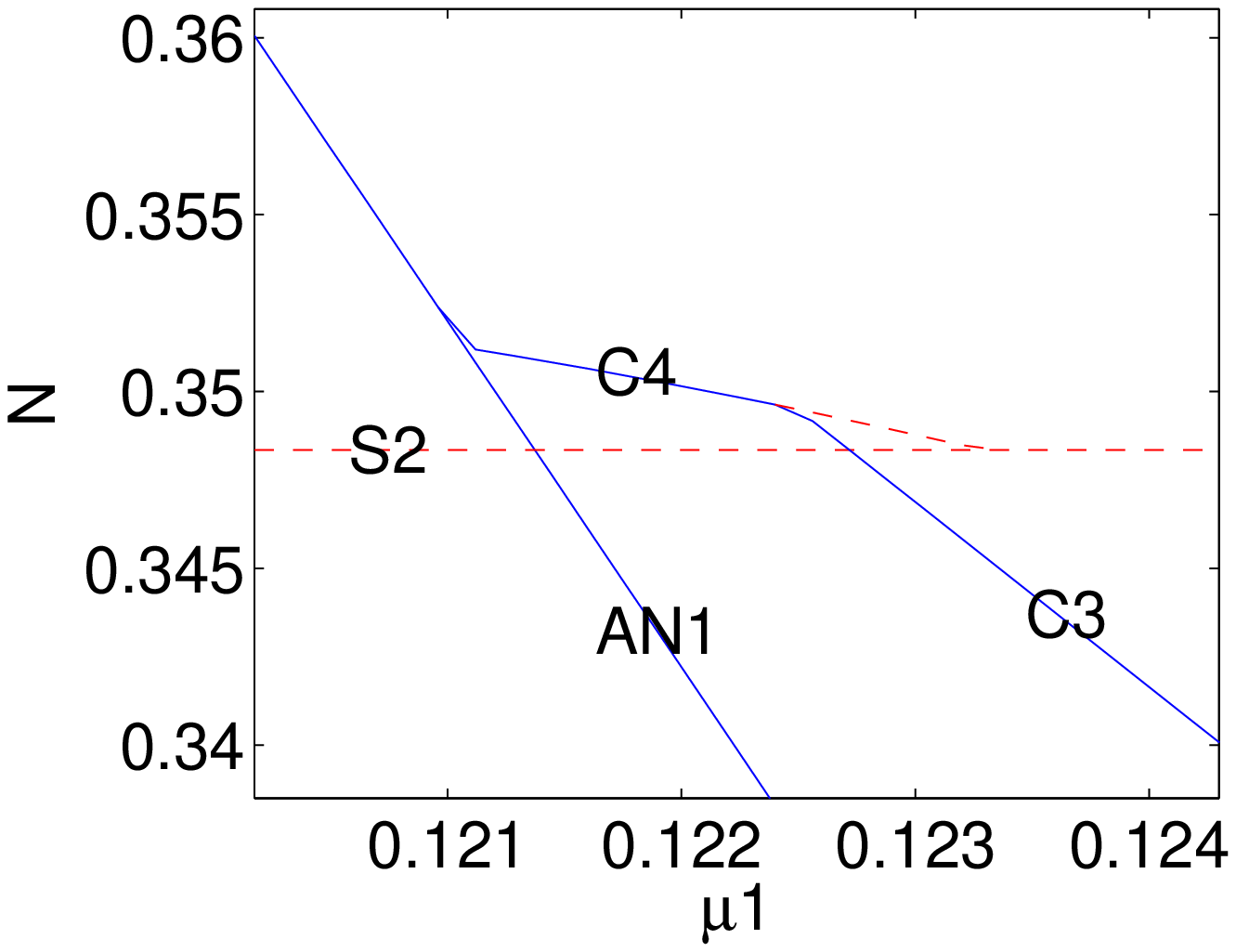}%
\newline
\caption{(Color online) Top panel: The norm (corresponding to the number of
atoms in BEC and total power in optics) of the numerically found solutions
of Eq. (\protect\ref{eq1}) (left) and their counterparts predicted by the
two-mode approximation (right) for the self-attractive nonlinearity ($%
\protect\sigma =-1$), as a function of $\protect\mu _{1}$, for $\protect\mu %
_{2}=0.1$. Here and in other figures, the (blue) solid lines and (red)
dashed ones depict stable and unstable solutions, respectively. The bottom
panels are blowups of segments in the top left panel where bifurcations
involving the new combined solutions occur. S1, AN1, AS1 and S2, AN2, AS2
mark, respectively, symmetric, antisymmetric, and asymmetric
single-component solutions belonging to fields $u$ or $v$. Symbols C1, C2,
C3, C4 mark branches of the new combined (two-component) solutions, which
are defined in the text.}
\label{fig1}
\end{figure}

We describe the most complex of these scenarios in Fig. \ref{fig1} and
subsequently summarize differences from this case in the remaining three
simpler regimes. In Fig. \ref{fig1}, the top panel presents the full
numerically generated bifurcation diagram, in terms of total norm $%
N=\int_{-\infty }^{+\infty }\left( |u|^{2}+|v|^{2}\right) dx$ (which is
proportional to the total number of atoms in both species in the BEC model,
or the total beam power in optics) as a function of $\mu _{1}$ for fixed $%
\mu _{2}=0.1$. The horizontal lines in this diagram represent
single-component solutions AS2, S2 and AN2 in the second field (in order of
the increasing norm), which are, obviously, unaffected by the variation of $%
\mu _{1}$ of the other component, that remains ``empty" in these
solutions. As the asymmetric branch AS2 exists at this value of $\mu
_{2}$ (in fact, at all values in region $\mu _{2}<\mu _{2}^{\mathrm{(cr)}}$%
), it is stable (as the one generated by the symmetry-breaking
bifurcation in the single-component equation \cite{theo,kirr}),
while branch S2 is destabilized by the same bifurcation. In addition
to these branches represented solely by the second component, with
the variation of $\mu _{1}$ we observe the emergence of the
single-component branches, S1 and AN1, in the first field, from the
corresponding linear limit, at $\mu _{1}=\omega _{0}$ and $\mu
_{1}=\omega _{1}$. In addition, past critical point $\mu
_{1}^{\mathrm{(cr)}}=0.1214$, the symmetry-breaking-induced branch
AS1 arises in the first component, destabilizing its symmetric
``parent branch" S1, in agreement with the known results for the
one-component model.

An additional explanation is necessary about the special case of $\mu
_{1}=\mu _{2}$ ($=0.1$ in the case of Fig. \ref{fig1}). In that case, we
observe that branches S1 and S2, as well as AN1 and AN2, and \emph{also} AS1
and AS2, collide. At this value of the parameter, stationary equations (\ref%
{eq1}) (with $u_{t}=v_{t}=0$) are degenerate (recall we consider the case of
$g=0$), making \textit{any} solution in the form of $au=bv$ is possible,
including pairs $(a,b)=(0,1)$ (which corresponds to the branches considered
above that contain only the first component), $(a,b)=(1,0)$ (i.e., the
second-component branches), and $(a,b)=(1,1)$ (where the two components are
identical). Therefore, in this degenerate case, the points of collisions of
the branches in the bifurcation diagram of Fig. \ref{fig1} actually denote a
one-parameter family of solutions, rather than an isolated solution.

The key new feature of this bifurcation diagram in comparison to its
one-component counterpart is the presence of the four new branches
of solutions, C1 through C4. Two of them are found to connect
(``bridge") two single-component branches of the solutions belonging
to \emph{different components}, and two other species of the new
solutions bifurcate from the bridging ones. More specifically, as
can be seen in the bottom left panel of Fig. \ref{fig1} (which is a
blowup of the\ one in the top left corner), branch C2 interpolates
between antisymmetric branch AN2 of the second component and
symmetric branch S1 of the first component (hence, one may naturally
expect that the solutions belonging to C2 have symmetric and
antisymmetric profiles in the first and second components,
respectively, see Figs. \ref{fig4}-\ref{fig7} below. Furthermore,
similarly to the bifurcation of the asymmetric branches from the
symmetric ones in the single-component models, there is such a
bifurcation which occurs to the C2 branch, destabilizing it and
giving rise to a stable branch C1 of combined solutions, with
asymmetric profiles instead of the symmetric and antisymmetric
waveforms in the first and second components of C2, respectively
(see Fig. \ref{fig6} below).

Naturally, there is a similar pair of two-component branches emerging due to
the bridging of S2 and AN1. As expected, the two-component branch C4, which
is responsible for the bridging \textit{per se}, features symmetric and
antisymmetric profiles in its second and first components. Finally, it is
clearly seen in the bottom right panel of Fig. \ref{fig1} (which is a blowup
of another relevant fragment of the top left panel) that a symmetry-breaking
bifurcation occurs on branch C4, which destabilizes it and gives rise to
branch C3 of stable asymmetric two-component solutions. It is worthy to note
that this emerging asymmetric branch C3 terminates through a collision with
the single-component asymmetric branch AS2, as can also be seen in the top
left panel of Fig. \ref{fig1}.

Similar to the emergence of AS1 from S1, the bifurcation of C1 from C2 and
of C3 from C4 are of the pitchfork type. Therefore, we actually have two
asymmetric branches of types C1 and C3, which are mirror images to each
other (accordingly, the additional asymmetric waveforms can be obtained from
those displayed in Fig. \ref{fig7} by the reflection around $x=0$).

Comparing this complex bifurcation picture with predictions of algebraic
equations (\ref{rho0})-(\ref{rho3}), we conclude that the picture is
captured \textit{in its entirety} by the two-mode approximation, as shown in
the top right panel of Fig. \ref{fig1}. Although some details may differ
(notice, in particular, slight disparity in the scales of the two diagrams
in the top panels of Fig. \ref{fig1}), the overall structure of the diagrams
is fully reproduced by the algebraic equations. As a characteristic
indicating the accuracy of the approximation, it is relevant to compare the
bifurcation points. In the finite-mode picture, the bifurcation of AS1 from
S1 occurs at $\mu _{1}=0.1211$, C1 bifurcates from C2 at $\mu _{1}=0.0780$,
and the bifurcation of C3 from C4 happens at $\mu _{1}=0.1225$. The full
numerical results yield the following values for the same bifurcation
points: $\mu _{1}=0.1214$, $\mu _{1}=0.0781$ and $\mu _{1}=0.1224$,
respectively, with the relative error $<0.25\%$.

\begin{figure}[tbph]
\centering
\includegraphics[width=.4\textwidth]{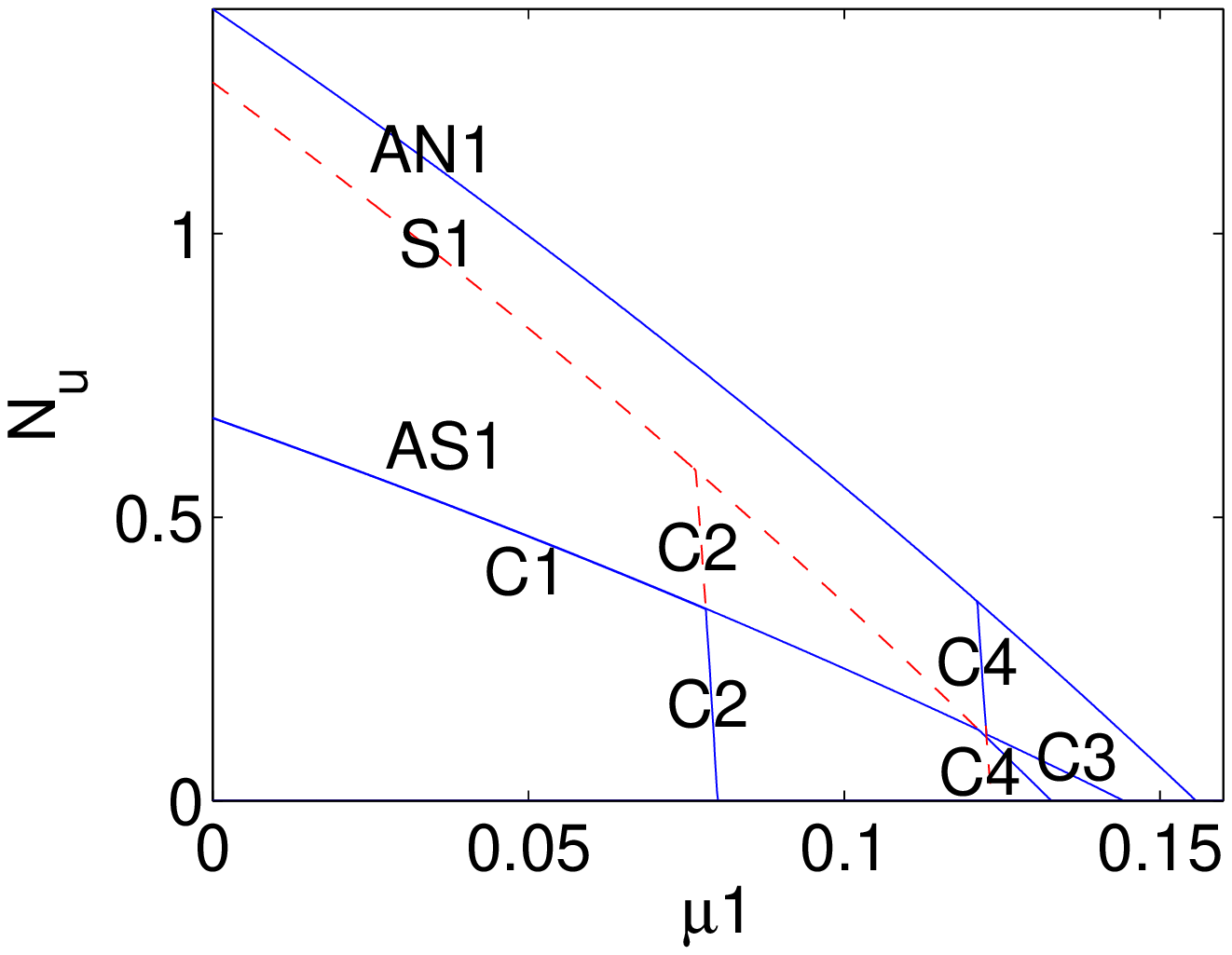} %
\includegraphics[width=.4\textwidth]{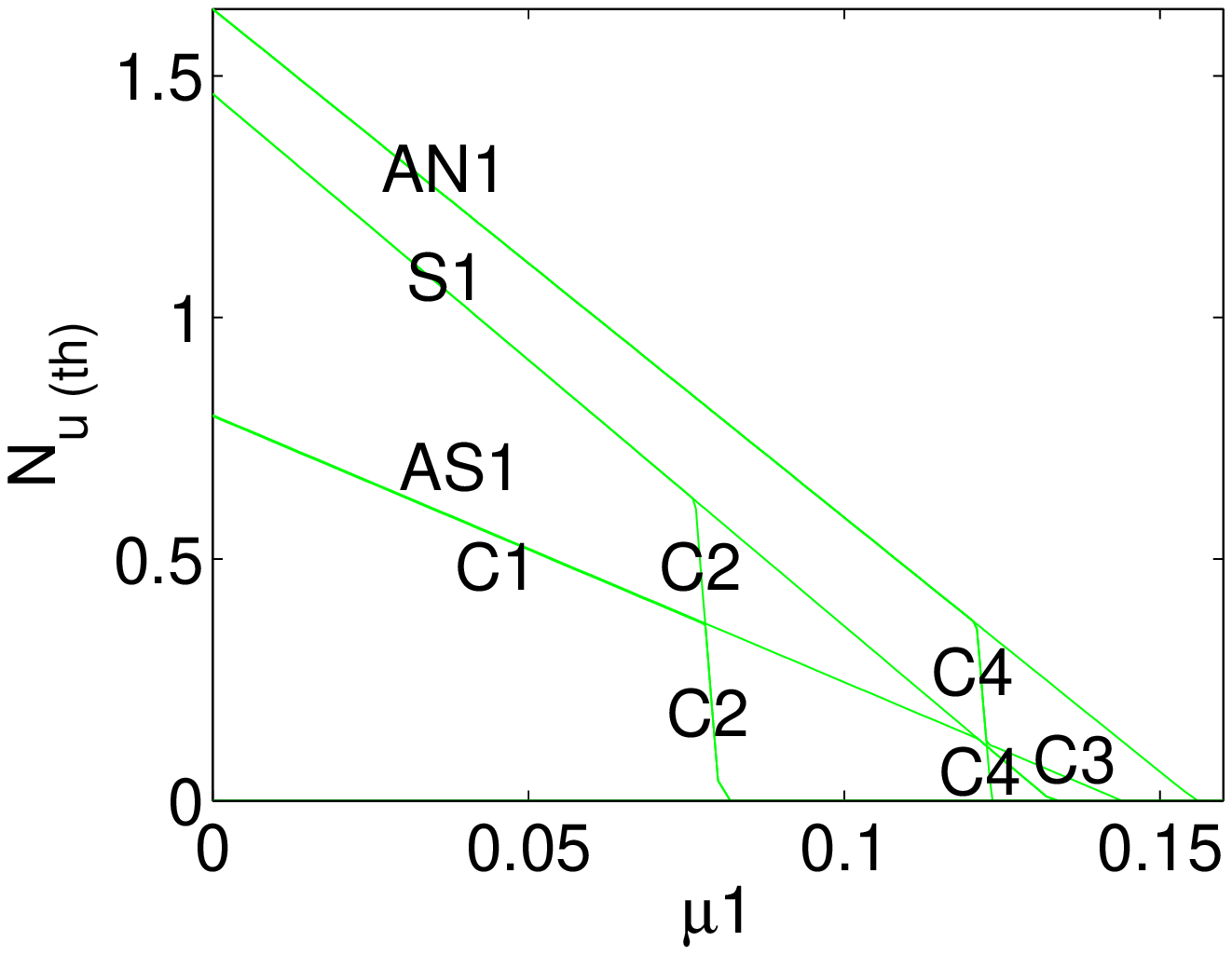}\newline
\includegraphics[width=.4\textwidth]{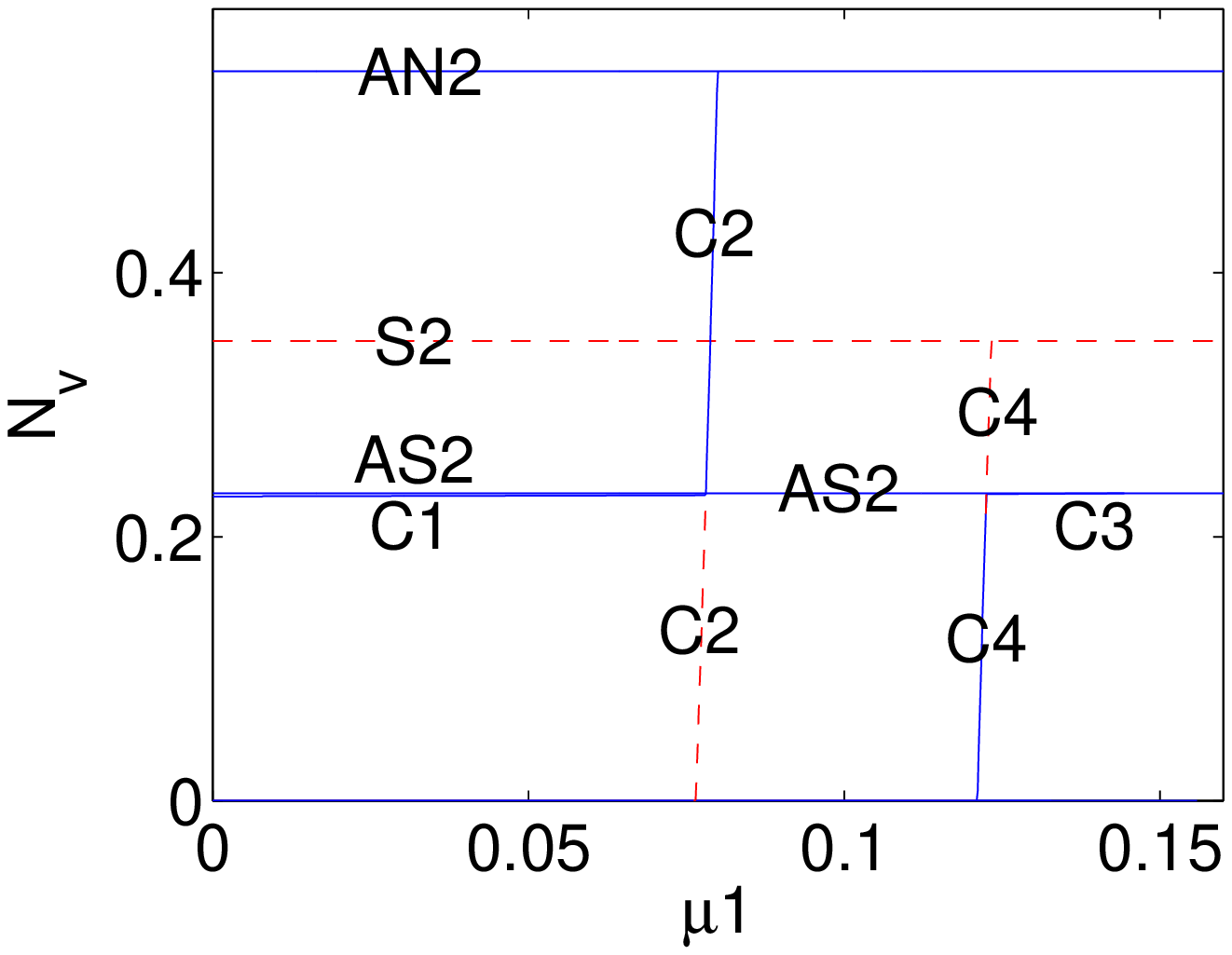} %
\includegraphics[width=.4\textwidth]{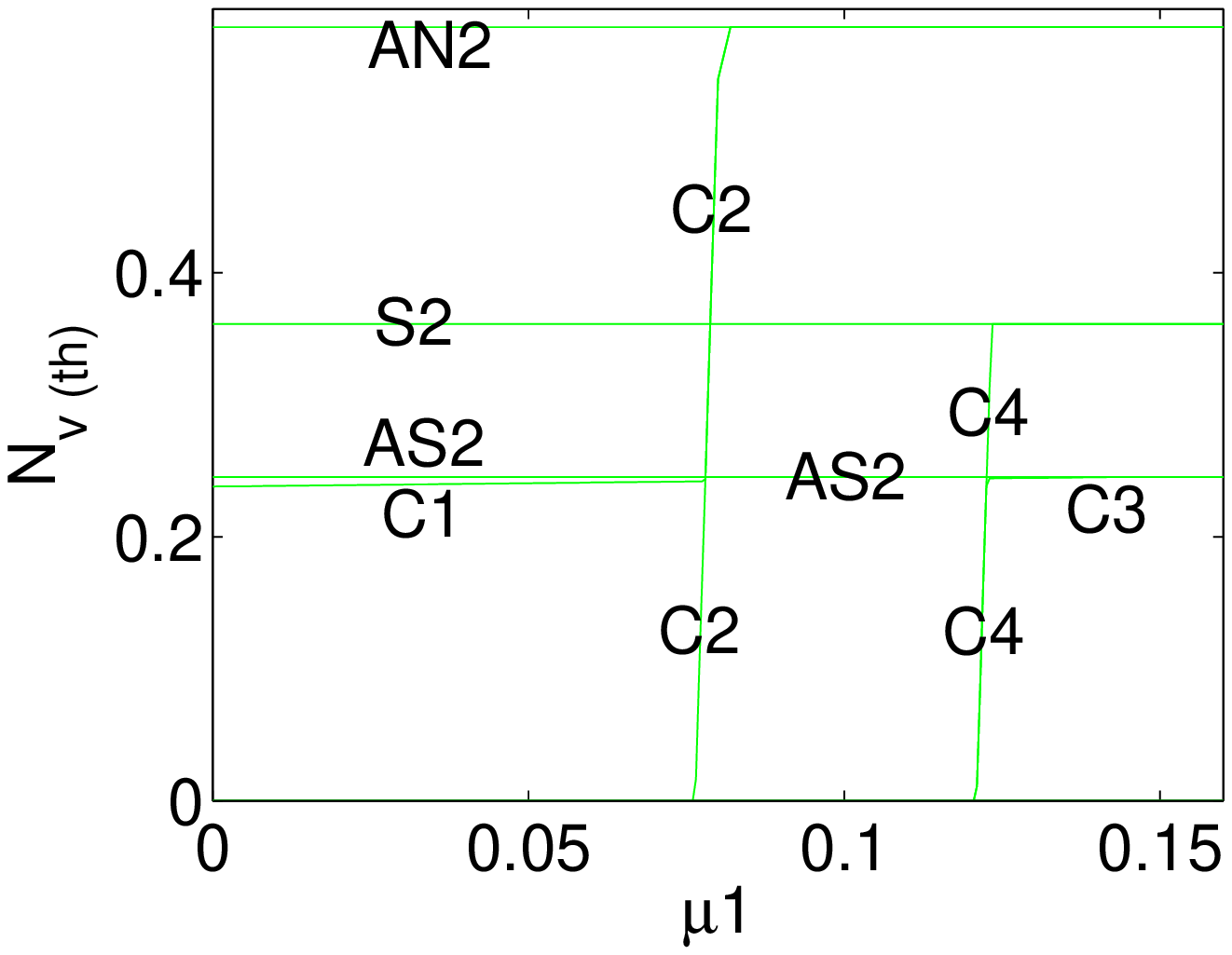}\newline
\caption{(Color online) The norm of the numerically found (left) and
approximate two-mode (right) wave functions $u$ (top) and $v$ (bottom) of
solutions to the stationary version of Eq. (\protect\ref{eq1}) with the
attractive nonlinearity ($\protect\sigma =-1$), as a function of $\protect%
\mu _{1}$, for $\protect\mu _{2}=0.10$. The notation is the same as in Fig.
\protect\ref{fig1}.}
\label{fig2}
\end{figure}

\begin{figure}[tbph]
\centering
\includegraphics[width=.4\textwidth]{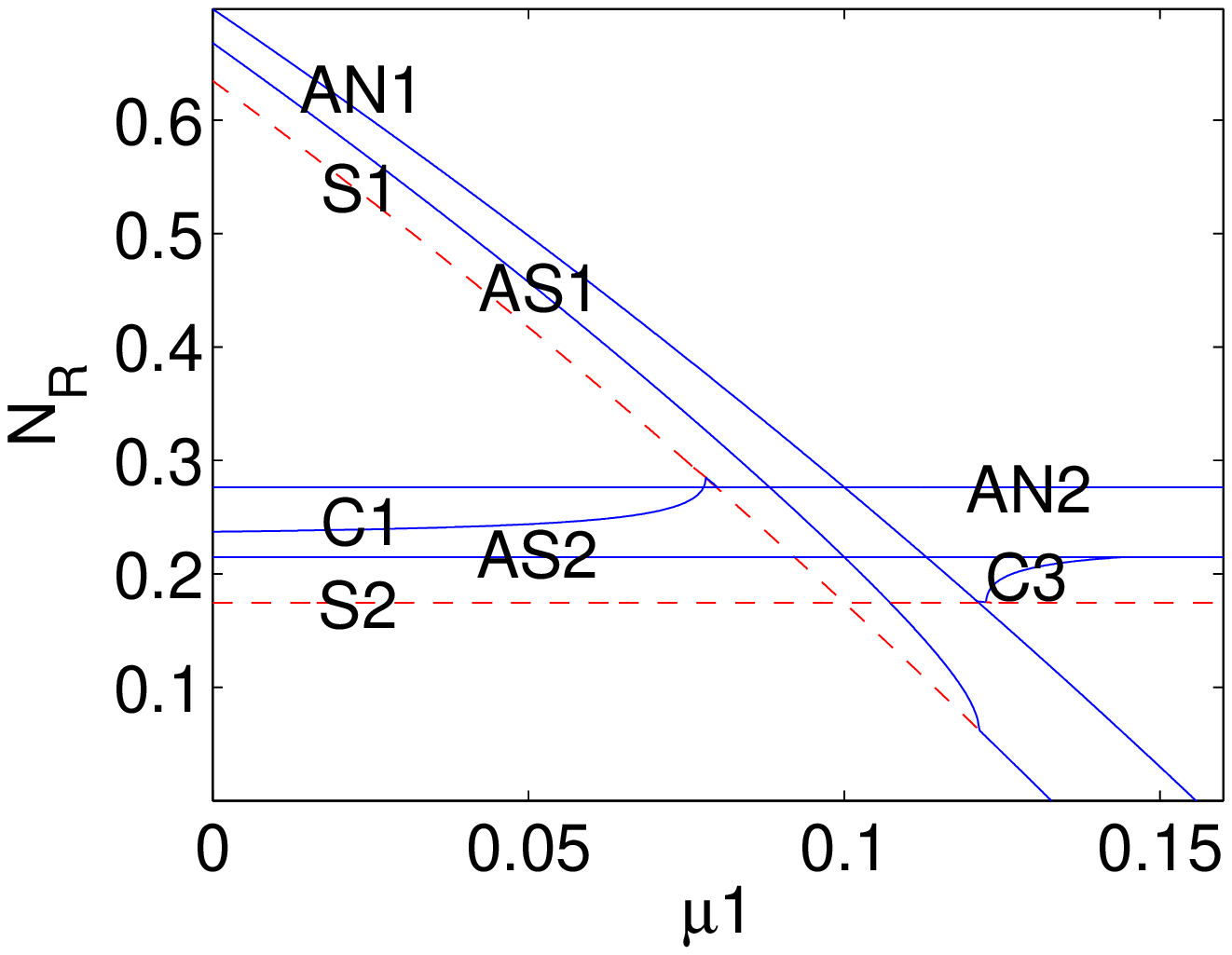} %
\includegraphics[width=.4\textwidth]{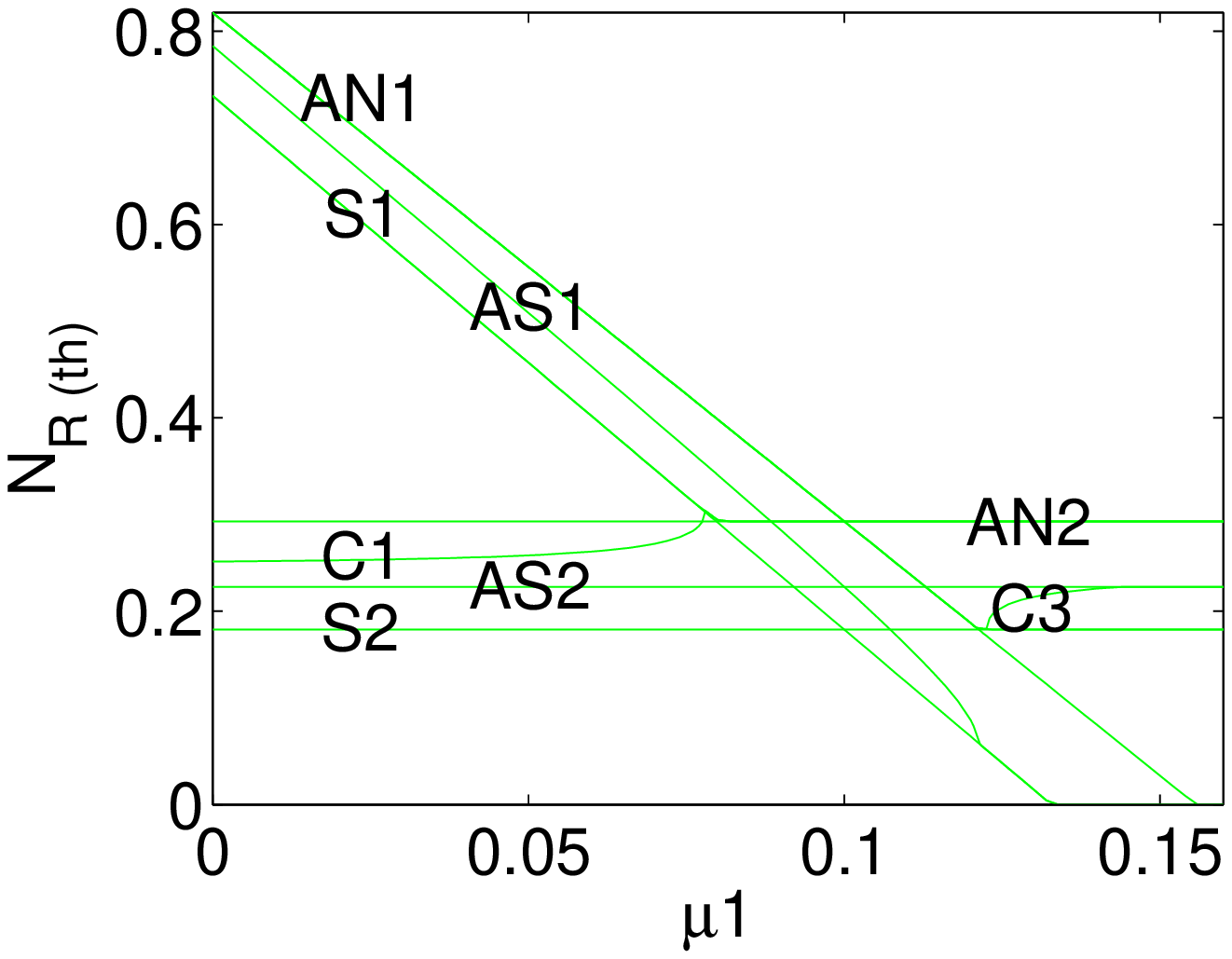}\newline
\includegraphics[width=.4\textwidth]{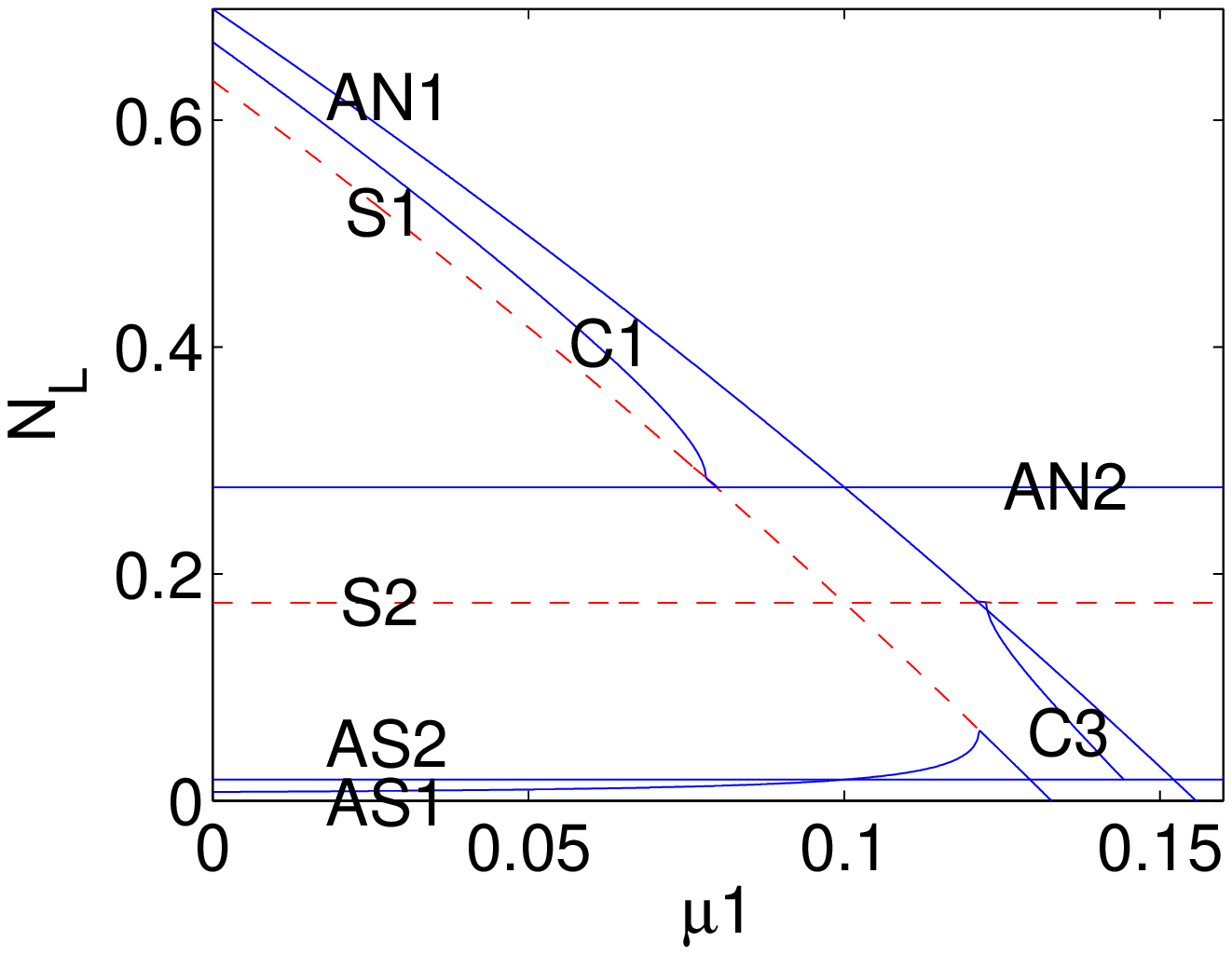} %
\includegraphics[width=.4\textwidth]{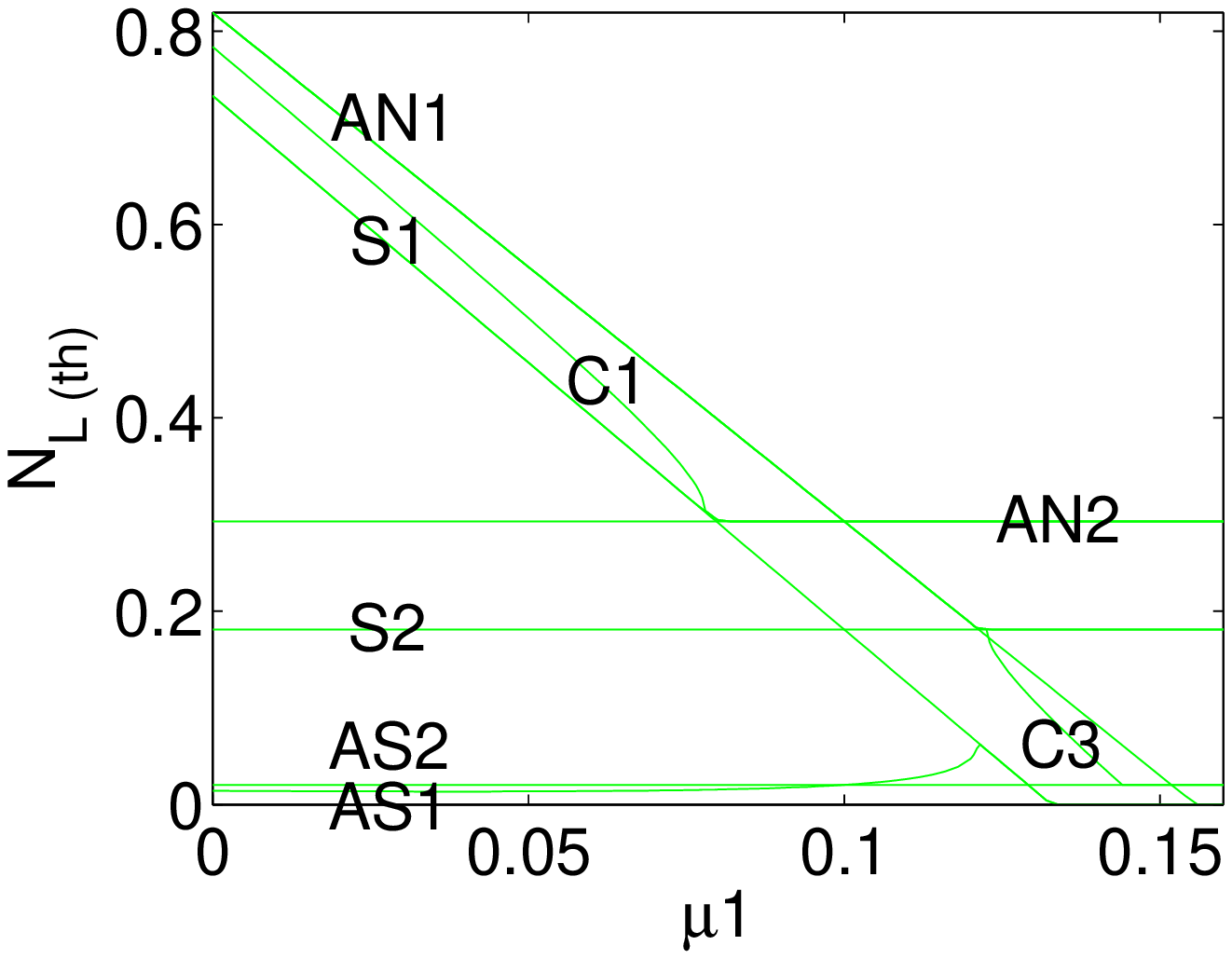}\newline
\caption{(Color online) The norms of the right (top) and left (bottom) parts
of the wave functions, $N_{R}=\protect\int_{0}^{+\infty }|u|^{2}+|v|^{2}dx$,
and $N_{L}=\protect\int_{-\infty }^{0}|u|^{2}+|v|^{2}dx$, as obtained from
the numerical (left) and approximate two-mode (right) solutions of Eq. (%
\protect\ref{eq1}) with the attractive nonlinearity ($\protect\sigma =-1$),
as a function of $\protect\mu _{1}$, for $\protect\mu _{2}=0.10$. The
notation is the same as in Fig. \protect\ref{fig1}. Notice that, as in the
top left panel of Fig. \protect\ref{fig1}, the branches C2 and C4 are
actually indiscernible.}
\label{figadd}
\end{figure}

Two additional descriptions of these bifurcations, which help to understand
the emergence of the combined branches, are presented in Figs. \ref{fig2}
and \ref{figadd}. The former one shows norms $N_{u}=\int_{-\infty }^{+\infty
}|u|^{2}dx$ and $N_{v}=\int_{-\infty }^{+\infty }|v|^{2}dx$, and the latter
figure shows the total number of atoms in the left and right wells, $%
N_{L}=\int_{-\infty }^{0}\left( |u|^{2}+|v|^{2}\right) dx$ and $%
N_{R}=\int_{0}^{\infty }\left( |u|^{2}+|v|^{2}\right) dx$, as functions of
the chemical potential $\mu _{1}$. The left panels are produced by the
numerical solution of the stationary version of underlying GPEs, Eqs. (\ref%
{eq1}), while the right panels correspond to the finite-mode approximation.
Notice a sharp (near vertical) form of the combined branches C2 and C4 in
the former figure, which indicates a short interval of their existence, and
their change of stability upon the collision with (or, more appropriately,
the bifurcation of) C1 and C3, respectively. In the latter figure, it is
worthwhile to notice the similarity of the curves for $N_{L}$ and $N_{R}$
appertaining to branches S1, AN1, S2 and AN2, in contrast with the
differences between the respective curves for AS1, AS2 and combined branches
C1 and C3, which indicates the asymmetry between the states trapped in the
two wells in the latter case.

\begin{figure}[tbph]
\centering
\includegraphics[width=.3\textwidth]{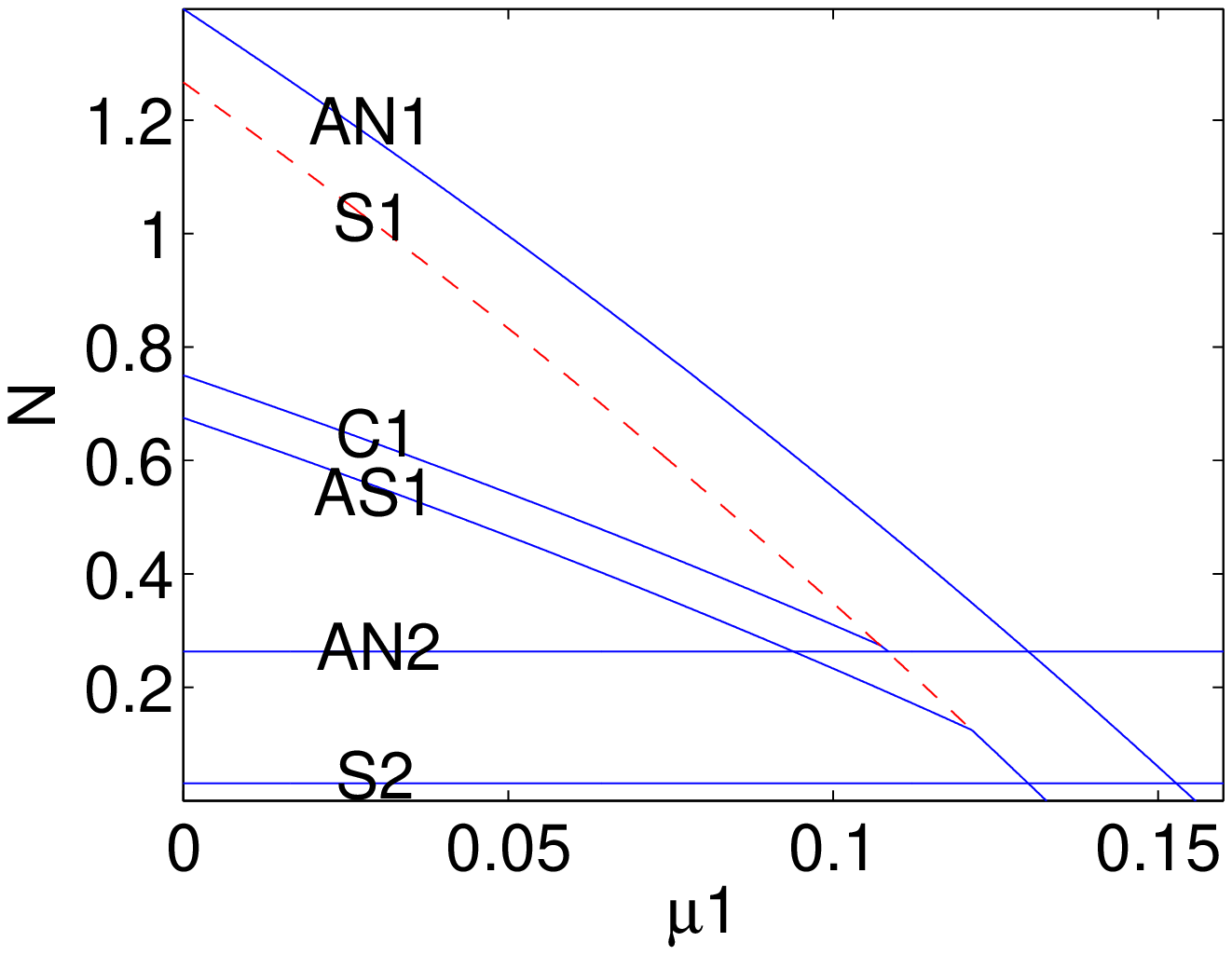} %
\includegraphics[width=.3\textwidth]{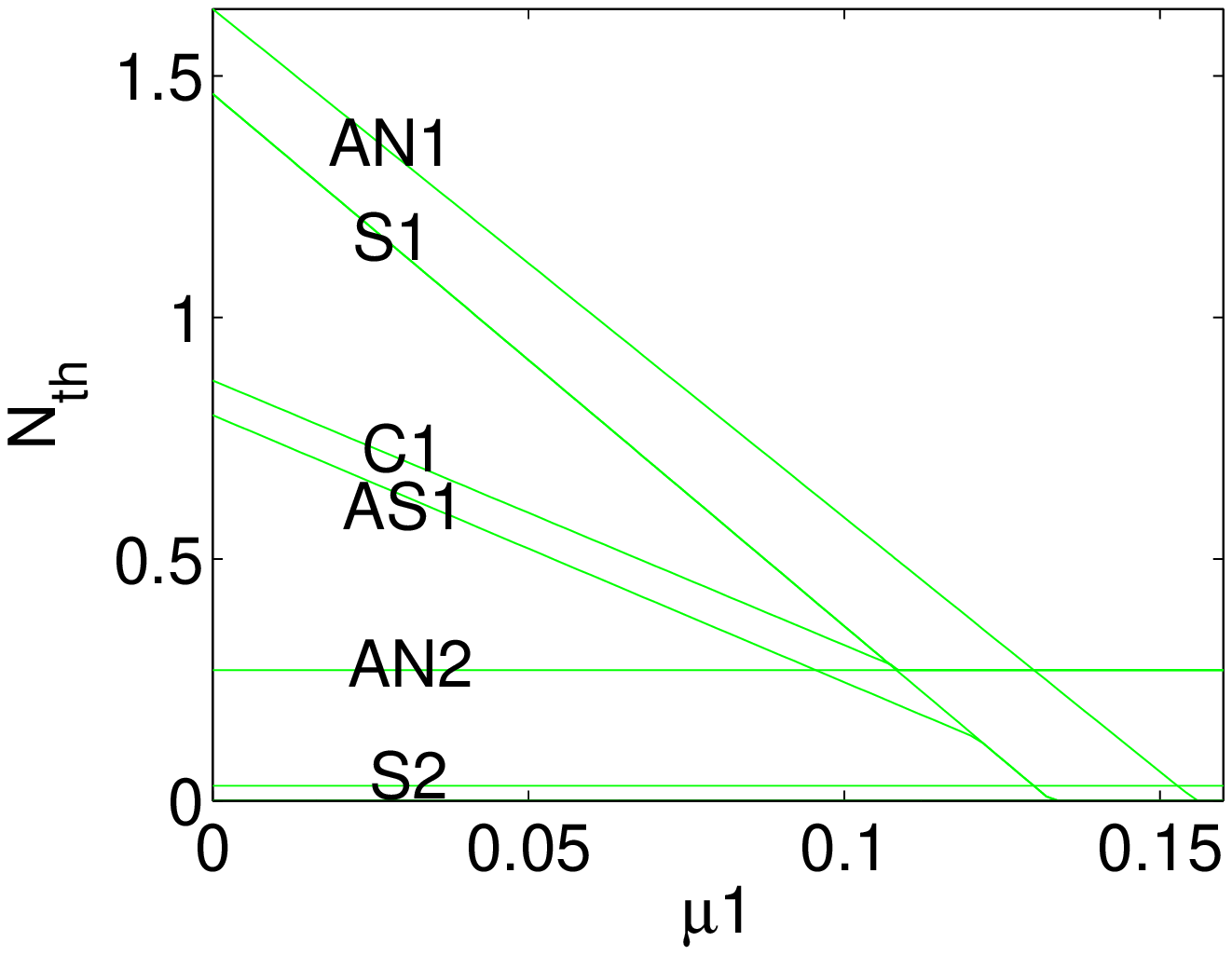}\newline
\includegraphics[width=.3\textwidth]{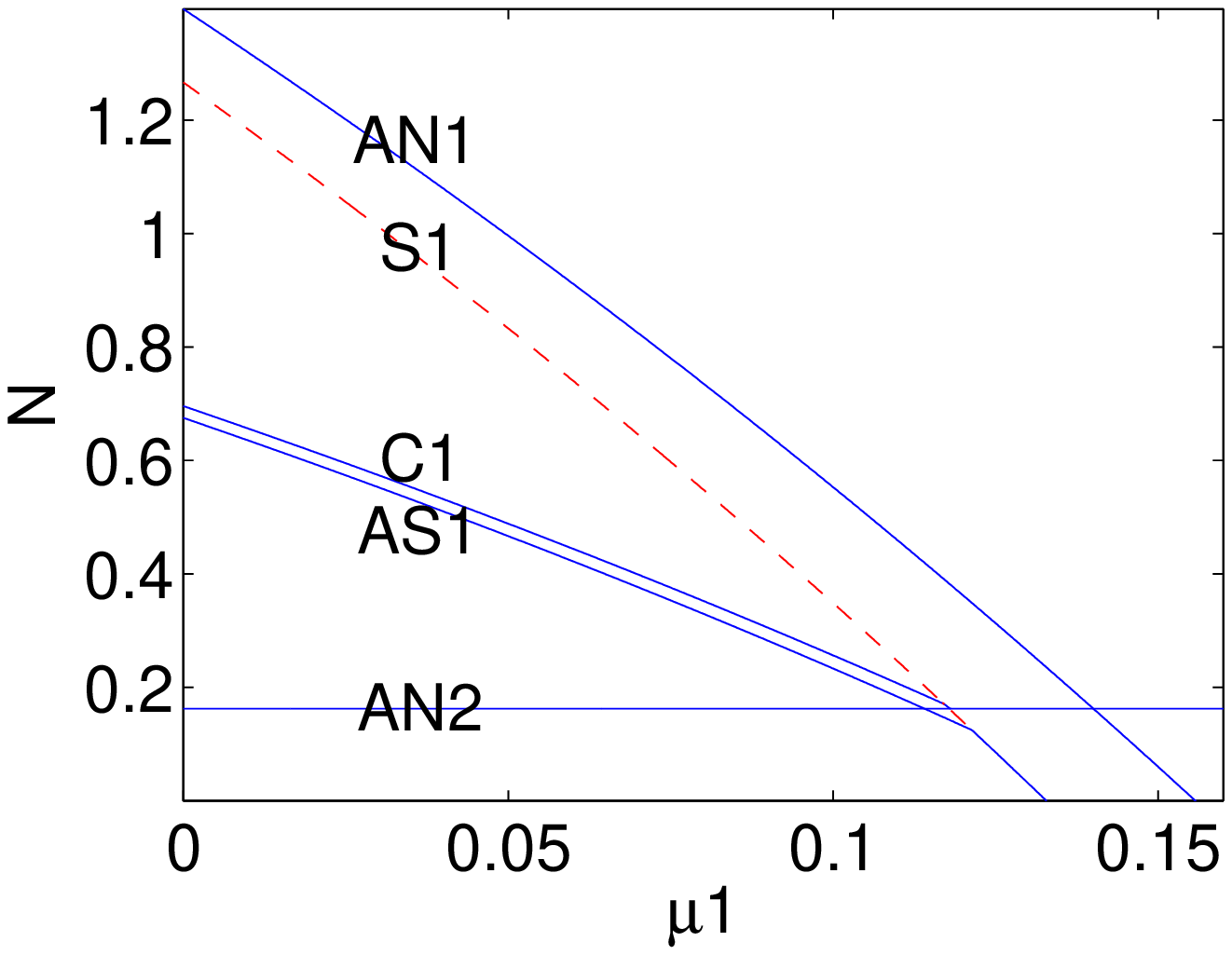} %
\includegraphics[width=.3\textwidth]{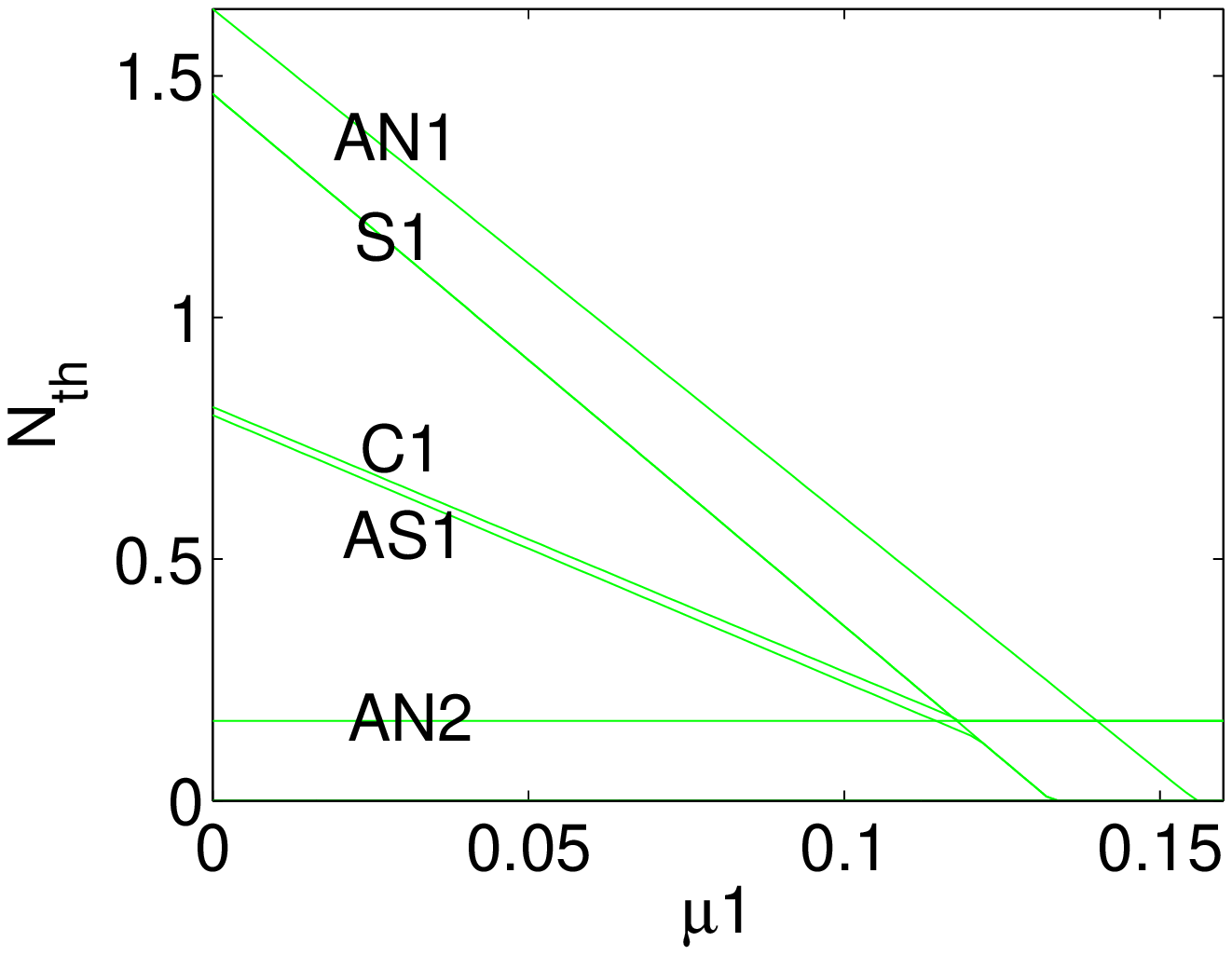}\newline
\includegraphics[width=.3\textwidth]{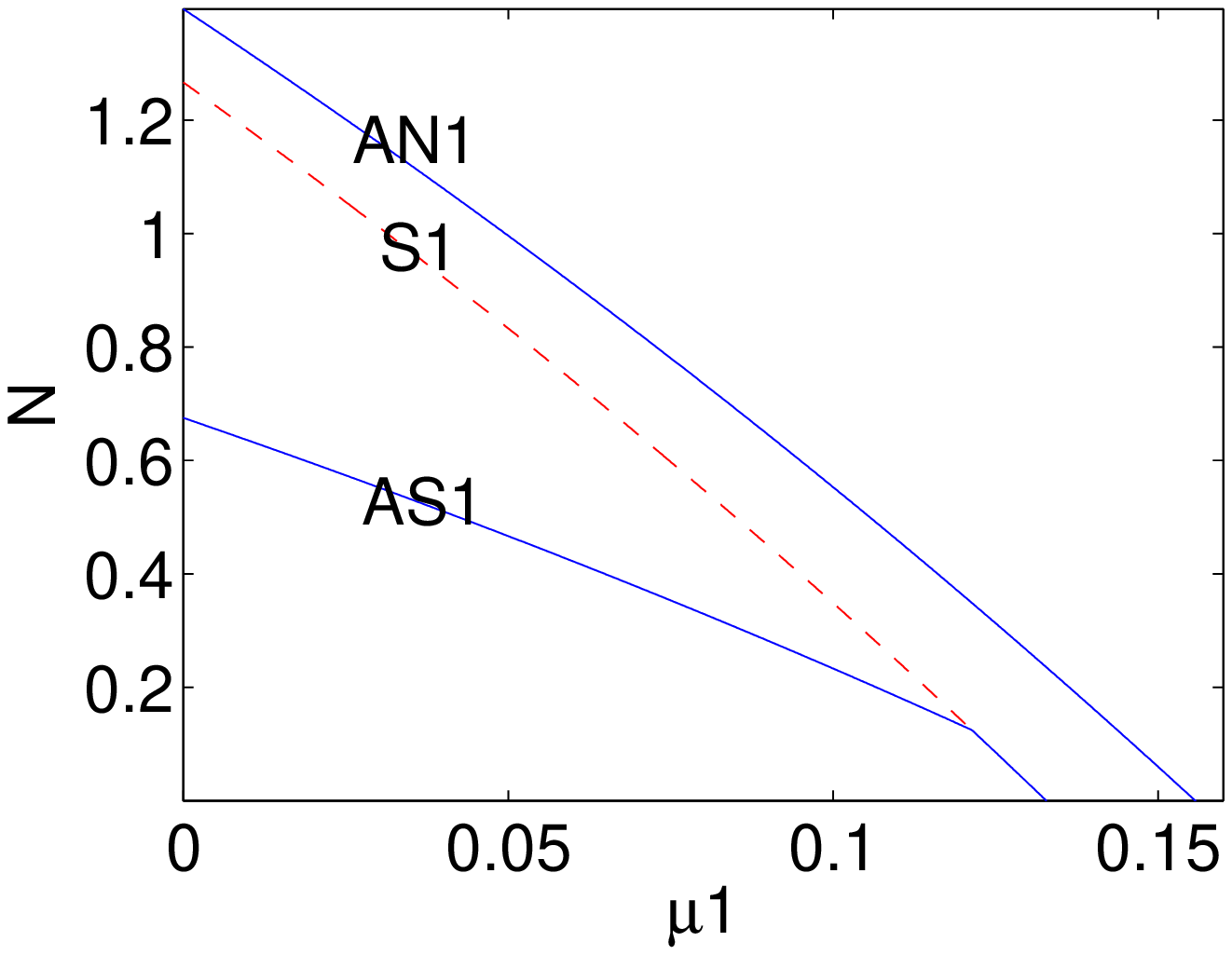} %
\includegraphics[width=.3\textwidth]{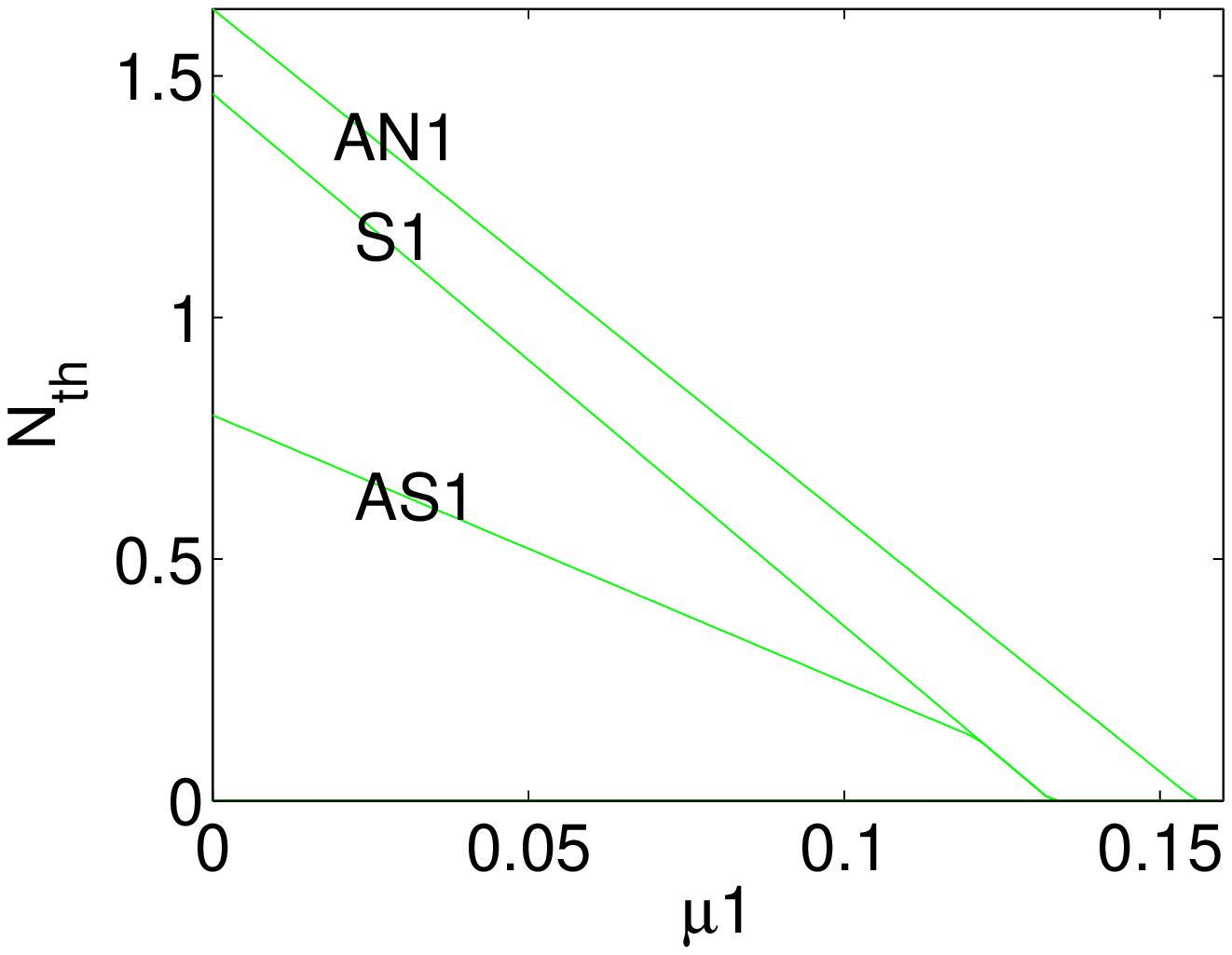}\newline
\caption{(Color online) The norm of the numerical (left) and approximate
two-mode (right) stationary solutions of Eq. (\protect\ref{eq1}) for the
attractive nonlinearity ($\protect\sigma =-1$) as a function of $\protect\mu %
_{1}$ for $\protect\mu _{2}=0.13$ (top), $\protect\mu _{2}=0.14$ (middle),
and $\protect\mu _{2}=0.16$ (bottom). The notation is the same as in Fig.
\protect\ref{fig1}.}
\label{fig3}
\end{figure}

Having addressed the most complex bifurcation scenario observed at $\mu
_{2}<\mu _{2}^{\mathrm{(cr)}}$, we now turn to the three remaining cases,
namely, $\mu _{2}^{\mathrm{(cr)}}<\mu _{2}<\omega _{0}$, $\omega _{0}<\mu
_{2}<\omega _{1}$, and $\omega _{1}<\mu _{2}$. Full numerically generated
bifurcation diagrams for each of these cases can be found in the left panels
of Fig. \ref{fig3}, while the corresponding results produced by algebraic
equations (\ref{rho0})-(\ref{rho3}) are presented in the right panels. Once
again, we notice a very good agreement between the two sets of the results.
In interval $\mu _{2}^{\mathrm{(cr)}}<\mu _{2}<\omega _{0}$, the main
difference from the case shown in Fig. \ref{fig1} is that, as concerns the
single-component branches belonging to the second field, the asymmetric
branch AS2 has not bifurcated from the symmetric one S2, therefore S2 is a
stable branch. As a result, the two-component branches C4 and C3 do not
exist in this case (in particular, the existence of C3 is not possible
topologically, since it would destabilize C4 which would subsequently have
to merge with stable branch S2).

Nevertheless, the behavior of branches C1 and C2 remains the same as before.
In the middle panels of Fig. \ref{fig3}, which represents the case of $%
\omega _{0}<\mu _{2}<\omega _{1}$, the situation is simpler in that branch
S2 does not exist (in this regime, only branch AN2 exists in the second
component), hence the presence of C3 and C4, that would connect S2 to AN1,
is impossible. Finally, in the case of $\omega _{1}<\mu _{2}$, there are no
single-component branches in the second field; as a result, even the
two-component branch C2, joining AN2 and S1, has to disappear (hence C1
bifurcating from it cannot exist either), and we are therefore left solely
with the single-component solutions for $u$ (only the branches S1, AN1 and
AS1 are present).

Examples of solutions representing all the branches considered above are
shown in Figs. \ref{fig4}-\ref{fig7}. Figure \ref{fig4} represents the three
branches that have only the $u$ component, namely S1, AN1 and AS1 in the
left, middle and right panels of the figure, for $\mu _{1}=0.04$ and $\mu
_{2}=0.1$. Notice that, since solution AS1 exists for this parameter sets,
solution S1 is unstable, while AN1 and AS1 are stable.

The results of the linear-stability analysis around the solutions are shown
in the bottom panels of the figure. For the purpose of this analysis,
perturbed versions of stationary solutions $\left\{
u_{0}(x),v_{0}(x)\right\} $ are taken as
\begin{eqnarray}
u(x,t) &=&u_{0}(x)+\epsilon \left( U_{1}(x)e^{\lambda t}+U_{2}^{\star
}(x)e^{\lambda ^{\star }t}\right) ,  \label{stab1} \\
v(x,t) &=&v_{0}(x)+\epsilon \left( V_{1}(x)e^{\lambda t}+V_{2}^{\star
}(x)e^{\lambda ^{\star }t}\right) ,  \label{stab2}
\end{eqnarray}%
where $\epsilon $ is an infinitesimal amplitude of perturbations, and the
resulting linearized equations for eigenvalue $\lambda $ and eigenvector $%
(U_{1},U_{2},V_{1},V_{2})^{T}$ are solved numerically. As usual, instability
is manifested by the existence of eigenvalue(s) $\lambda $ with a non-zero
real part. The stability results are shown in terms of the spectral plane $%
(\lambda _{r},\lambda _{i})$ for $\lambda =\lambda _{r}+i\lambda _{i}$,
hence the instability corresponds to the presence of an eigenvalue in the
right-hand half-plane (for example, for branch S1 in Fig. \ref{fig4}). Very
similar results to those displayed in Fig. \ref{fig4} can be obtained for
solutions S2, AN2 and AS2, in which only the second component is nonzero
(because they are direct counterparts to those in Fig. \ref{fig4}, they are
not shown here).

The combined solutions emerging due to the bridging of the single-component
branches are shown in Figs. \ref{fig6} and \ref{fig7}. The former figure
shows two instances of branch C2 (right and middle panels), before and after
the bifurcation of branch C1 (the latter one is shown in the left panel).
The stability features of these solutions are displayed in the bottom panels
of the figure. Similar features are shown in Fig. \ref{fig7} for
two-components branches C4 and C3.

\begin{figure}[tbph]
\centering
\includegraphics[width=.3\textwidth]{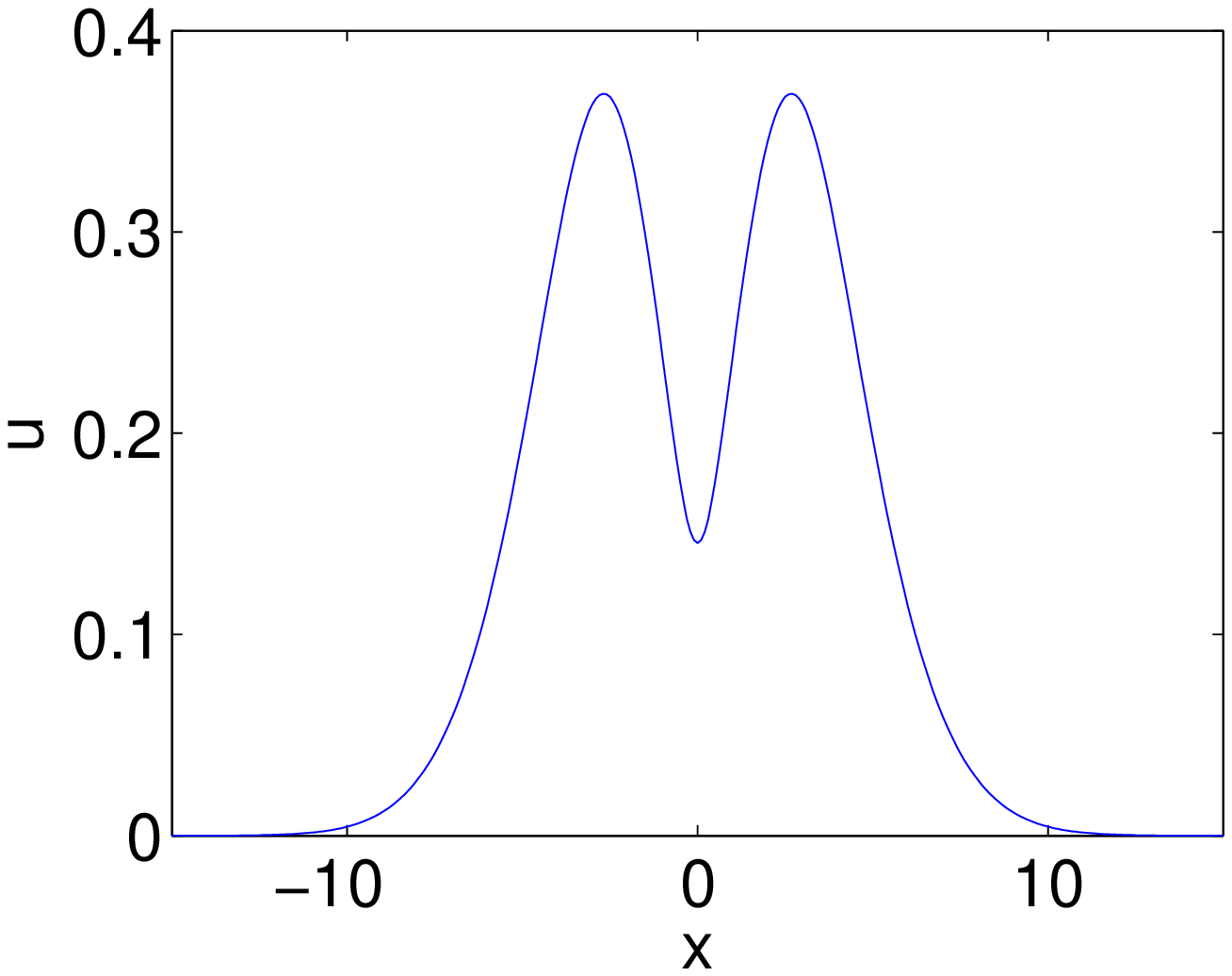} %
\includegraphics[width=.3\textwidth]{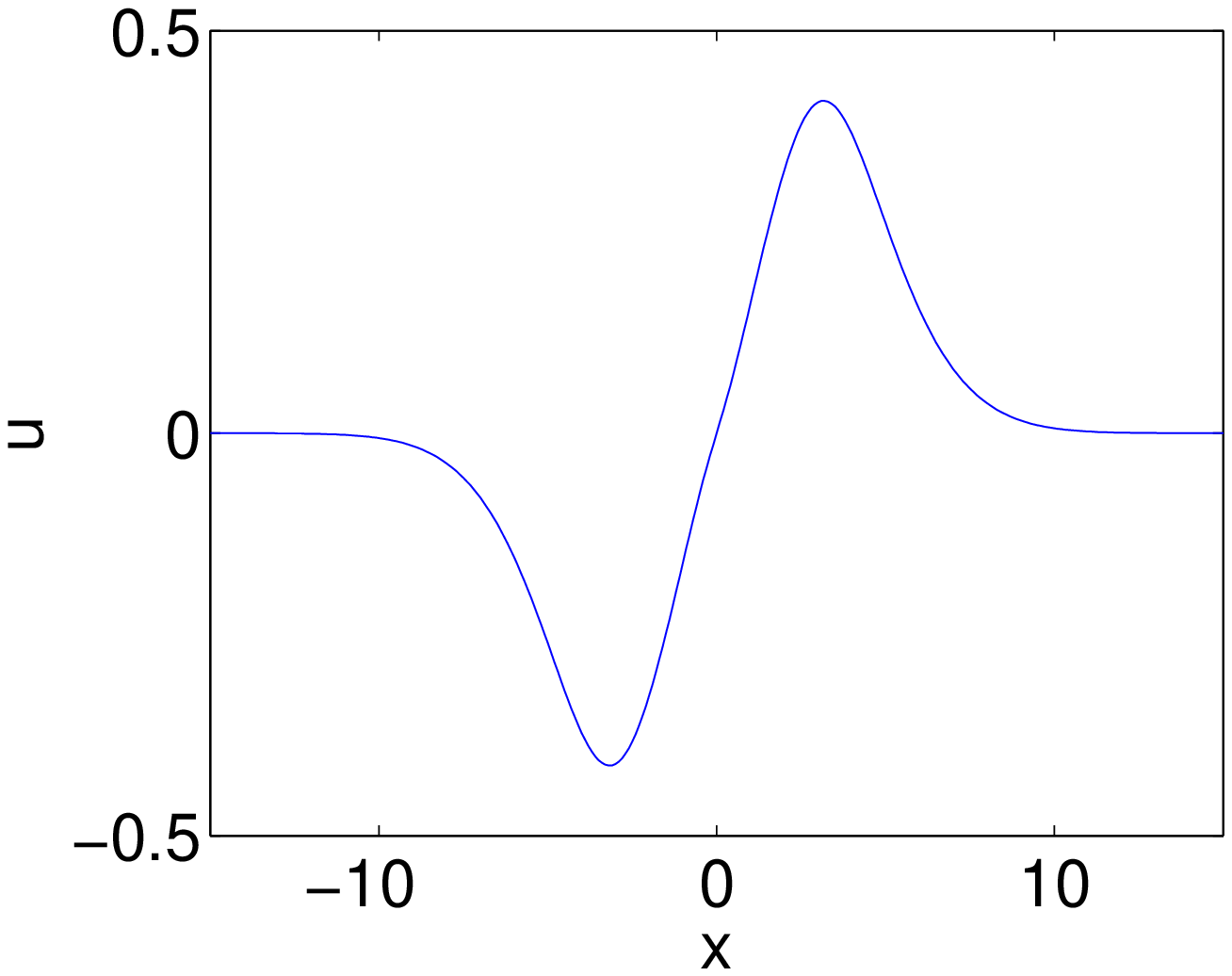} %
\includegraphics[width=.3\textwidth]{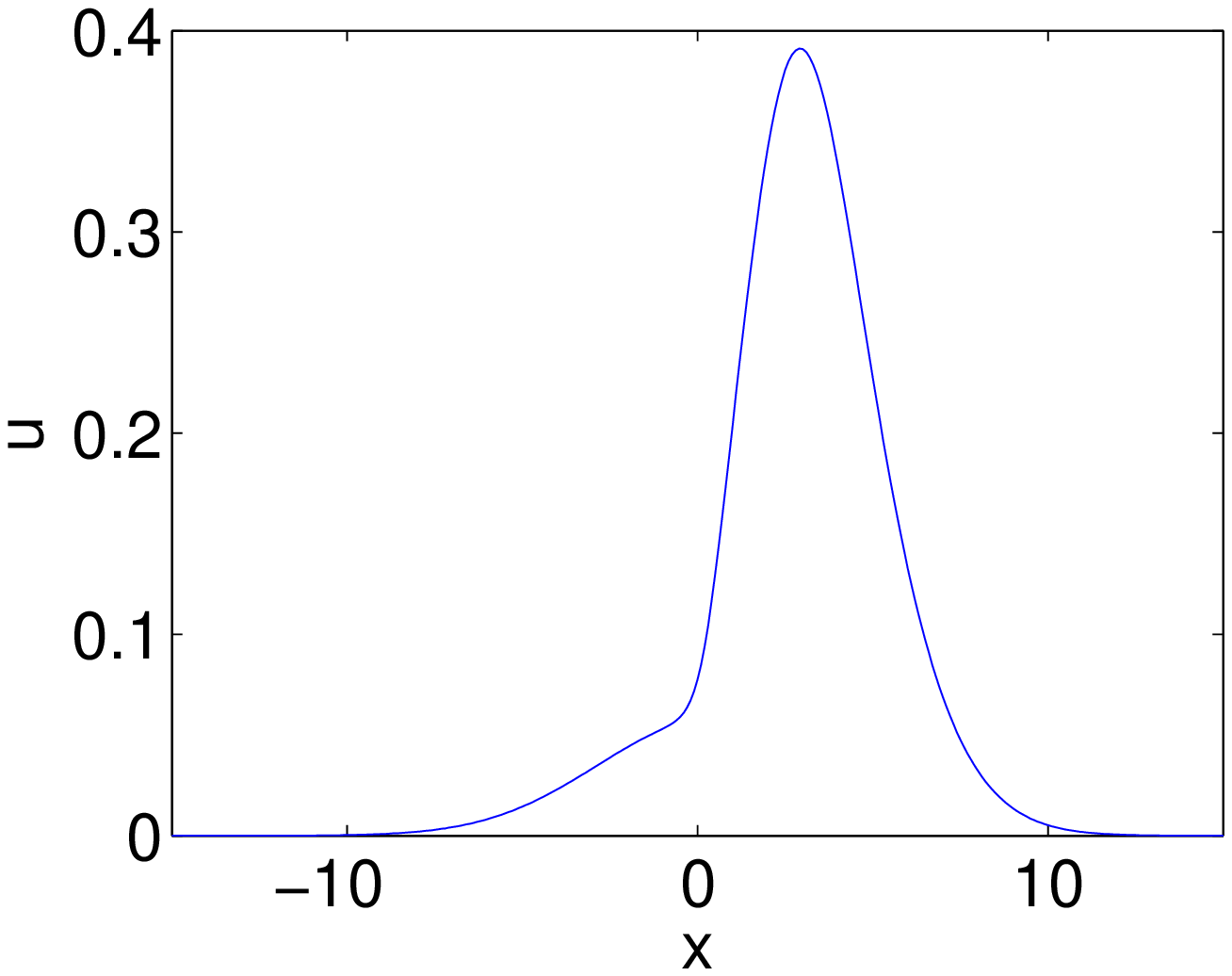}\newline
\includegraphics[width=.3\textwidth]{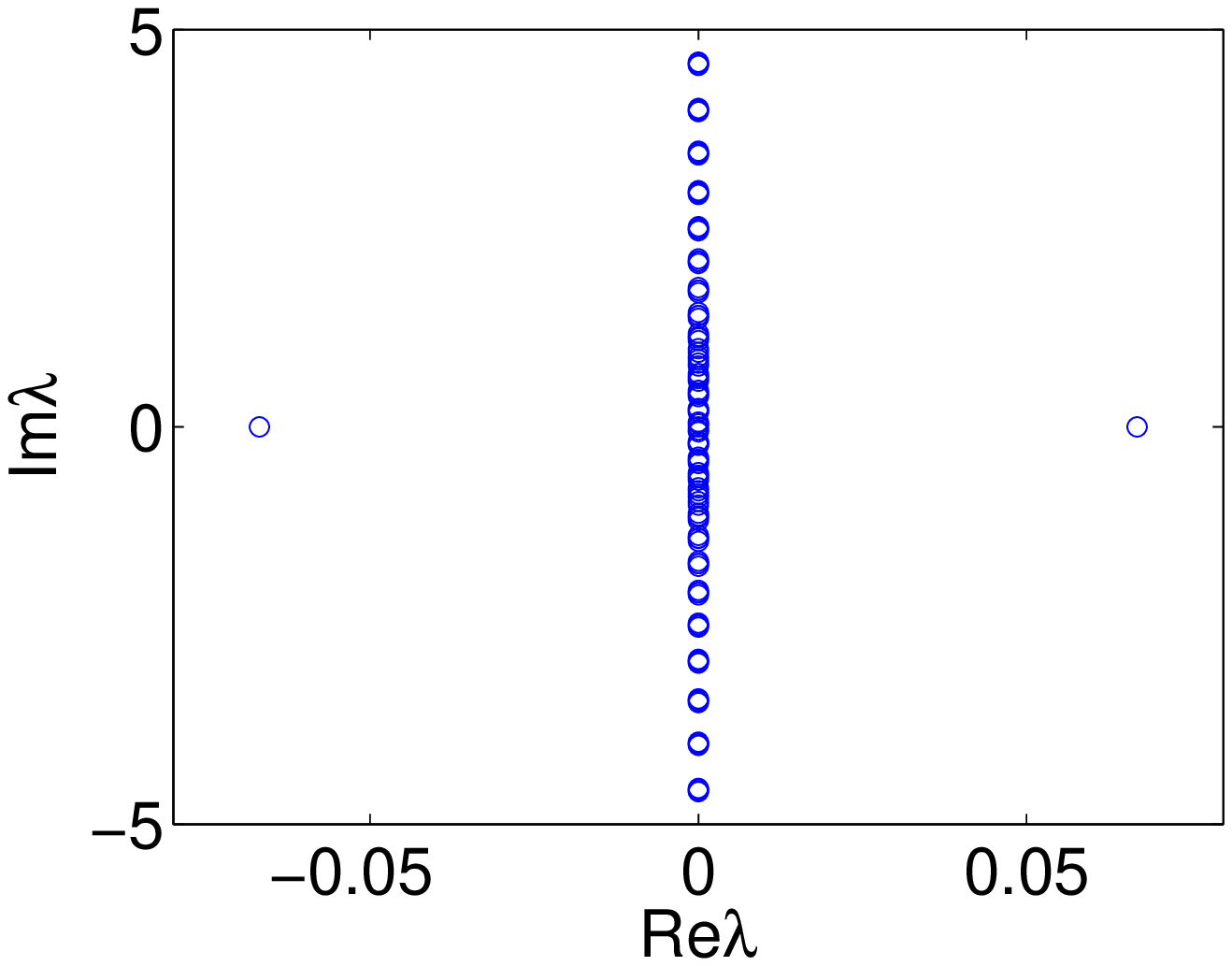} %
\includegraphics[width=.3\textwidth]{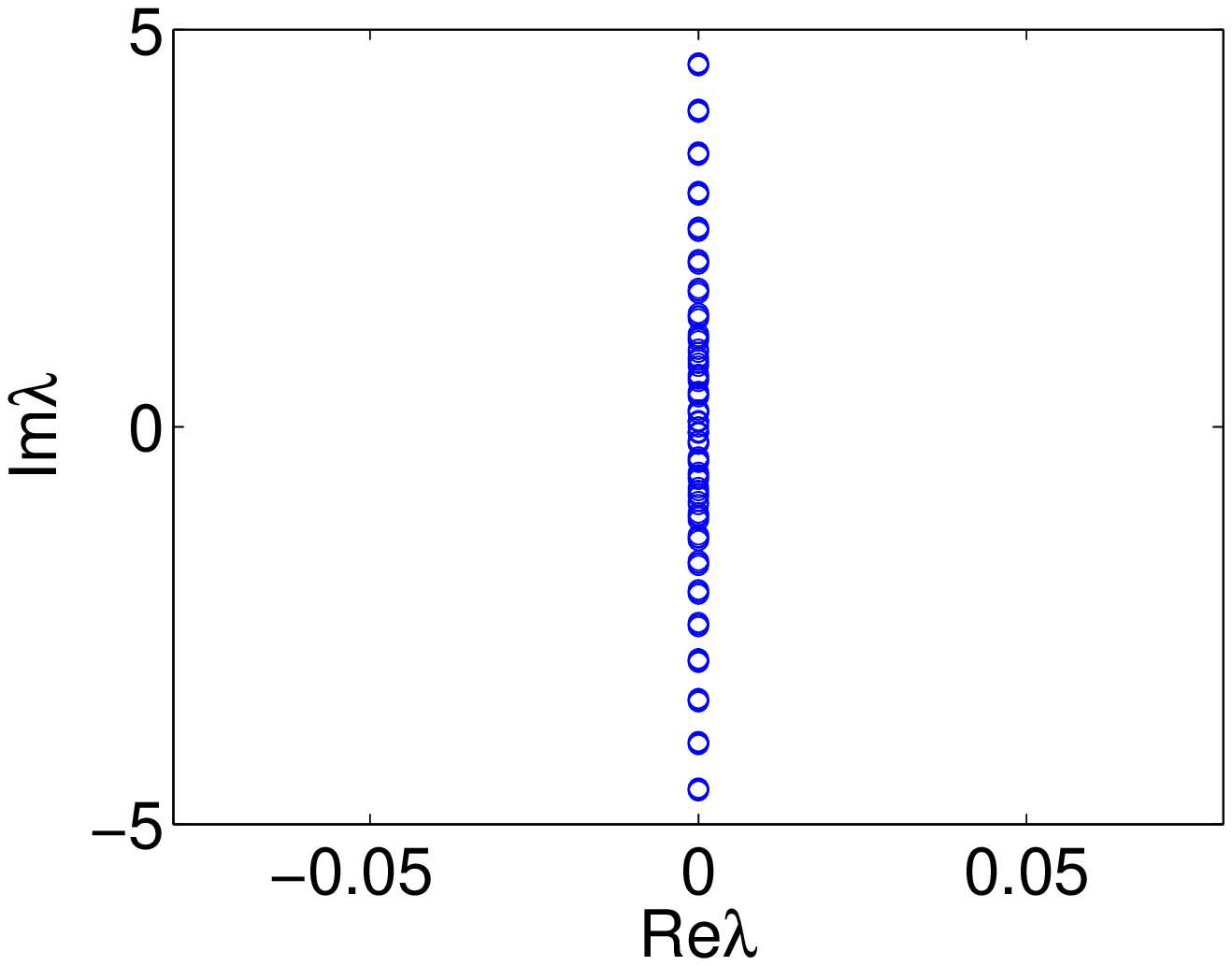} %
\includegraphics[width=.3\textwidth]{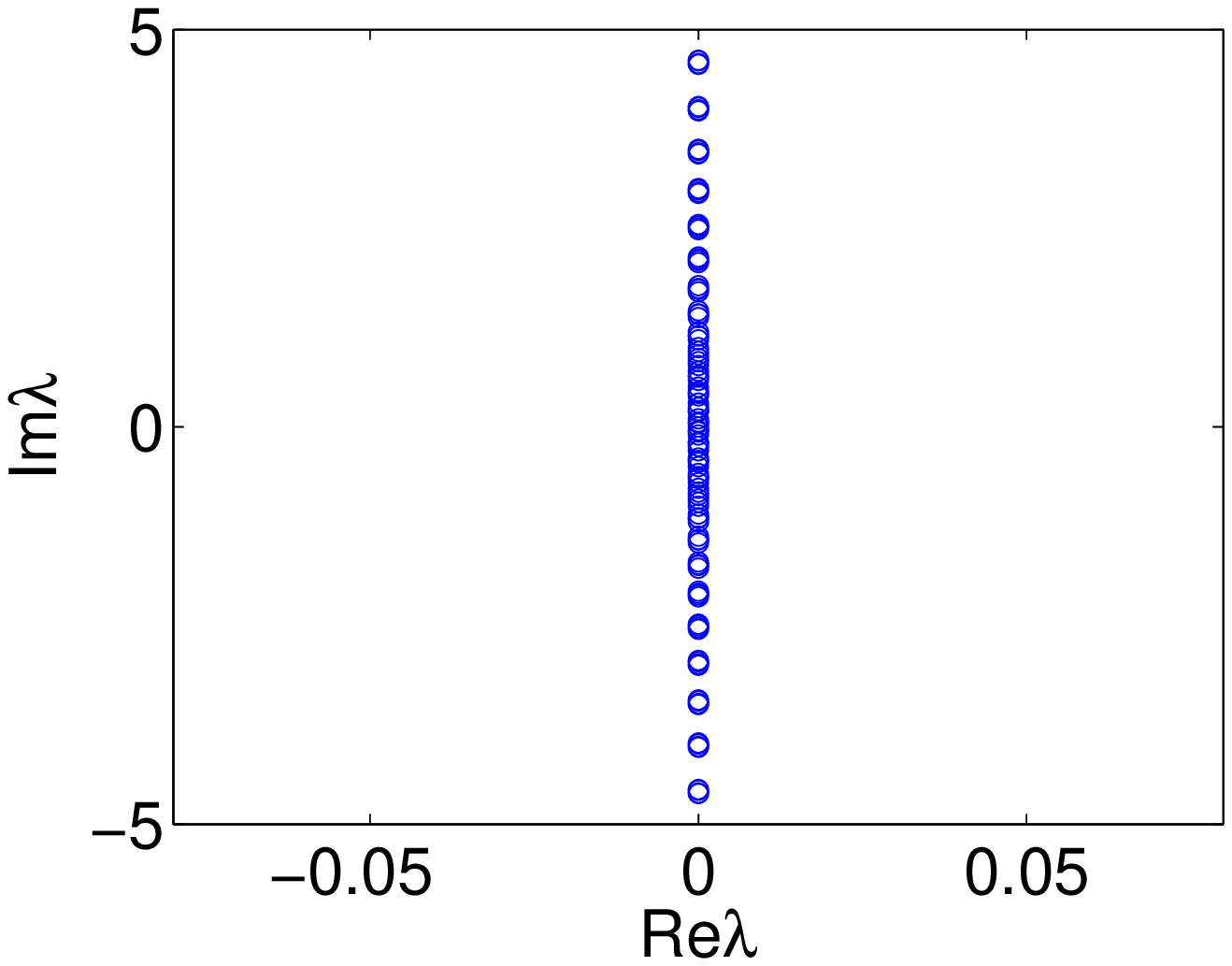}\newline
\caption{(Color online) Top panel: profiles of wave function $u$
corresponding to the symmetric (left), antisymmetric (middle) and asymmetric
(right) branches of the single-component solutions, S1, AN1, and AS1,
respectively, in Fig. \protect\ref{fig1} for $\protect\mu _{1}=0.04$. Bottom
panel: the result of the linear-stability analysis around S1 (left), AN1
(middle) and AS1 (right), in the complex plane of (\textrm{Re}$\protect%
\lambda $, \textrm{Im}$\protect\lambda $). The existence of an eigenvalue
with a positive real part implies instability of the solution.}
\label{fig4}
\end{figure}


\begin{figure}[tbph]
\centering
\includegraphics[width=.3\textwidth]{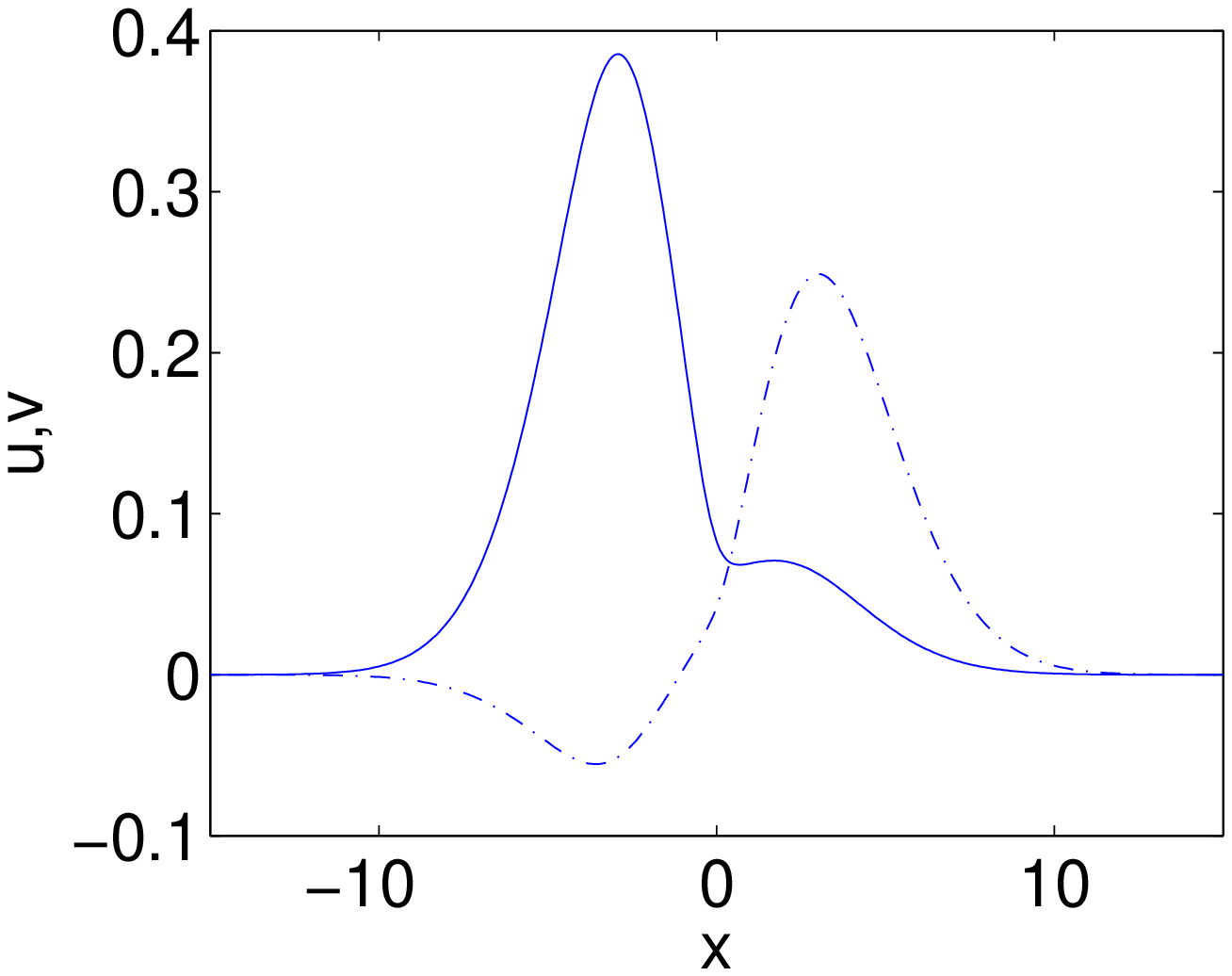} %
\includegraphics[width=.3\textwidth]{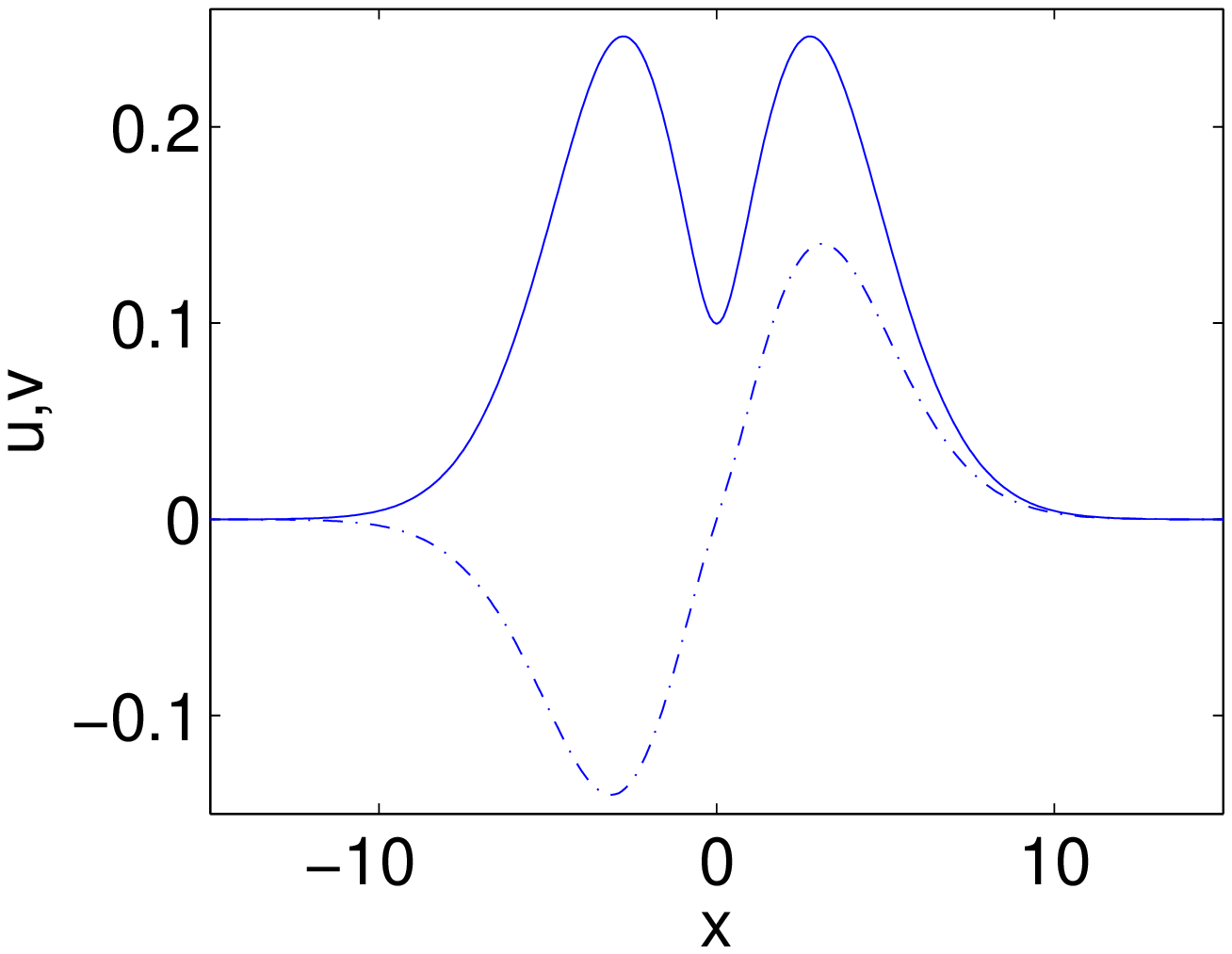} %
\includegraphics[width=.3\textwidth]{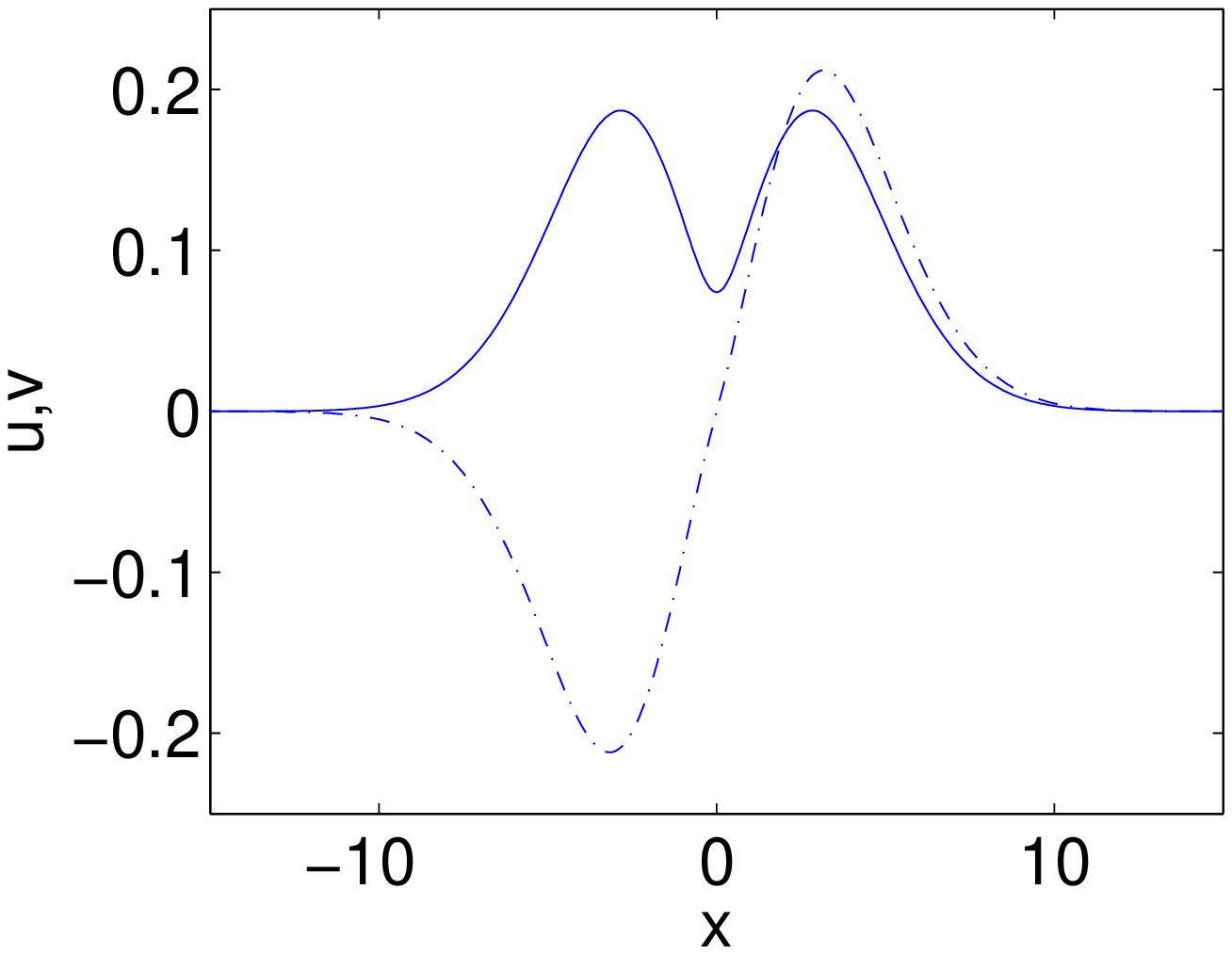}\newline
\includegraphics[width=.3\textwidth]{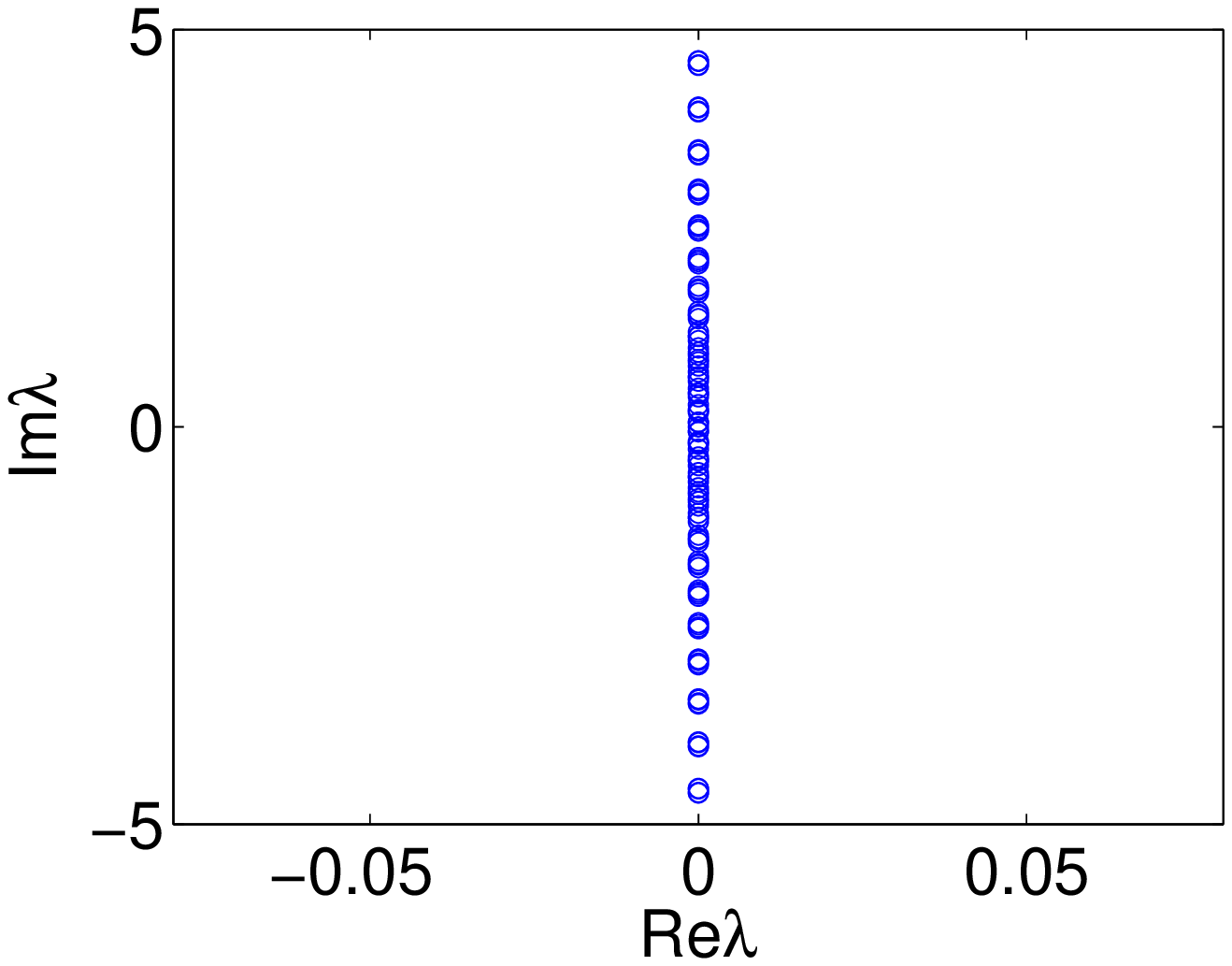} %
\includegraphics[width=.3\textwidth]{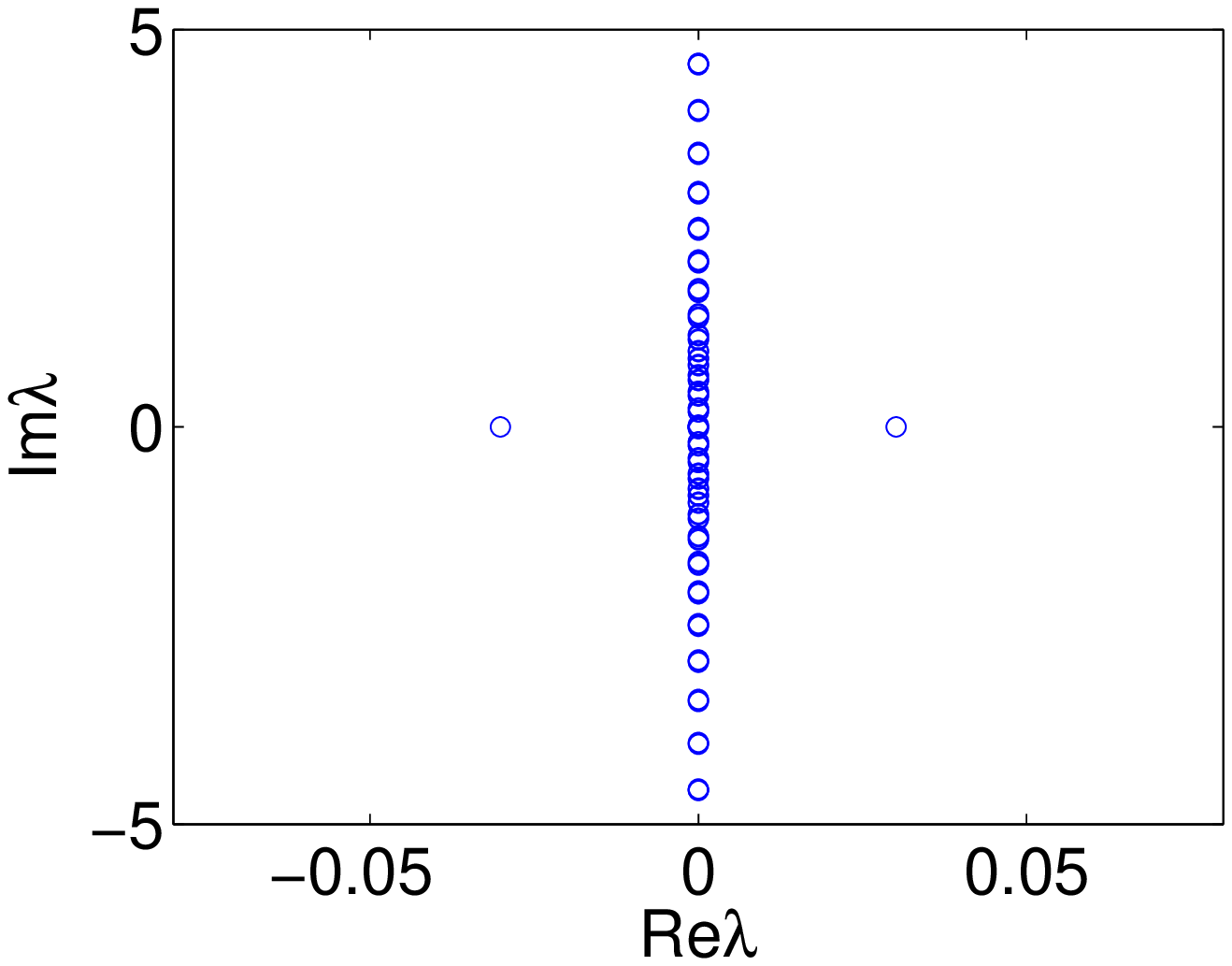} %
\includegraphics[width=.3\textwidth]{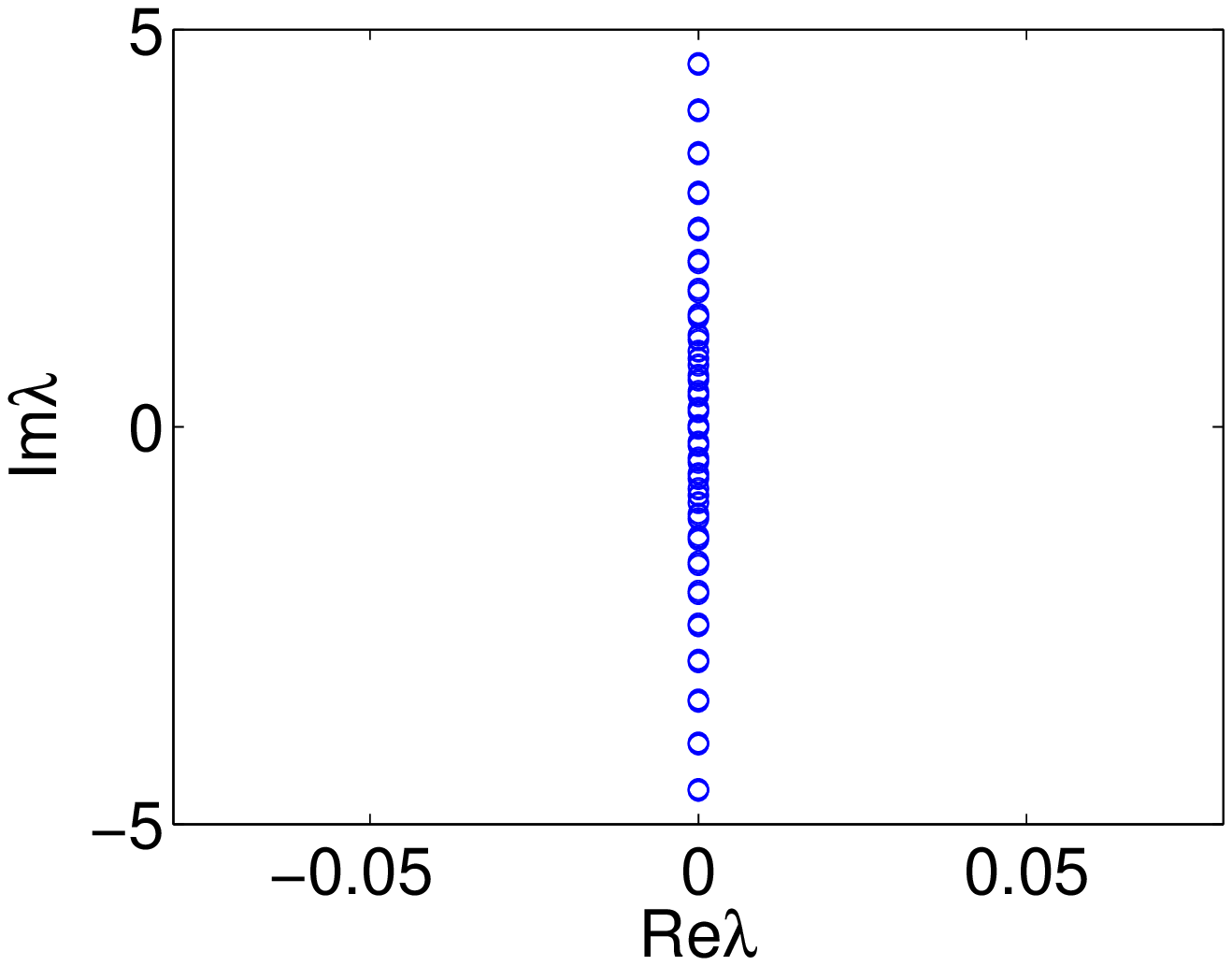}\newline
\caption{(Color online) Profiles of wave functions $u$ (top), $v$ (middle)
and respective stability eigenvalues (bottom) which correspond to
two-component solutions C1 and C2 (branches of these solutions are shown in
Fig. \protect\ref{fig1}): C1 for $\protect\mu _{1}=0.04$ (left), C2 for $%
\protect\mu _{1}=0.0774$ (middle), and C2 for $\protect\mu _{1}=0.0786$
(right).}
\label{fig6}
\end{figure}

\begin{figure}[tbph]
\centering
\includegraphics[width=.3\textwidth]{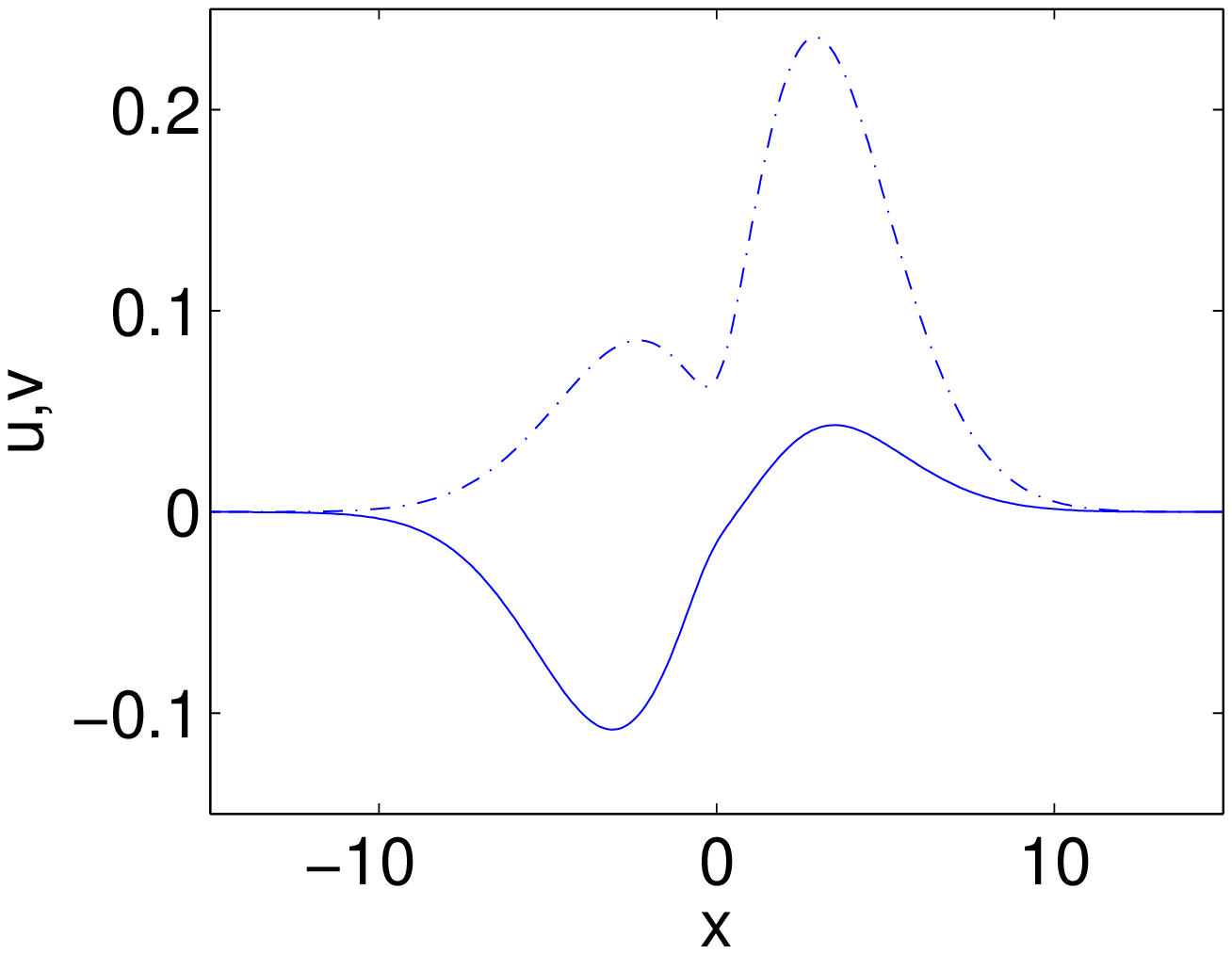} %
\includegraphics[width=.3\textwidth]{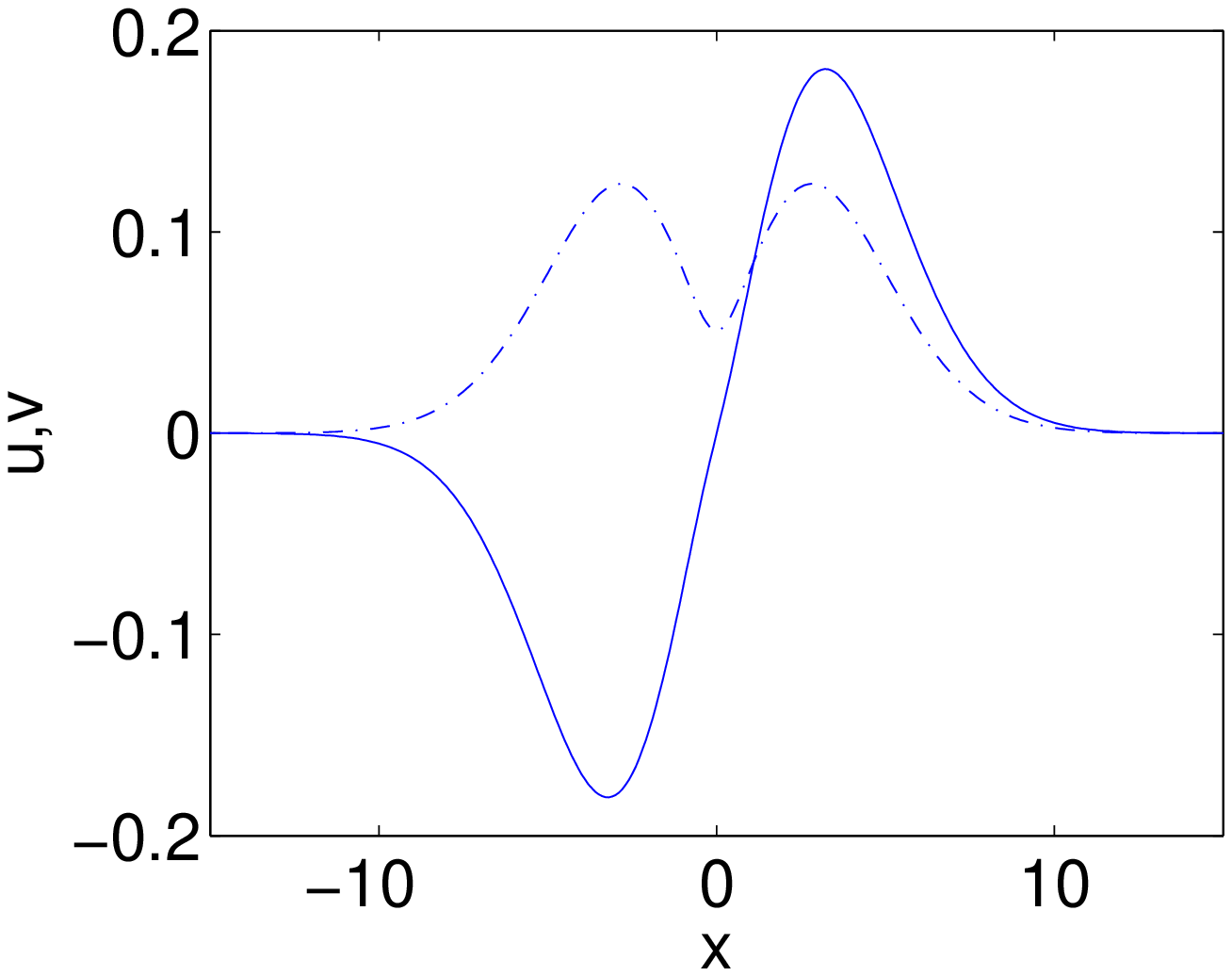} %
\includegraphics[width=.3\textwidth]{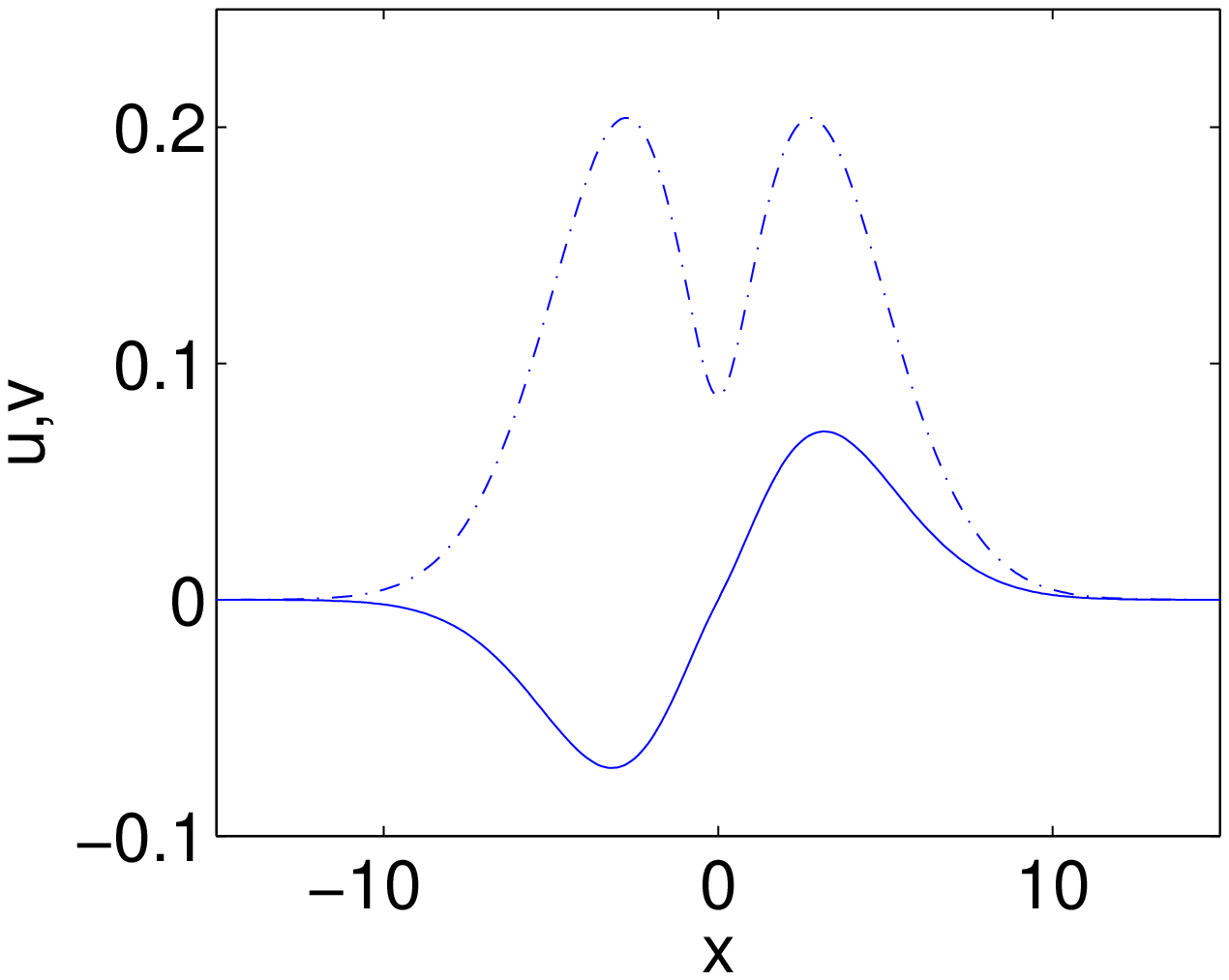}\newline
\includegraphics[width=.3\textwidth]{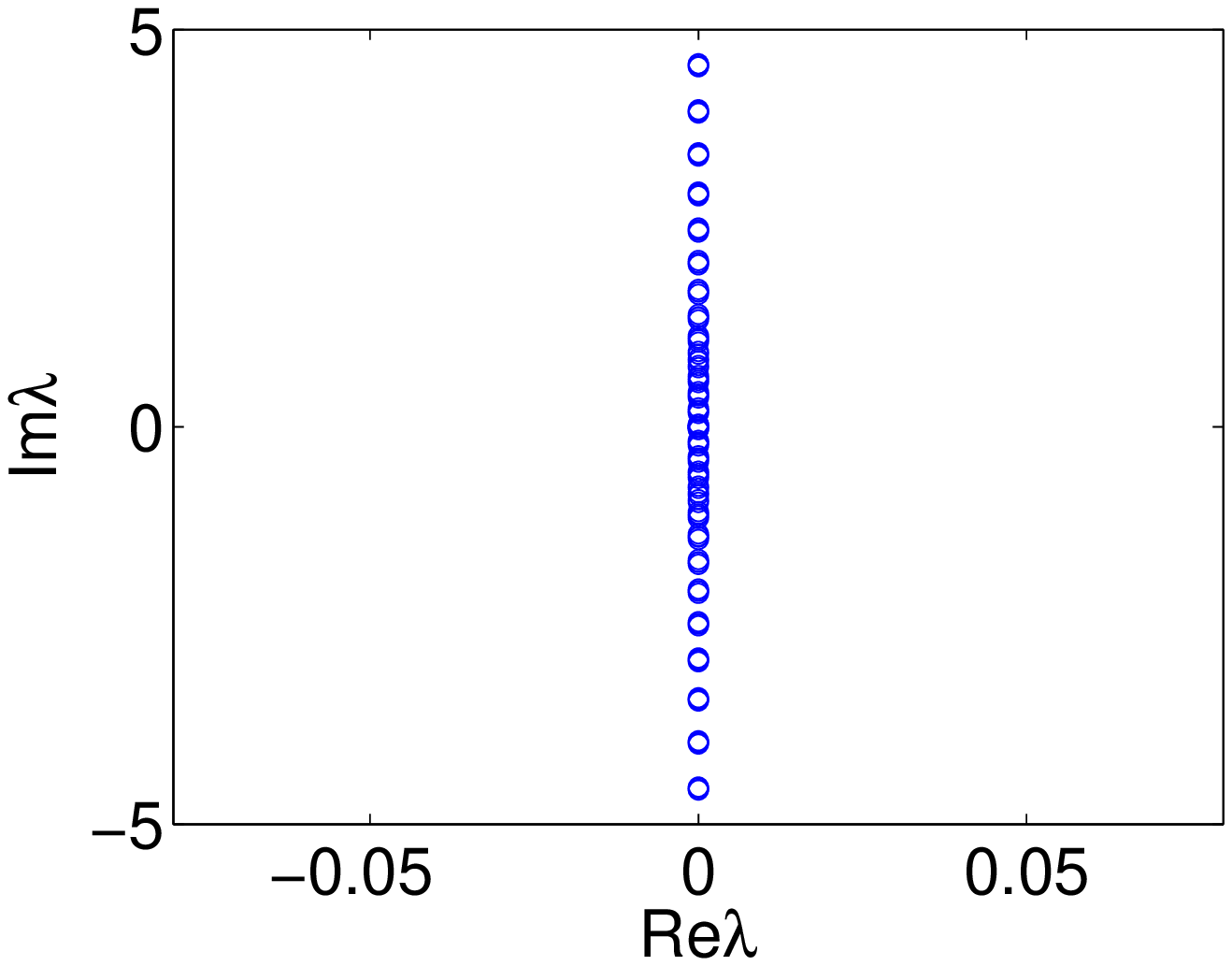} %
\includegraphics[width=.3\textwidth]{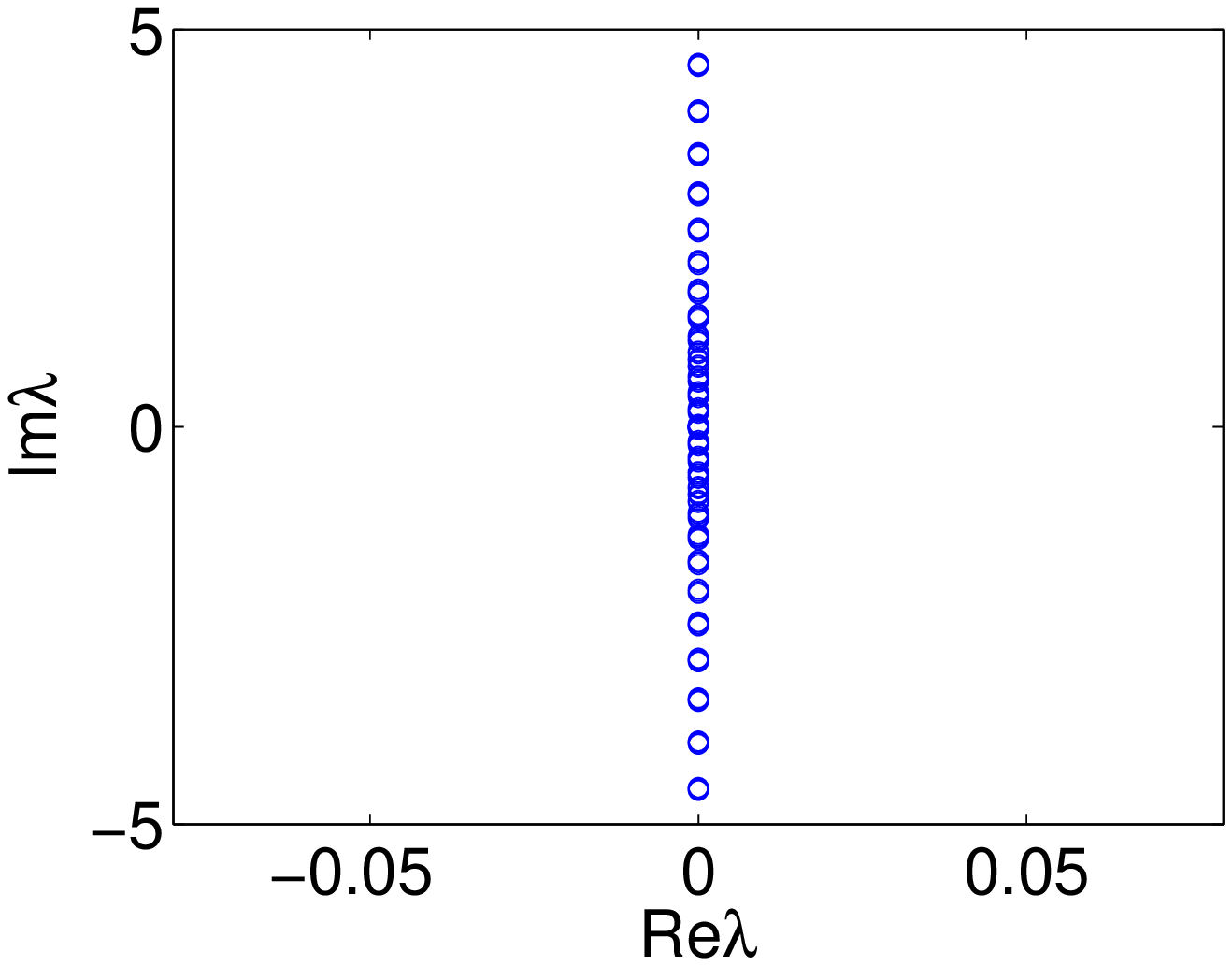} %
\includegraphics[width=.3\textwidth]{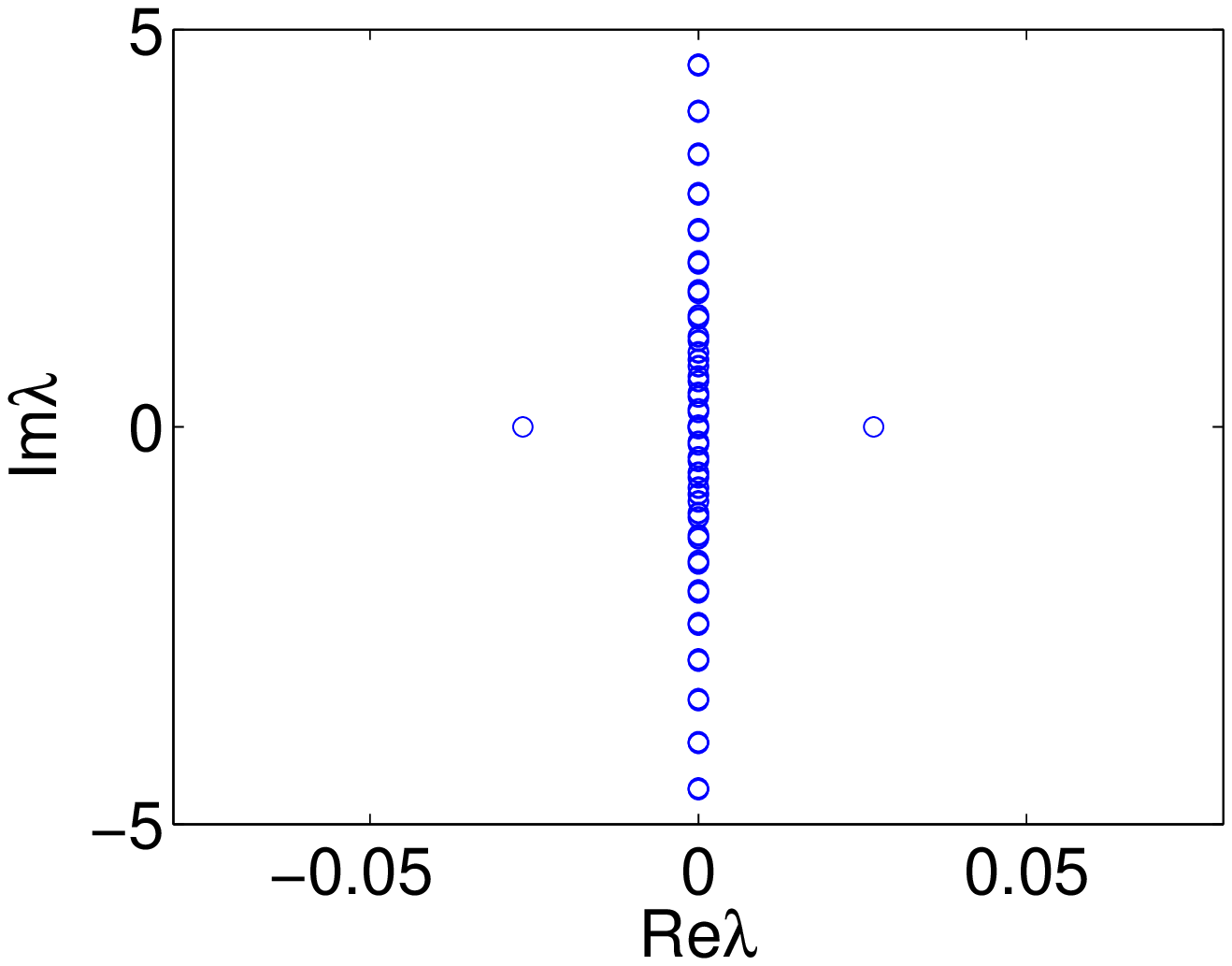}\newline
\caption{Profiles of wave functions $u$ and $v$ (top panel's solid and
dash-dotted line, respectively) and stability eigenvalues (bottom)
corresponding to two-component solutions C3 and C4 (the branches of these
solutions are shown in Fig. \protect\ref{fig1}): C3 for $\protect\mu %
_{1}=0.135$ (left), C4 for $\protect\mu _{1}=0.122$ (middle), and C4 for $%
\protect\mu _{1}=0.123$ (right).}
\label{fig7}
\end{figure}

\begin{figure}[tbph]
\centering
\includegraphics[width=.4\textwidth]{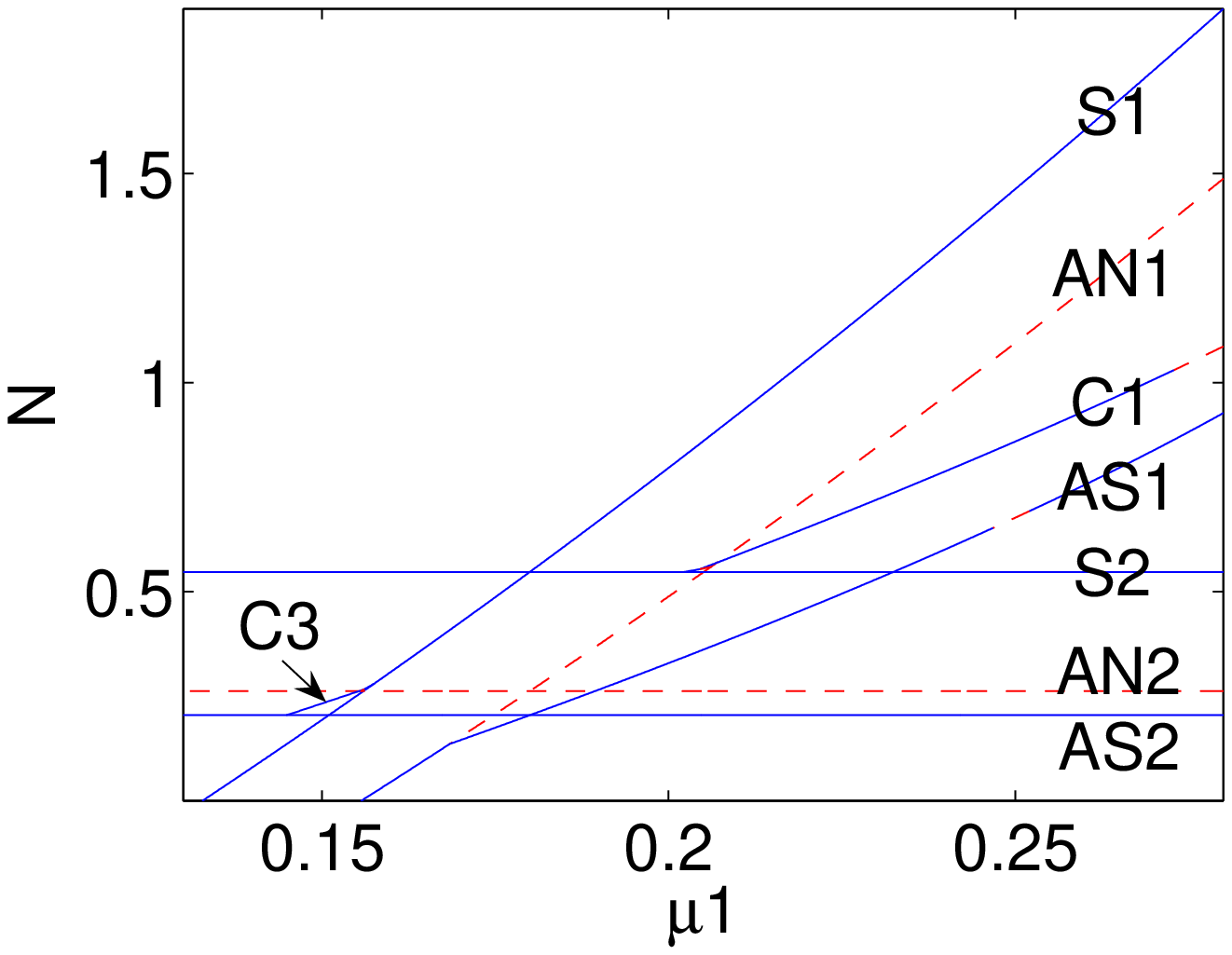} %
\includegraphics[width=.4\textwidth]{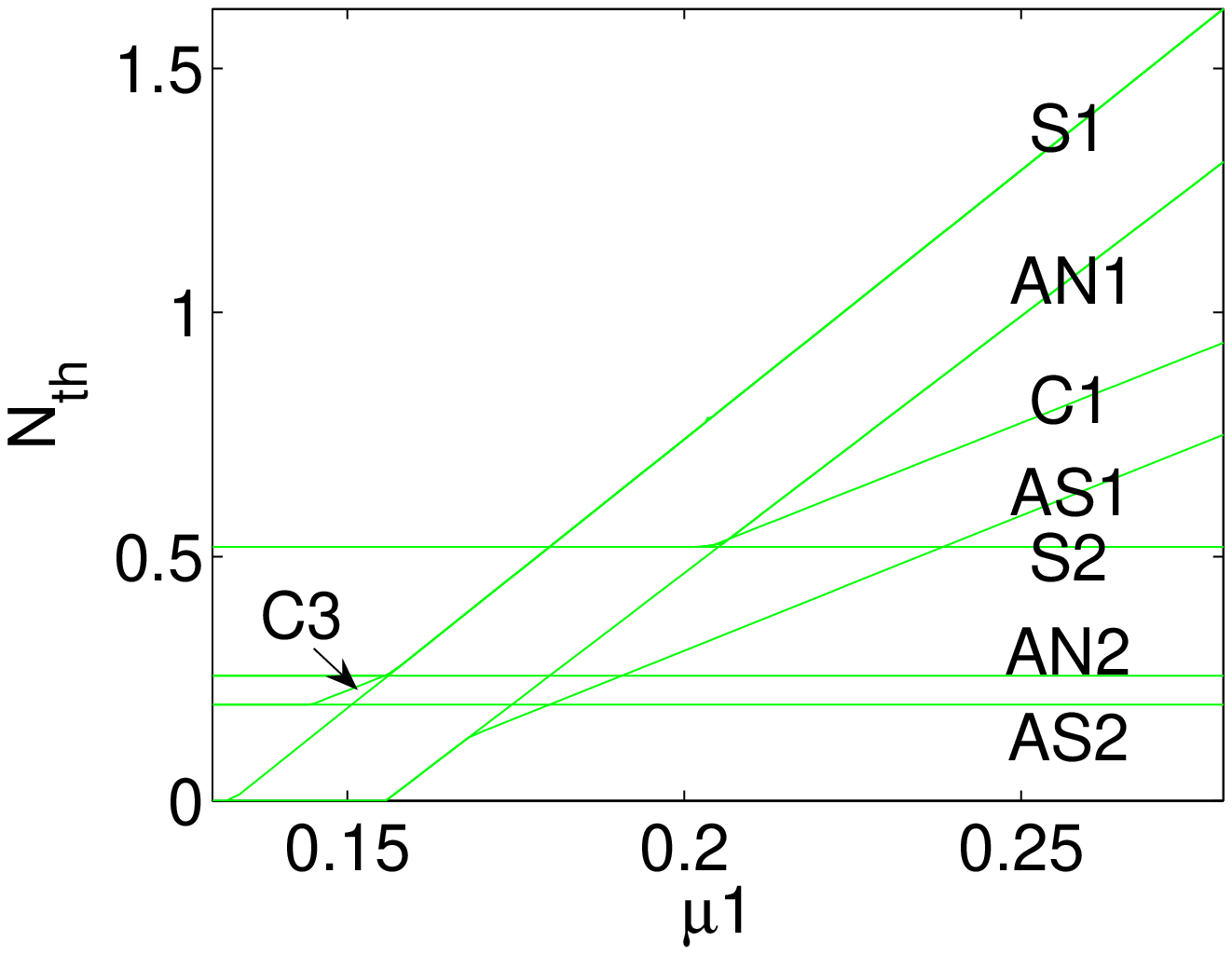}\newline
\includegraphics[width=.4\textwidth]{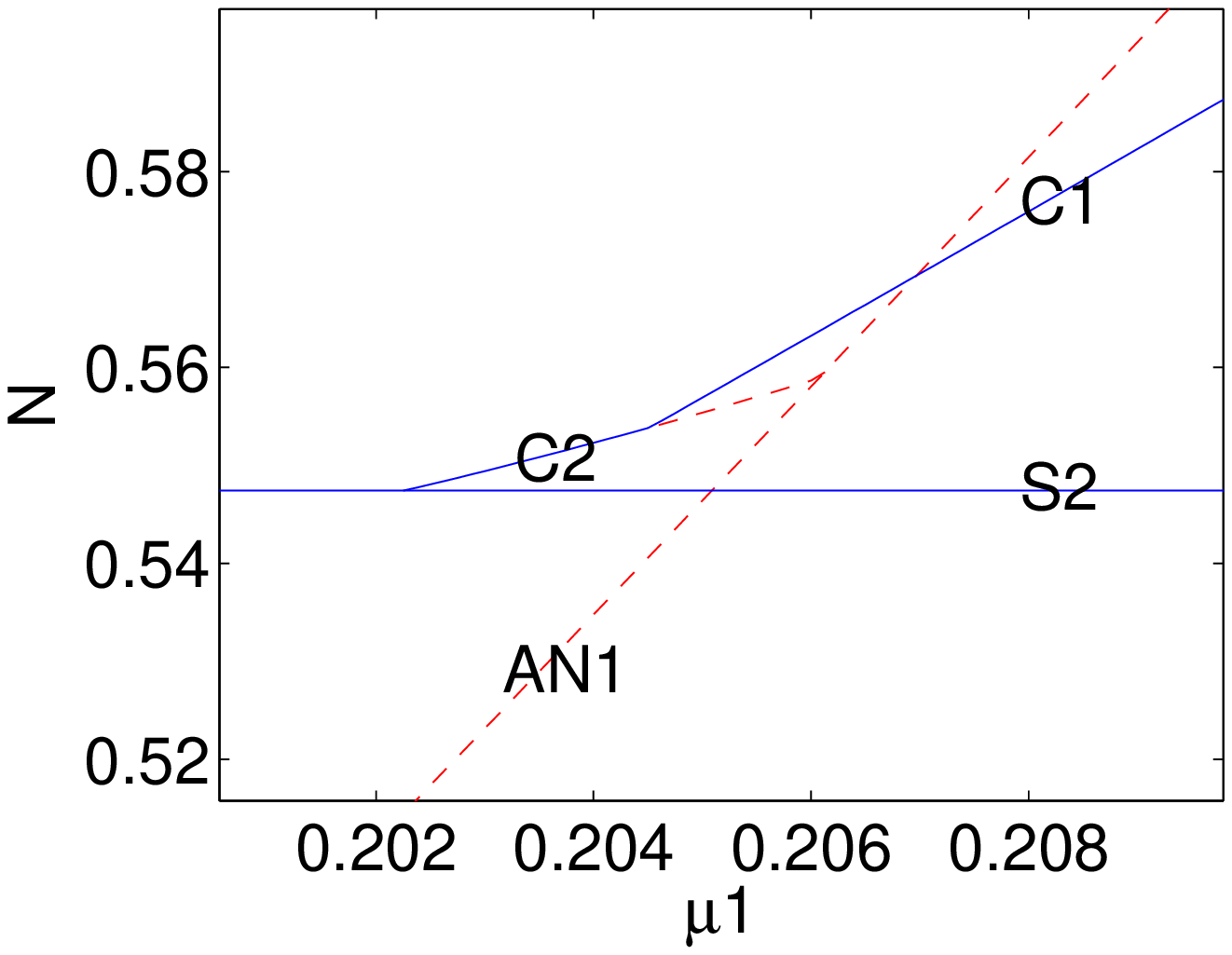} %
\includegraphics[width=.4\textwidth]{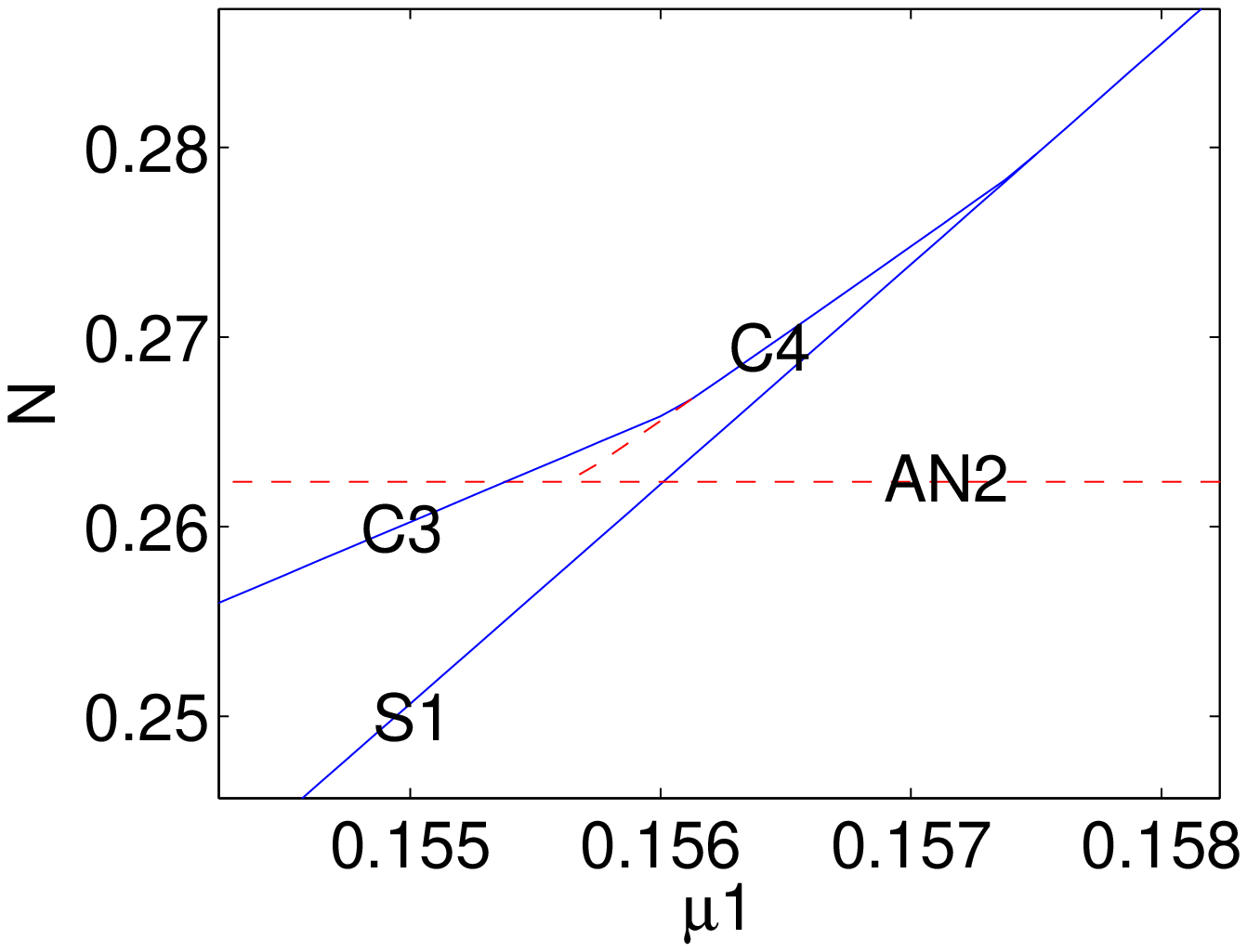}\newline
\caption{(Color online) Top panel: the norm of the numerical (left) and
approximate (right) stationary solutions to Eq.(\protect\ref{eq1}) with the
self-repulsive nonlinearity ($\protect\sigma =1$) as a function of $\protect%
\mu _{1}$ for $\protect\mu _{2}=0.18$. Bottom panel: blowups of segments of
the top left panel where bifurcations of two-components solutions happen.}
\label{fig8}
\end{figure}

\begin{figure}[tbph]
\centering
\includegraphics[width=.3\textwidth]{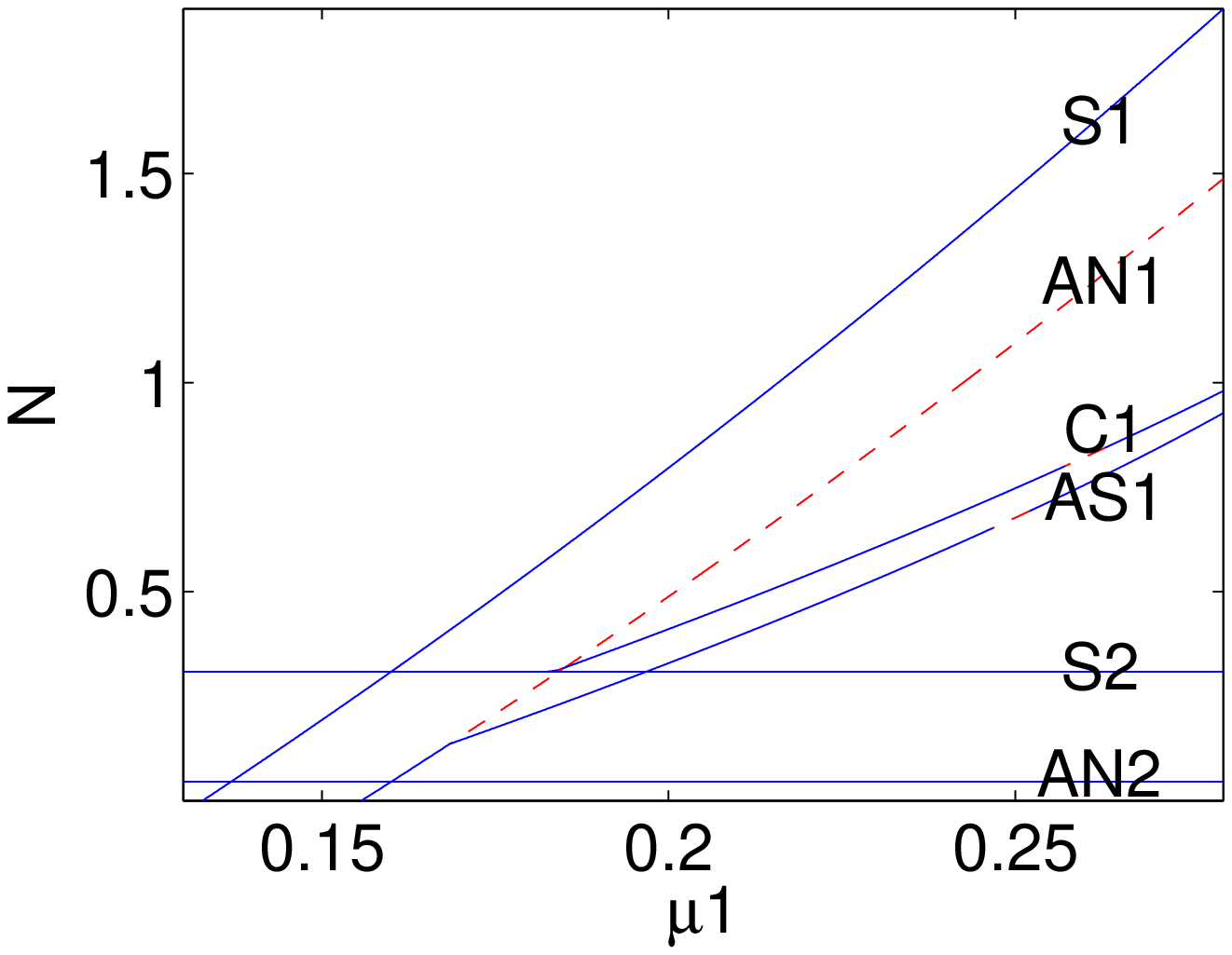} %
\includegraphics[width=.3\textwidth]{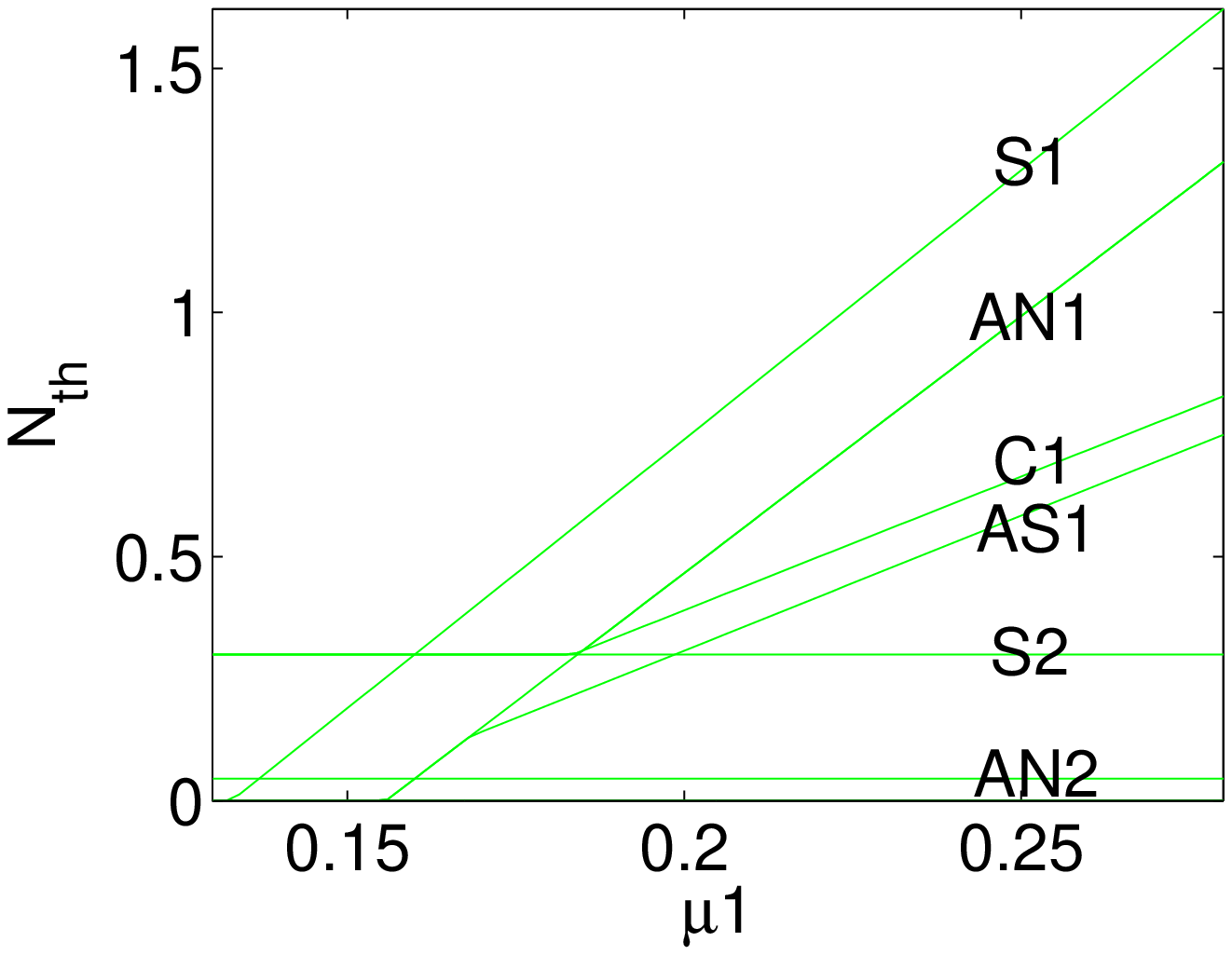}\newline
\includegraphics[width=.3\textwidth]{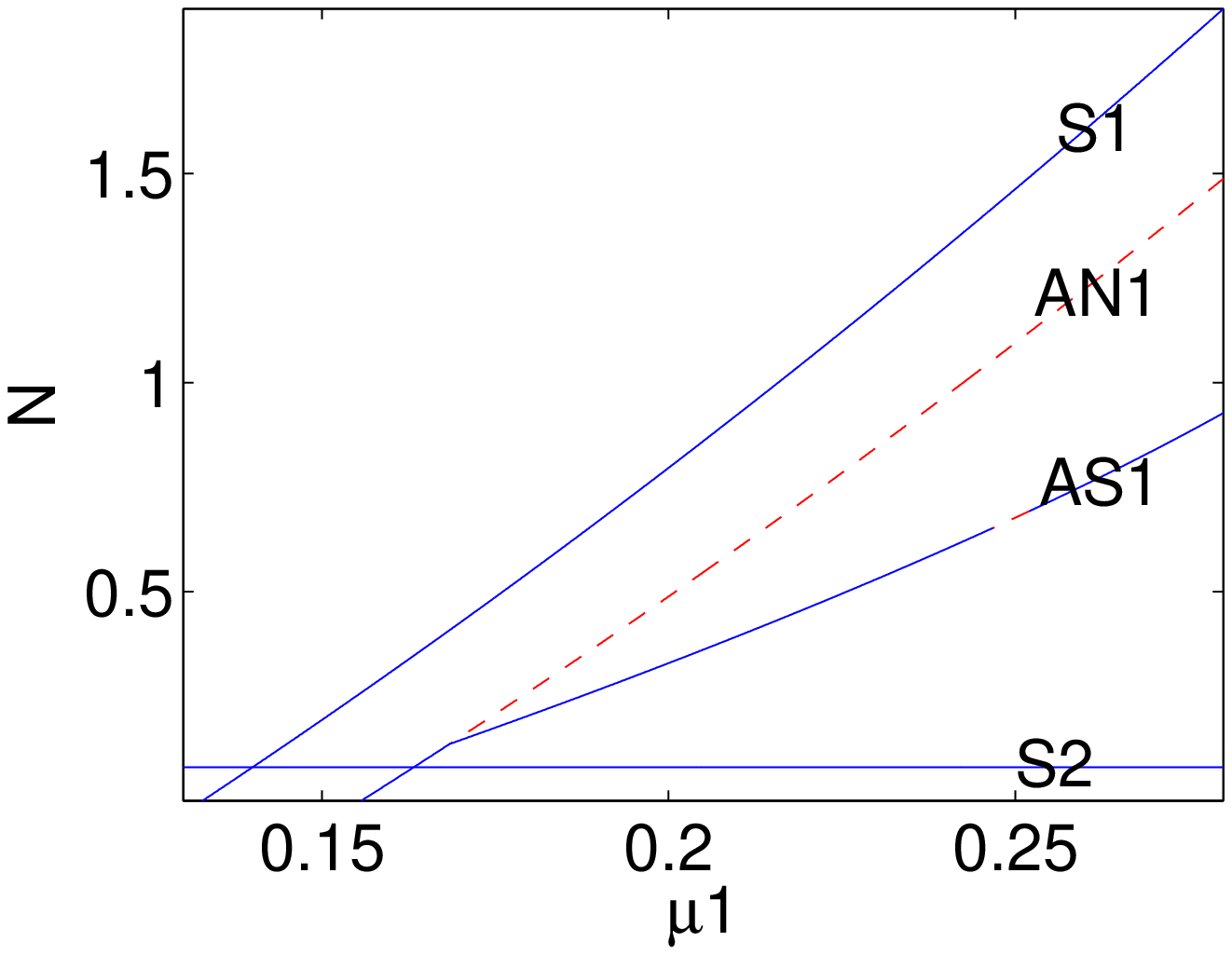} %
\includegraphics[width=.3\textwidth]{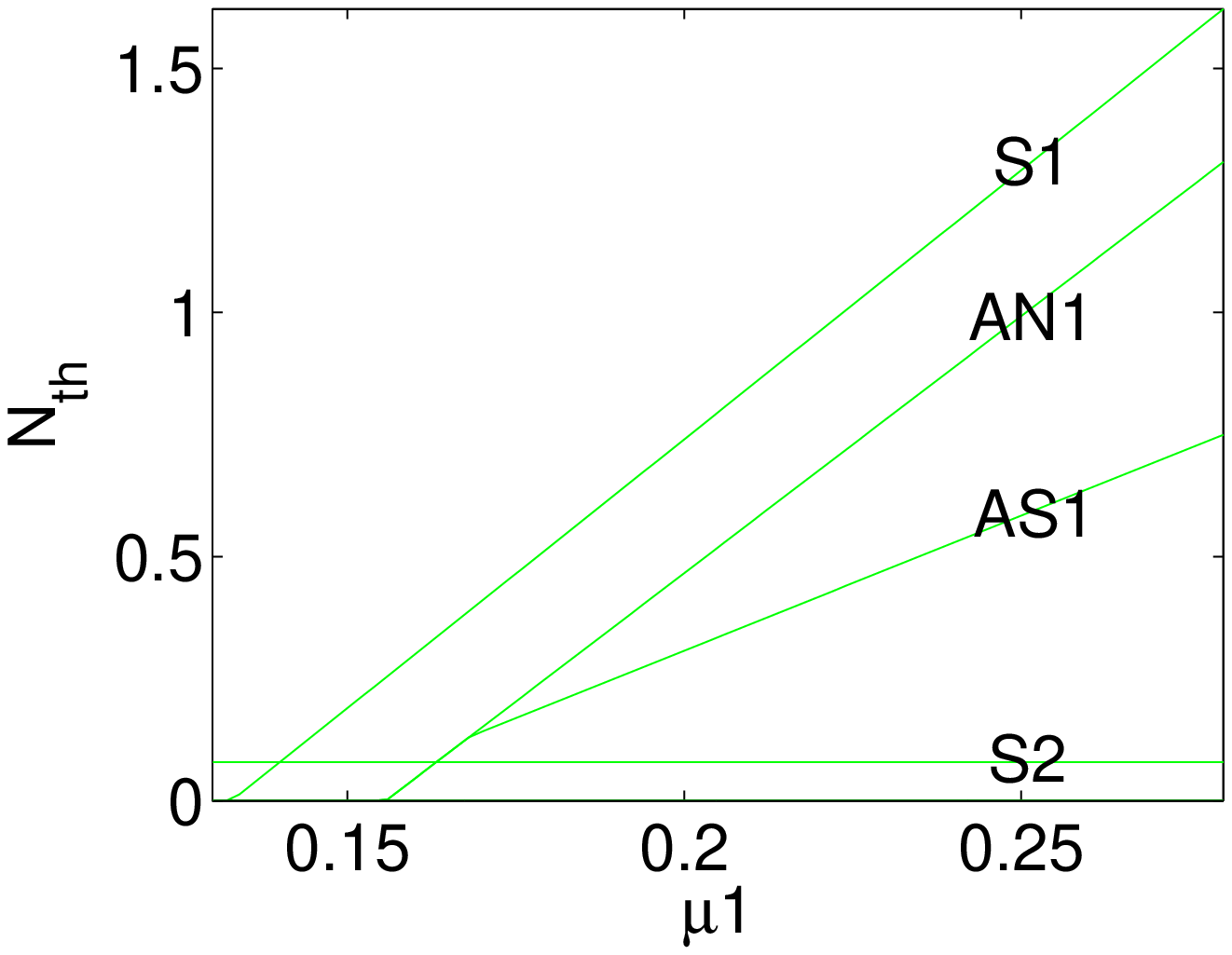}\newline
\includegraphics[width=.3\textwidth]{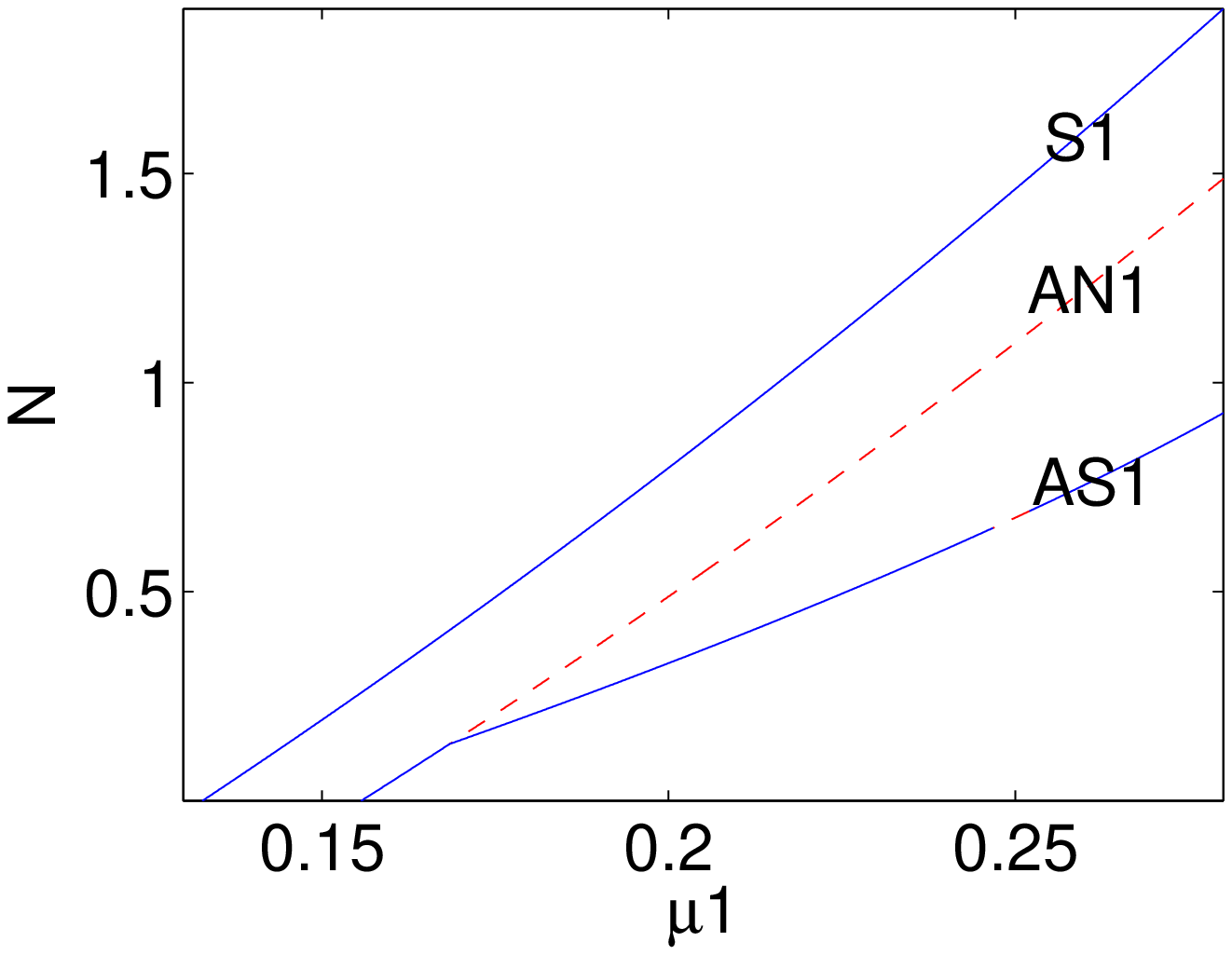} %
\includegraphics[width=.3\textwidth]{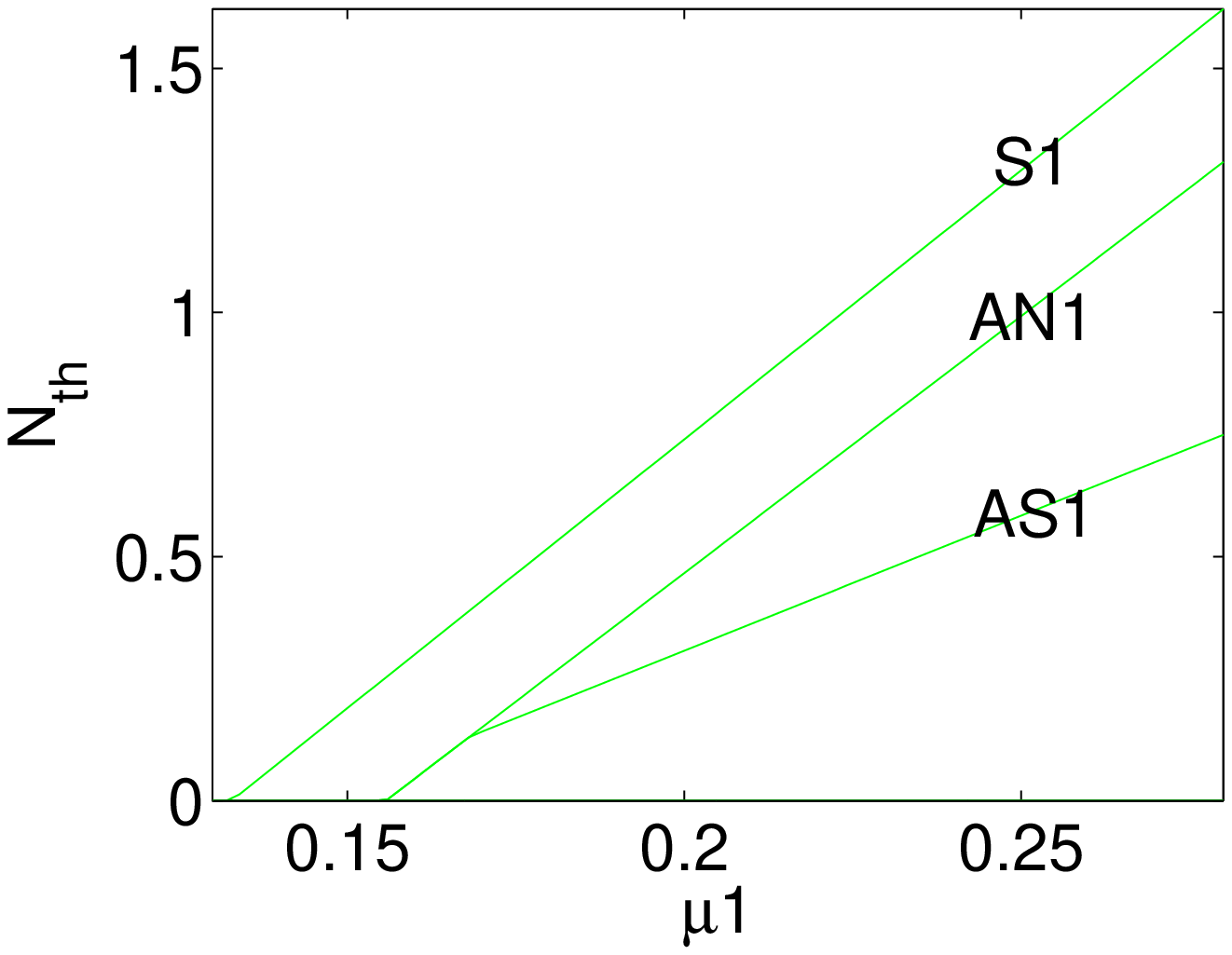}\newline
\caption{(Color online) The norm of the numerical (left) and semi-analytical
(right) solutions of Eq. (\protect\ref{eq1}) for the self-repulsive
nonlinearity ($\protect\sigma =1$) as a function of $\protect\mu _{1}$, with
$\protect\mu _{2}=0.16$ (top), $\protect\mu _{2}=0.14$ (middle), $\protect%
\mu _{2}=0.12$ (bottom). The notation is the same as in Fig. \protect\ref%
{fig8}.}
\label{fig10}
\end{figure}

We now turn to the self-defocusing case (corresponding to repulsive
interatomic interactions in BEC), with $\sigma =1$ in Eqs. (\ref{eq1}). The
results are shown in Figs. \ref{fig8}-\ref{fig10}. In this case, we do not
show profiles of solutions belonging to various branches, as they are very
similar to those obtained in the model with the self-attraction, the only
difference being that their spatial size is larger, due to the
self-repulsive nature of the nonlinearity.

We again distinguish the main regimes, namely, $\mu _{2}^{\mathrm{(cr)}}<\mu
_{2}$, when all three branches S2, AN2 and AS2 exist; $\omega _{1}<\mu
_{2}<\mu _{2}^{\mathrm{(cr)}}$, for which only S2 and AN2 exist; $\omega
_{0}<\mu _{2}<\omega _{1}$, when only S2 is present; and $\mu _{2}<\omega
_{0}$, for which there is no branch of single-component solutions in the $v$
field. The first (and most complex) of these regimes is shown in Fig. \ref%
{fig8}, in direct analogy to Fig. \ref{fig1} for the focusing case. Once
again, we observe the presence (in addition to the three single-component
branches in the $u$ field, and three such branches in the $v$ field) of four
families of the combined solutions. In Fig. \ref{fig8}, C2 (with symmetric $%
u $ and anti-symmetric $v$ components) bridges stable branch S2 and unstable
one AN1, while branch C1 of stable two-component asymmetric solutions
bifurcates from C2, making it unstable. There is also branch C4 which links
S1 and AN2, and C3, which bifurcates from C4 and terminates by merging into
AS2. Details of this picture are shown in the bottom panels of Fig \ref{fig8}%
. To highlight the accuracy of the approximation based on algebraic
equations (\ref{rho0})-(\ref{rho3}), we display the bifurcation diagram
predicted by this approximation in the top right panel of Fig. \ref{fig8},
which confirms that all the bifurcations and branches are captured within
the two-mode framework. As a measure of the agreement between the
approximate and numerical results, we again present coordinates of the
bifurcation points: AS1 emerges from AN1 at $\mu _{1}=0.1684$ according to
the numerical results, while the approximation predicts this to happen at $%
\mu _{1}=0.1682$; further, the bifurcations of C1 from C2 and C3 from C4 are
found numerically to occur at $\mu _{1}=0.2045$ and $\mu _{1}=0.1561$,
respectively, while Eqs. (\ref{rho0}) - (\ref{rho3}) predict the
corresponding values $\mu _{1}=0.2045$ and $\mu _{1}=0.1561$.

The remaining three regimes and the bifurcation scenarios predicted by the
finite-mode approximation are shown in the left and right panels of Fig. \ref%
{fig10}. The top panel corresponds to the case of $\omega _{1}<\mu _{2}<\mu
_{2}^{\mathrm{(cr)}}$, when the absence of asymmetric branch AS2 (upon
merging into which, C3 would terminate) prevents the existence of the pair
of two-component branches, C3 and C4, while C1 and C2 persist in this case.
In the middle panel, corresponding to $\omega _{0}<\mu _{2}<\omega _{1}$, we
observe that only S2 survives among the single-mode branches in field $v$,
and none of the combined solutions is present. Finally, as might be
naturally expected, no branches with $u=0$ exist at $\mu _{2}<\omega _{0}$,
hence no bifurcations of two-component branches are possible (i.e., we get
back to the one-component picture). In all these three cases, we observe
good agreement of the numerically found scenarios with the bifurcation
diagrams predicted by the two-mode approximation, which are displayed in the
right panels.

An interesting difference between the self-focusing and defocusing
nonlinearities is that, as seen in both Figs. \ref{fig8} and \ref{fig10}, in
the latter case the bifurcating branches (such as AS1 and C1) may become
unstable due to a Hamiltonian Hopf bifurcation and the resulting oscillatory
instability associated with a quartet of complex eigenvalues. This
instability, which was known in the single-component setting (see, e.g.,,
Ref. \cite{todd}), affects a short dashed (red-colored) interval within
branches AS1 and C1 in Figs. \ref{fig8} and \ref{fig10}. 
An example of such an unstable configuration and the
associated spectral plane of the stability eigenvalues are shown in Fig. \ref%
{fig15}.

\begin{figure}[tbph]
\par
\begin{center}
\begin{tabular}{ll}
\includegraphics[width=.3\textwidth]{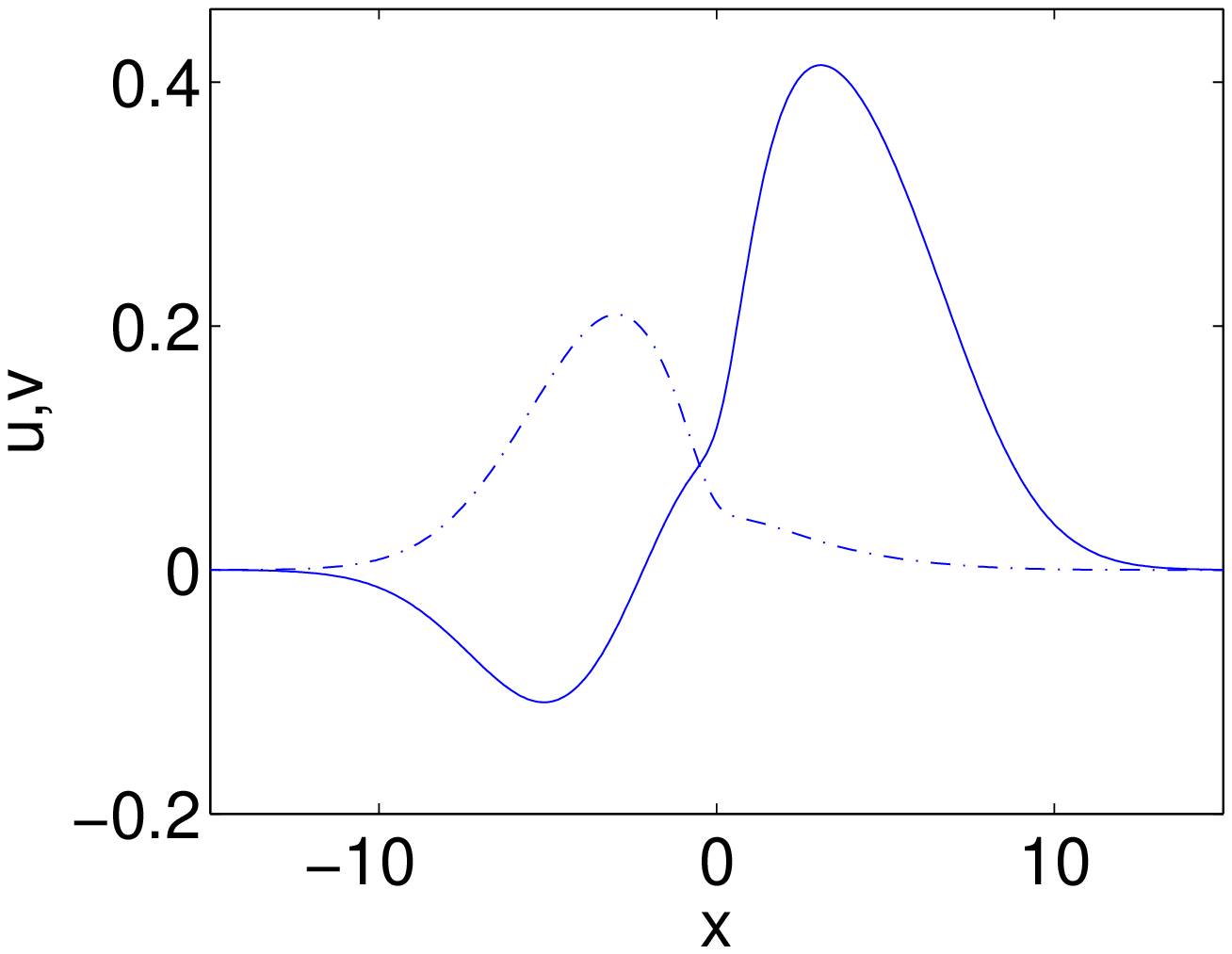} & %
\includegraphics[width=.3\textwidth]{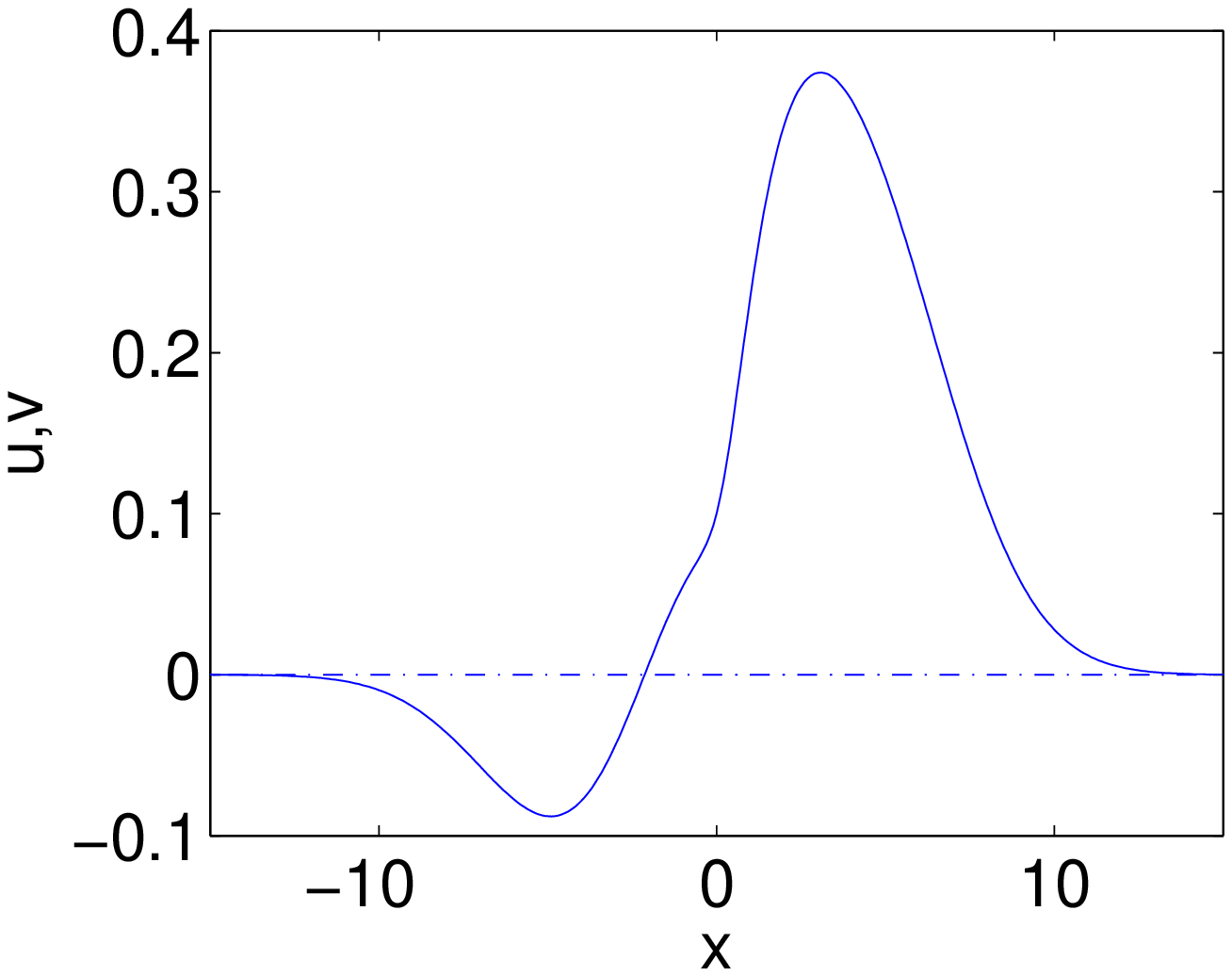} \\
\includegraphics[width=.3\textwidth]{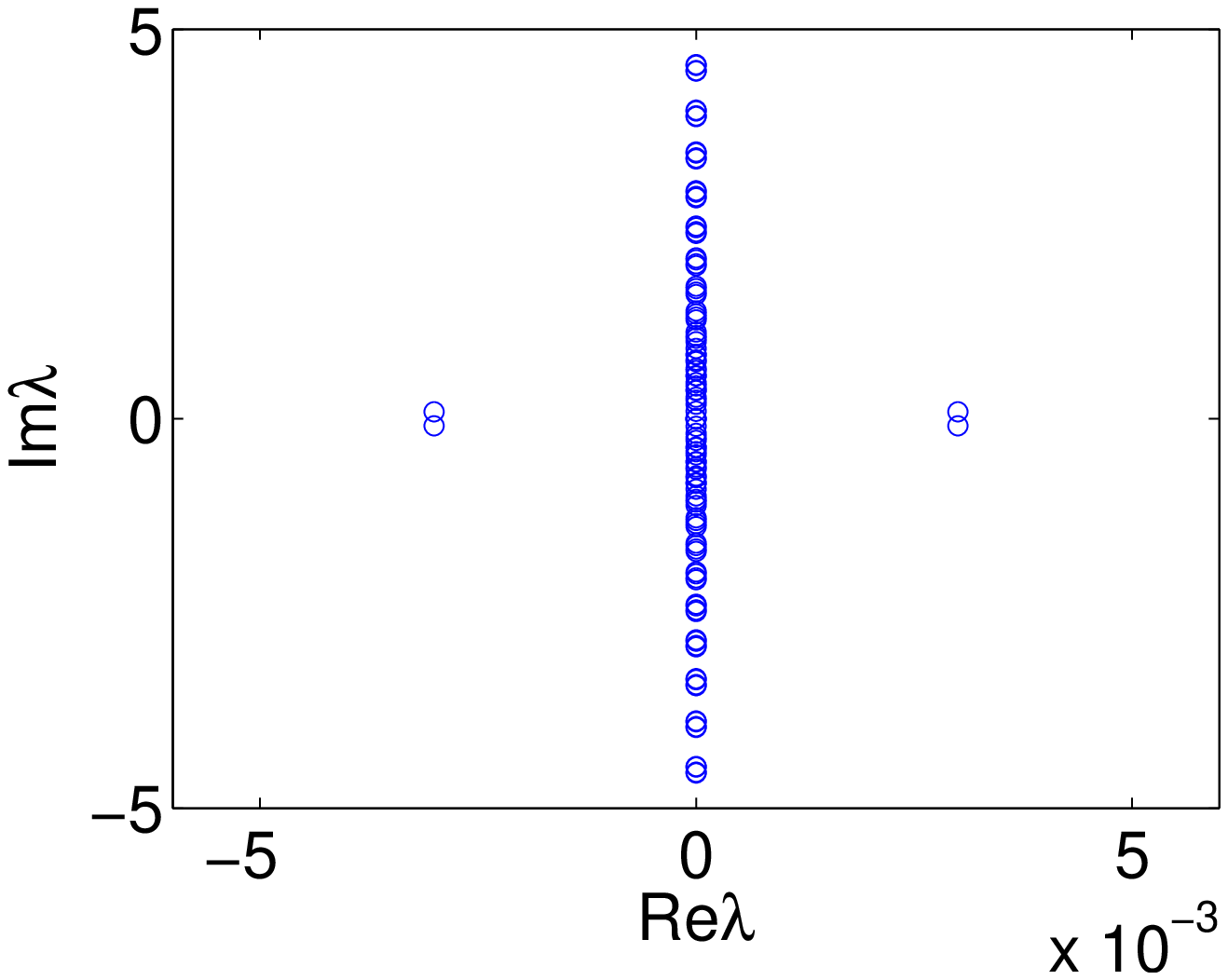} & %
\includegraphics[width=.3\textwidth]{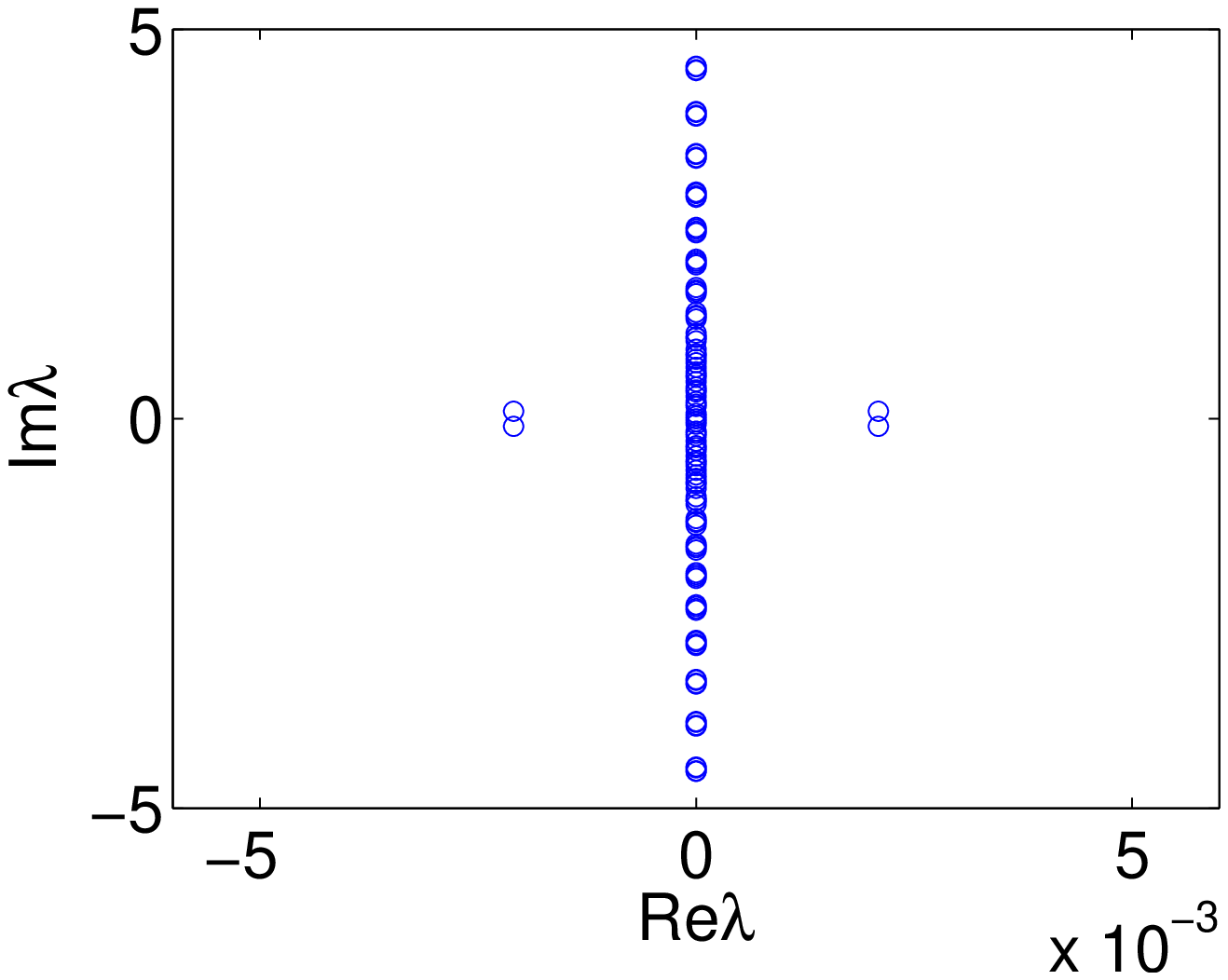} \\
&
\end{tabular}%
\end{center}
\caption{(Color online) Profiles of wave functions $u$ and $v$ (solid and
dash-dotted lines, respectively, in the top panel), and the respective
stability eigenvalues (bottom) corresponding to solutions of types C1 at $%
\protect\mu _{2}=0.276$ (left), and AS1 at $\protect\mu _{2}=0.25$ (right).
Branches C1 and AS1 are shown in Fig. \protect\ref{fig8}. The existence of
the quartet of complex stability eigenvalues implies the oscillatory
instability of the solution.}
\label{fig15}
\end{figure}

\begin{figure}[tbph]
\centering
\includegraphics[width=.4\textwidth]{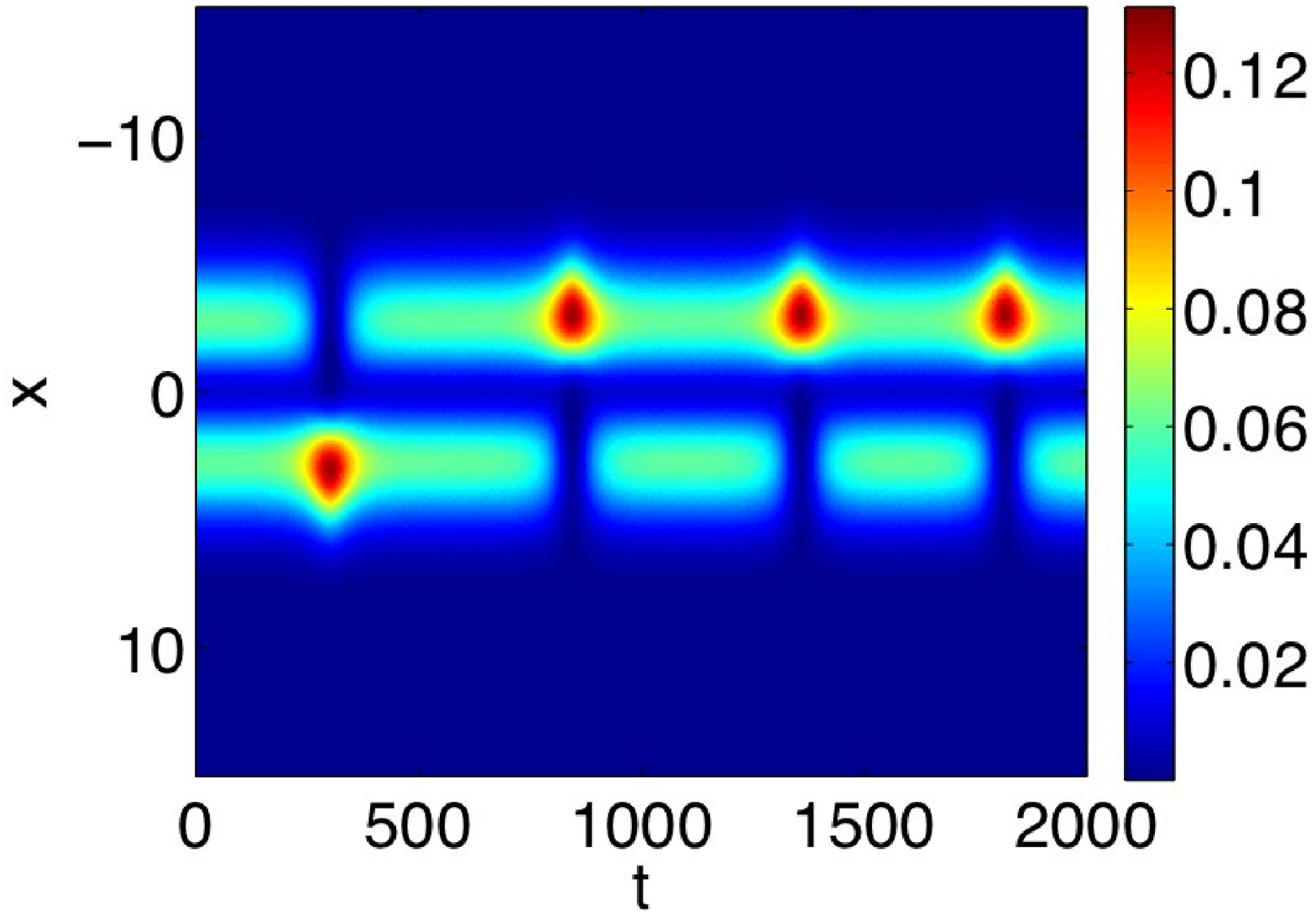} %
\includegraphics[width=.4\textwidth]{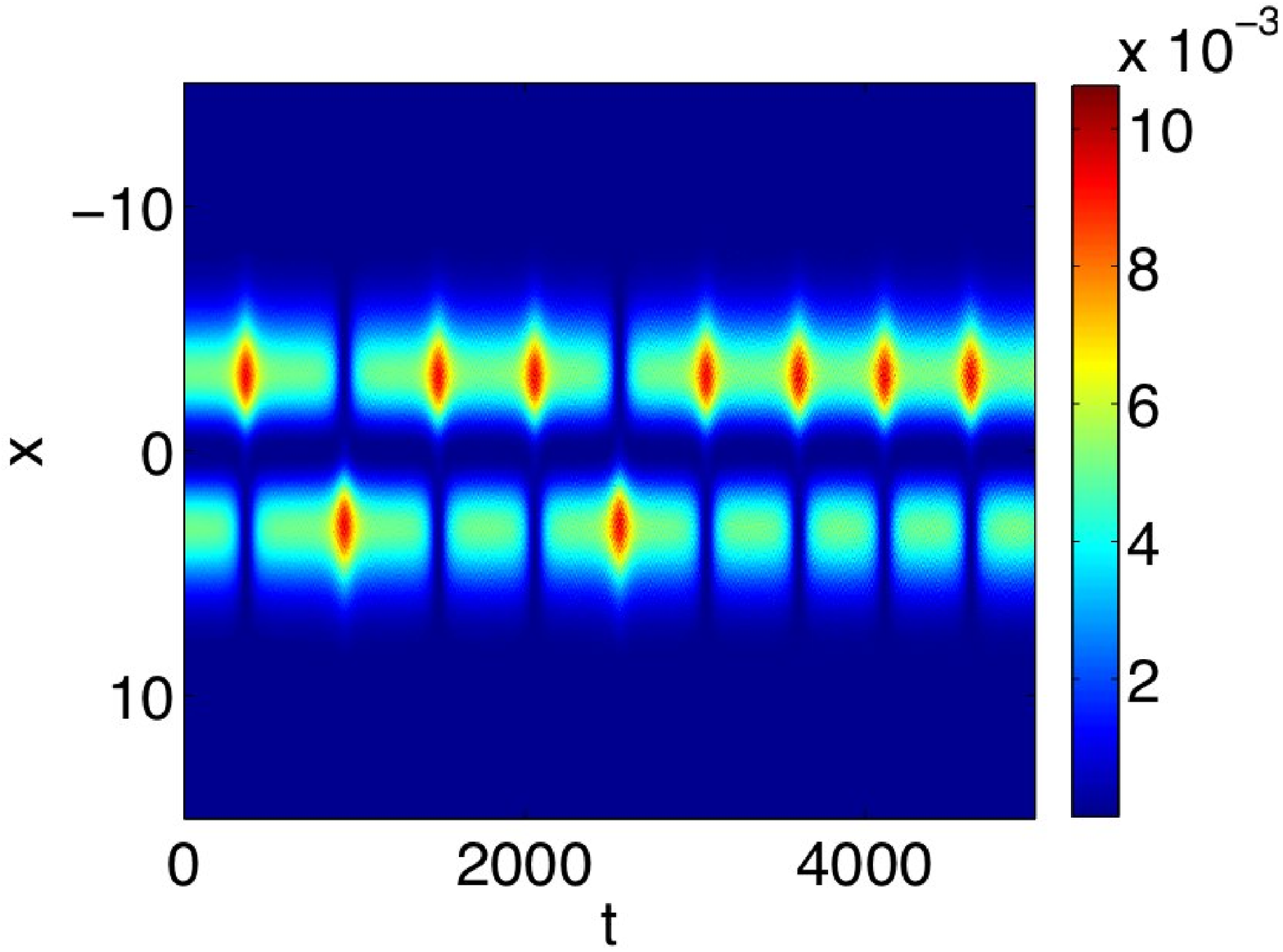}%
\newline
\includegraphics[width=.4\textwidth]{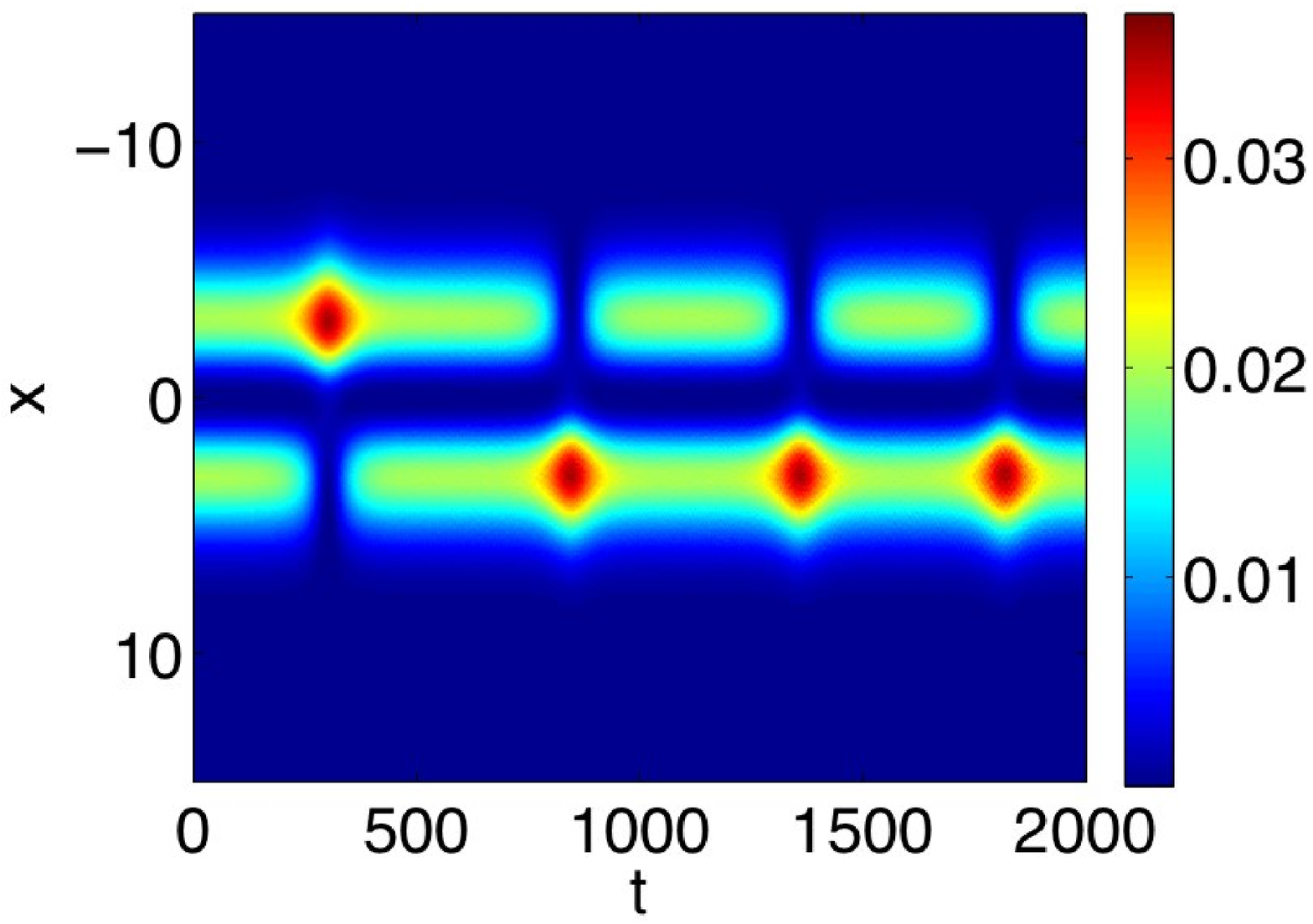} %
\includegraphics[width=.4\textwidth]{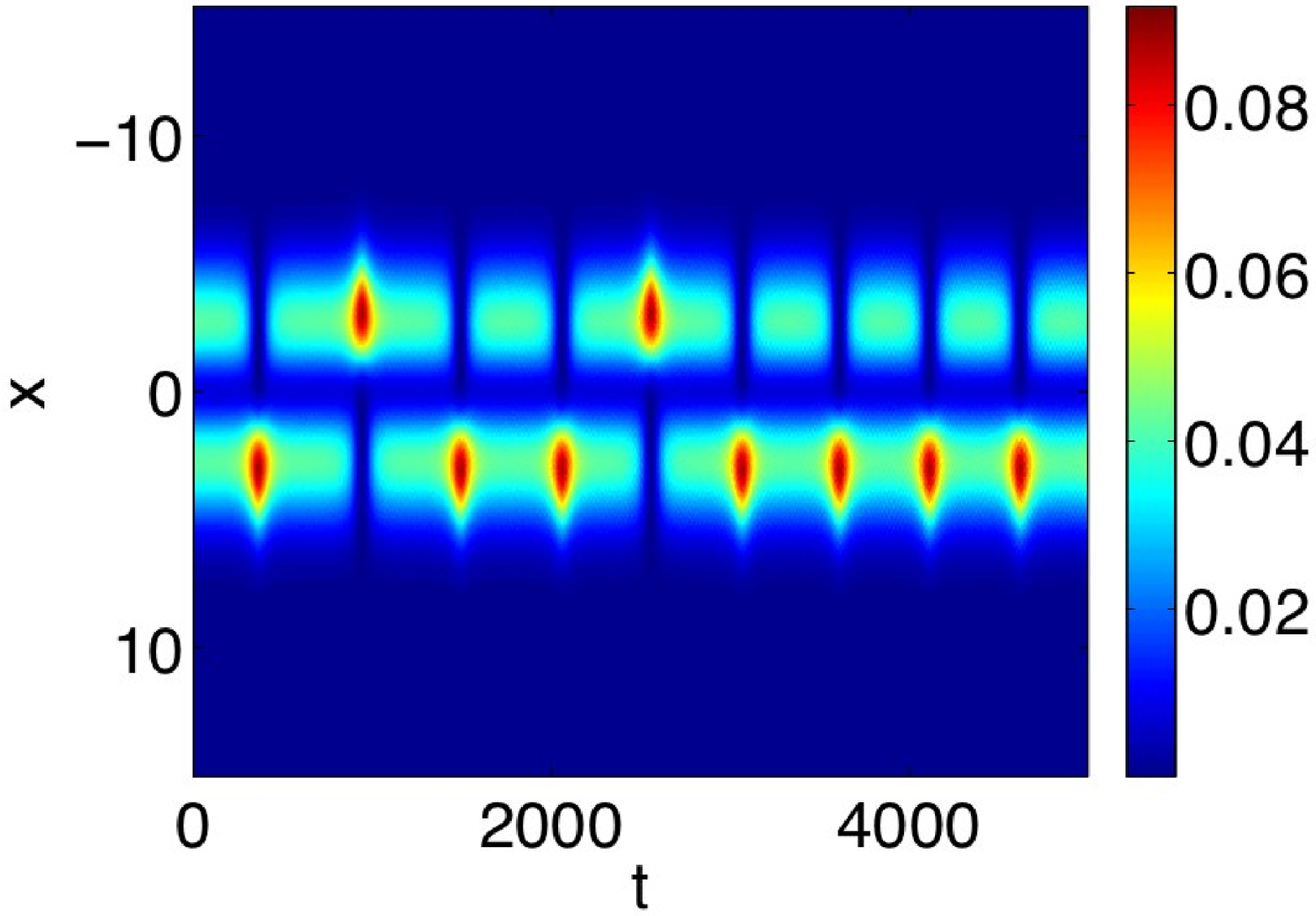}%
\newline
\caption{(Color online) Spatiotemporal contour plots of the densities, $%
|u|^{2}$ and $|v|^{2}$, of unstable two-components (combined) solutions for
the self-attractive nonlinearity ($\protect\sigma =-1$). The left and right
panels depict the simulated evolution of wave functions $u$ (top) and $v$
(bottom) in the unstable solutions of types C2 and C4, respectively, from
Fig. \protect\ref{fig1}.}
\label{fig16}
\end{figure}
\begin{figure}[tbph]
\centering
\includegraphics[width=.4\textwidth]{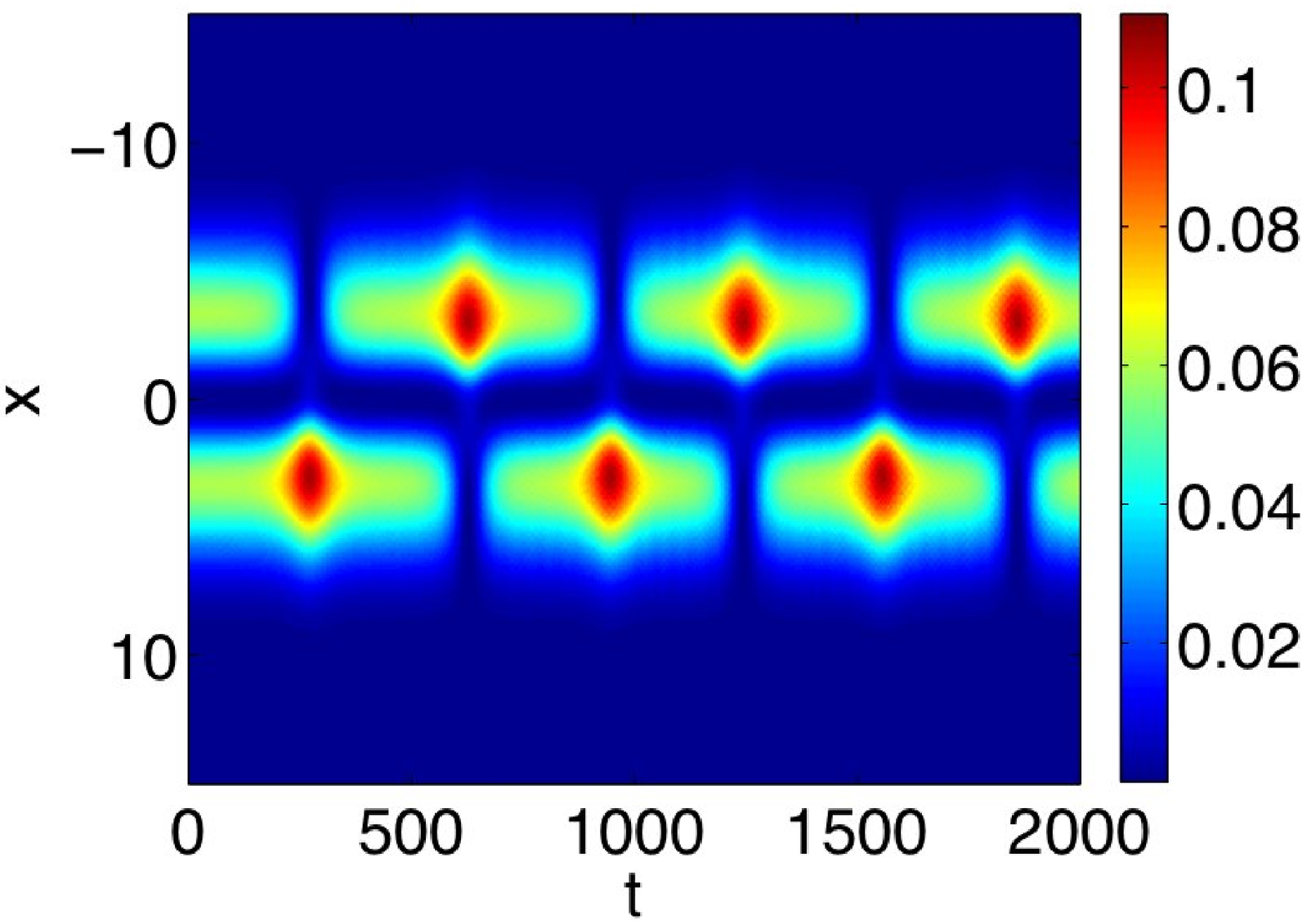} %
\includegraphics[width=.4\textwidth]{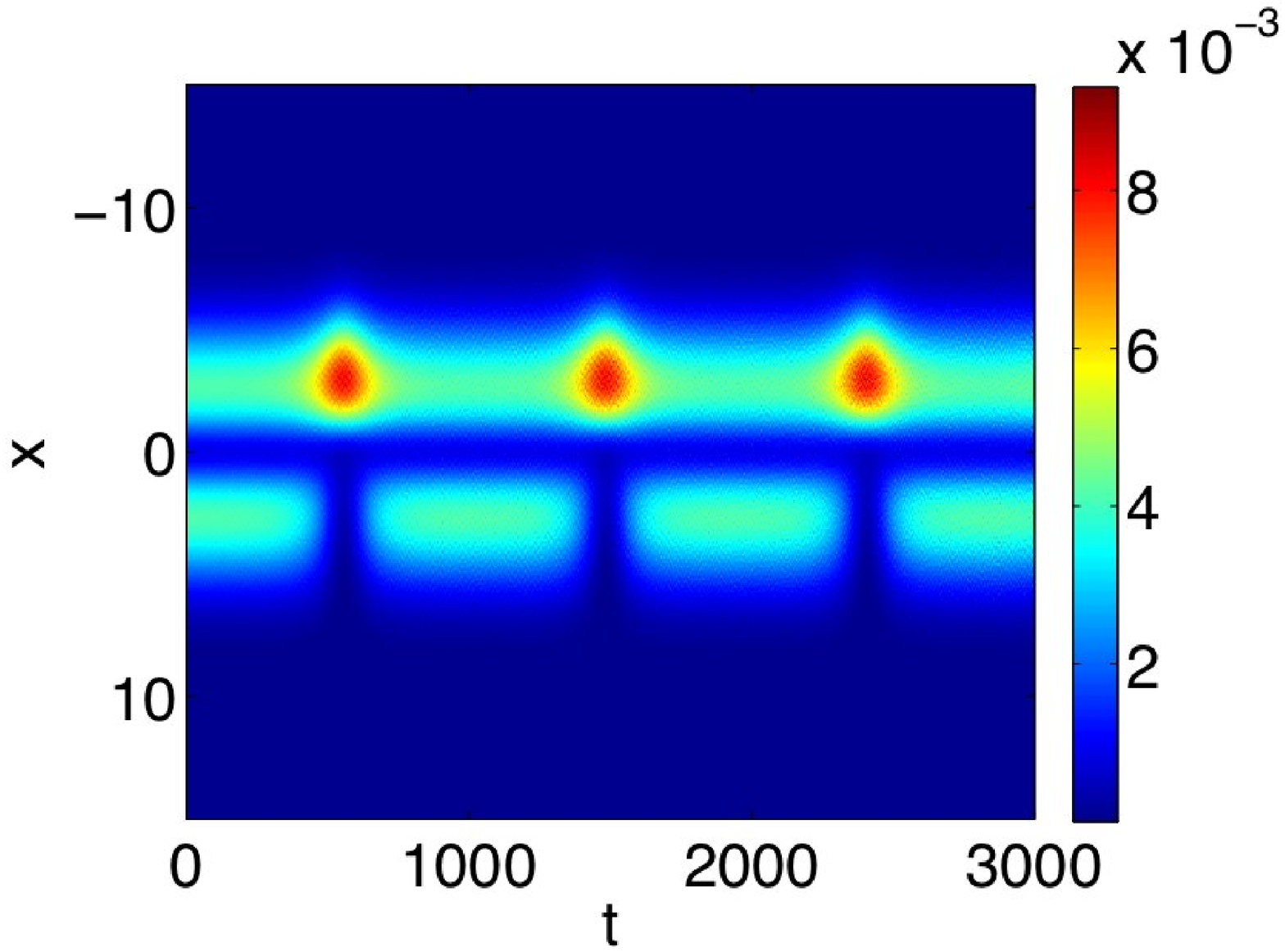}%
\newline
\includegraphics[width=.4\textwidth]{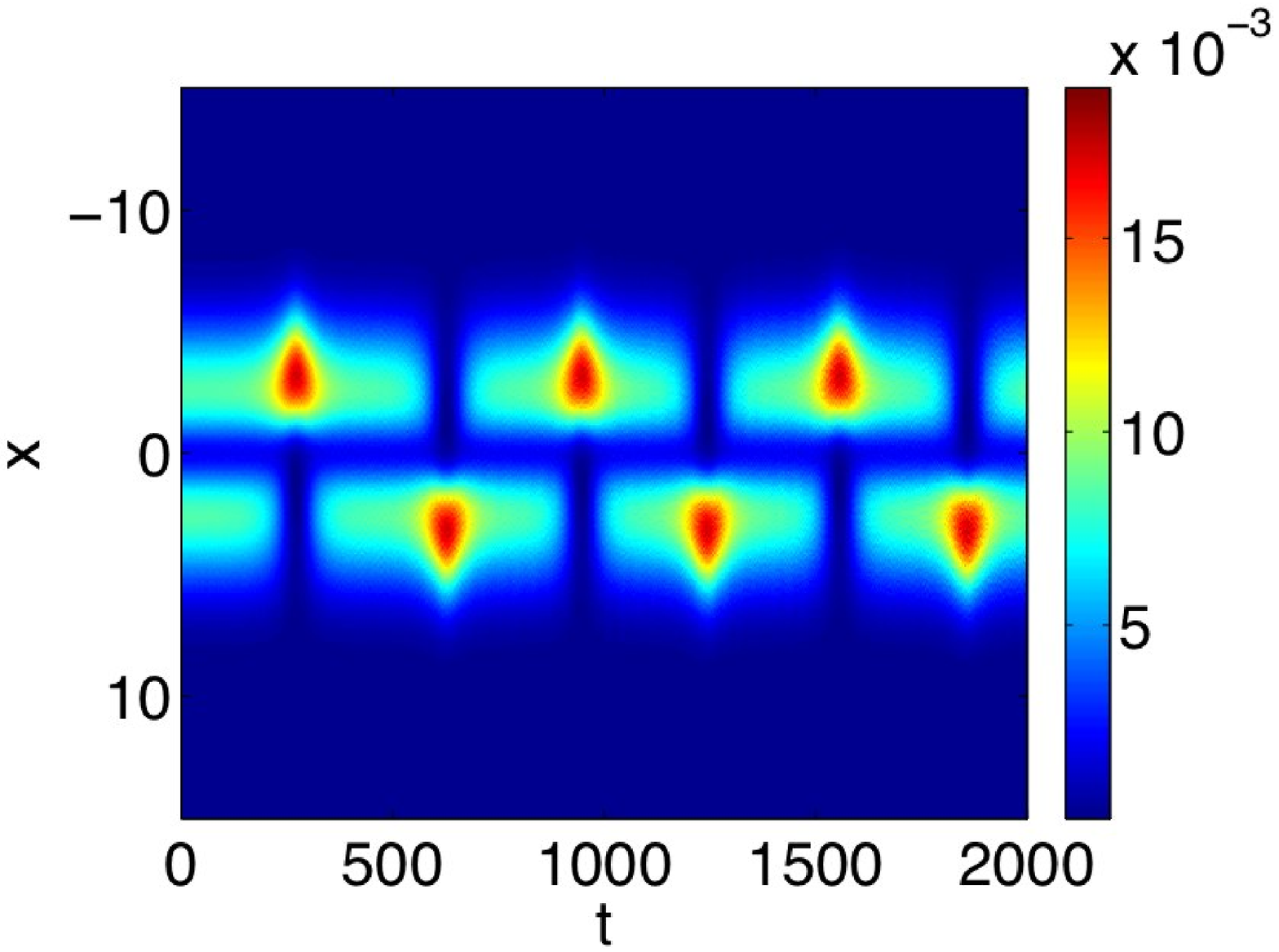} %
\includegraphics[width=.4\textwidth]{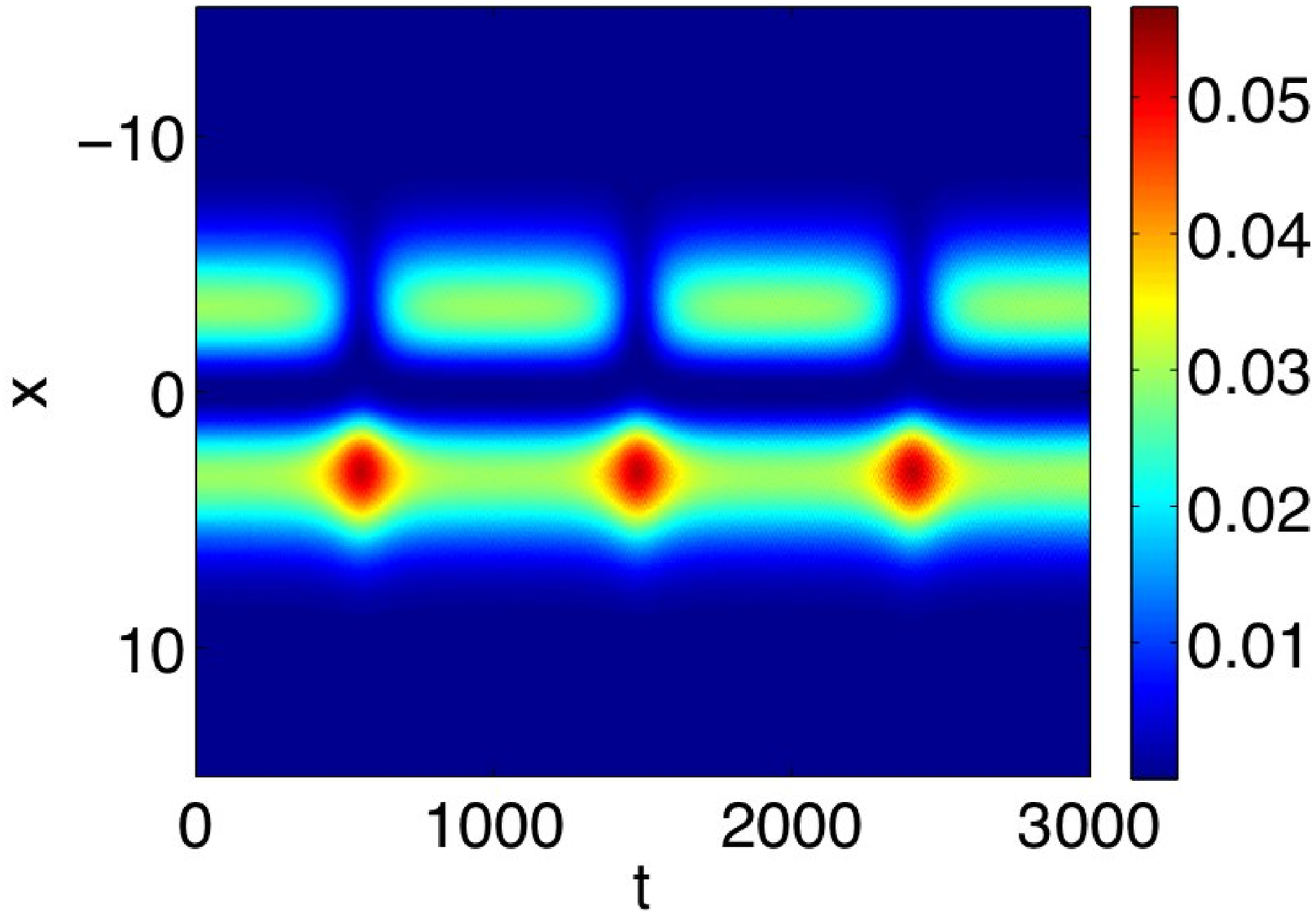}%
\newline
\caption{(Color online) The same as in Fig. \protect\ref{fig17}, but for the
self-repulsive nonlinearity ($\protect\sigma =1$). Left and right panels
show the evolution in the case of the unstable solutions of types C2 and C4,
respectively, from Fig. \protect\ref{fig8}.}
\label{fig17}
\end{figure}

Finally, we examine the dynamical evolution of the unstable two-component
solutions that were revealed by the above analysis. The evolution of a
typical unstable solution belonging to branches C2 (left panels) and C4
(right panels) past the bifurcation points (at which branches C1 and C3
emerge, respectively) is shown for the cases of the self-attractive and
self-repulsive nonlinearity in Figs. \ref{fig16} and \ref{fig17}. The main
dynamical feature apparent in the instability evolution is the growth of the
asymmetry of the wave functions between the two potential wells. In each
case, this leads to different wells trapping more atoms (or more power, in
terms of optics) in each one of the two components. For instance, in the
left panel of Fig. \ref{fig16}, we observe, at $t\approx 300$, that the
first component features a larger norm in the right well, while the second
component -- in the left one. Due to the Hamiltonian nature of the system,
the two components do not settle down into a static asymmetric
configuration, but rather oscillate between the two mirror-image asymmetric
states, which (as stationary solutions) are generated by the pitchfork
bifurcation. In particular, the instability of branch C2 causes oscillations
around the two asymmetric states belonging to branches C1, and, similarly,
the evolution of C4 leads to oscillations around the two states of type C3.

\section{Conclusions}

In this work, we have presented the phenomenology and full
bifurcation analysis of two-component mixtures trapped in the DWP\
(double-well potential). The model is of straightforward interest to
BEC and nonlinear optics (in the spatial domain). In our analytical
considerations we have developed a two-mode (in terms of each
component) approximation, that reduces the search for stationary
states to solving a set of algebraic equations. The bifurcation
diagrams obtained in this approximation, which involve all relevant
solutions, have been verified by the comparison with their
counterparts produced by the numerical solution of the full PDE
model. The novel feature of the two-component setting, in comparison
with its previously explored single-component counterpart, is the
existence of numerous branches of the ``combined" solutions that
involve both components. These branches emerge from and merge into
previously known single-component ones. The new branches may combine
a symmetric field profile in one component and an anti-symmetric one
in the other. In addition, asymmetric (in both components) combined
branches have been found too; they emerge from the
symmetric/anti-symmetric two-component states via pitchfork
bifurcations, similar to how bifurcations of the same type give rise
to asymmetric solutions in single-component models. The stability
analysis of all the considered branches confirms expectations
suggested by the general bifurcation theory, according to which the
pitchfork destabilizes the previously stable ``parent" branch, from
which the two new ones (mirror images of each other) emerge. Direct
numerical simulations of unstable symmetric two-component solutions
(past the bifurcation point) indicate that the instability leads
(quite naturally) to oscillations around the asymmetric profiles
emerging from the bifurcation.

This study can be extended in several directions. On the one hand, it would
be interesting to address this two-component setting using the phase-space
analysis, similarly to how it was done in \cite{smerzi}. Another possibility
is to develop an extension of the one-component shooting method of \cite%
{zezy1} to find an entire set of stationary solutions of the system;
this is similar to what has been done for two-component optical
lattices in the interesting, 
very recent work of \cite{zezy3} (also, many of the relevant solutions
appear quite similar to the ones obtained herein). It should nevertheless
be mentioned, in connection to \cite{zezy3}, that  in the
present problem there is a straightforward linear limit whose
eigenfunctions/eigenvalues allow
us to construct all the possible solutions that emerge in the
presence of nonlinearity.  It may
also be interesting to consider the two-component double-well system in the
presence of a spatially modulated nonlinearity in the spirit of Ref. \cite%
{zezy2}; note that both our semi-analytical and numerical approach
could be directly adapted to that setting. Finally, it would be
particularly interesting to extend this analysis to a four-well,
two-dimensional setting (which may naturally emerge from a
combination of a magnetic trap with a square 2D optical lattice in a
``pancake-shaped" BEC). Numerous additional solutions and a much
richer phenomenology may be expected in the latter case. Some of
these directions are currently under study.

\end{document}